\documentclass[12pt,preprint]{aastex}

\usepackage{amssymb}
\usepackage{graphicx}
\usepackage{subfigure}

\shorttitle{Comparison of Fitting Methods for M33 Star Clusters}
\shortauthors{Z. Fan, B.-Q. Chen, X.-Y., Pang, J.-J. Ren, S. Wang, J. Wang,
  K.-F. Tan, N. Song, C. Li, J. Zheng \& G. Zhao}

\begin{document}

\title{Comparisons of Different Fitting Methods for the Physical Parameters of A
  Star Cluster Sample of M33 with Spectroscopy and Photometry}

\author{Zhou Fan\altaffilmark{1}, Bingqiu Chen\altaffilmark{2},
  Xiaoying Pang\altaffilmark{3},Juanjuan Ren\altaffilmark{1}, Song
  Wang\altaffilmark{1}, Jing
  Wang\altaffilmark{1,4}, Kefeng Tan\altaffilmark{1}, Nan
  Song\altaffilmark{1}, Chun Li\altaffilmark{1}, Jie
  Zheng\altaffilmark{1}, Gang Zhao\altaffilmark{1,4}}

\altaffiltext{1}{CAS Key Laboratory of Optical Astronomy, National
  Astronomical Observatories, Chinese Academy of Sciences, Beijing
  100101, China}

\altaffiltext{2}{South-Western Institute For Astronomy Research,
  Yunnan University, Chenggong District, Kunming
  650500, China}

\altaffiltext{3}{Xi'an Jiaotong-Liverpool University, 111 Ren'ai Road
Suzhou Industrial Park, Suzhou, Jiangsu Province, 215123, China}

\altaffiltext{4}{School of Astronomy and Space Science, University of
  Chinese Academy of Sciences, Beijing 100049, China}

\email{zfan@bao.ac.cn}

\begin{abstract}
  Star clusters are good tracers for formation and evolution of
  galaxies. We compared different
  fitting methods by using spectra (or by combining photometry) to
  determine the physical parameters. We choose a sample of 17
  star clusters in M33, which previously lacked spectroscopic observations.
  The low-resolution spectra were taken with the Xinglong 2.16-m
  reflector of NAOC. The photometry used in the fitting includes
  $\rm u_{SC}$ and $\rm v_{SAGE}$ bands from the SAGE survey, as well
  as the published $UBVRI$ and $ugriz$
  photometry. We firstly derived ages and metallicities with the {\sc ULySS}
  (Vazdekis et al. and {\sc pegase-hr}) SSP model and the Bruzual \& Charlot
  (2003) (BC03) stellar population synthesis models for the
  full-spectrum fitting. The fitting results of both the BC03 and {\sc
    ULySS} models seem consistent with those of previous works as
  well. Then we add the SAGE $\rm u_{SC}$ and $\rm v_{SAGE}$ photometry in the
  spectroscopic fitting with the BC03 models. It seems the results
  become much better, especially for the Padova 2000+Chabrier IMF
  set. Finally we add more photometry data, $UBVRI$ and $ugriz$,
  in the fitting and we found that the results do not improve
  significantly. Therefore, we conclude that the photometry is useful
  for improving the fitting results, especially for the blue bands
  ($\lambda <4000$ {\AA}),
  e.g., $\rm u_{SC}$ and $\rm v_{SAGE}$ band. At last, we discuss the
  ``UV-excess'' for the star clusters and we find five star clusters
  have UV-excess, based on the $GALEX$ FUV, NUV photometry.
\end{abstract}


\section{Introduction}
\label{intro.sec}

Over the last decades, study of the Local Group (LG) has become more
and more important and exciting as near-field cosmology and galactic
archeology have made great achievements. One of the most important
discoveries is confirmation of the future collision and merger of
Andromeda (M31) and the Milky Way (MW) from {\sl
  Hubble Space Telescope} ({\sl HST}) data \citep[see,
e.g.,][]{vdm12}, and the latest results including Gaia's DR2 data
\citep{vdm19}. In addition, the Pan Andromeda Archeological Survey
(PAndAS) \citep{mccon09} discovered a large number of stellar
streams and substructures in the M31 halo. Numerous giant stellar
streams (e.g., northwest stream, Richardson et al. 2011) and
substructures have been discovered. With deep observations of the
Canada Canada-France-Hawaii Telescope (CFHT) $g = 26.0$ and $i =
24.8$), the furthest streams are located up to $\sim150$ kpc from
the M31 center \citep{mac19}. A large number of star clusters and
dwarf galaxies have also been discovered in the halo of M31 up to
150 kpc. Recently, \citet{mac19} systematically investigated the
density map of M31 with PAndAS. They found a relation between the
bright substructures in the metal-poor halo field and positions of
star clusters. At the same time, M31 has been having strong
interactions with M33 since 3.4 Gyr ago \citep{mccon09}. Therefore,
study of star cluster systems associated with M33 is also important
for the interaction and evolution of the M31-M33 system.

The integrated light (IL) spectroscopy is a useful and powerful tool
for the analysis of star clusters in extra-galaxies. The
astrophysical parameters, e.g., ages, chemical abundances (including
the [$\alpha$/Fe]), velocities and masses can be derived with the
stellar synthesis models, which provide crucial information about
their host galaxies \citep{sak19}. If the halo star clusters are
spatially related to the star stream or substructures of M31, the
study of these halo star clusters can reveal information about the
nature of interactions between M31 and M33. \citet{fy14} described
how to trace the substructures and the interaction of M31-M33 with
the associated star clusters via spectroscopic observations of the
Xinglong 2.16-m telescope \citep{fan11,fan12} and the Multiple
Mirror Telescope (MMT) 6.5-m telescopes \citep{fan16a}. The studies
of \citet{chen15,chen16} derived stellar parameters (e.g., radial
velocities, ages, metallicities, and masses) for star clusters in
M31, using the LAMOST (Large sky Area Multi-Object fiber
Spectroscopic Telescope) low-resolution spectroscopic survey with
R$\sim1800$ from 3700 {\AA} to 9100 {\AA} in wavelength
\citep{zhao12,cui12}.

Many different spectroscopic fitting techniques have been developed
to increase the accuracy of parameters derivation.  One of these
involves full-spectral fitting, e.g., using {\sc ULySS}
\citep{kol09,chen16}. Alternatively, one could pursue $\chi^2_{\rm
min}$ fitting of various Lick/IDS indices
\citep[e.g.,][]{fan11,fan12,chen16}. Note that the results and
precision are model-dependent. Inclusion of more useful information
obviously leads to higher precision. Therefore, $\chi^2_{\rm min}$
fitting combined with SED and Lick-index fitting is expected to
provide more reliable and higher-precision results than any of the
individual approaches \citep{fan16a}. \citet{lfd09} have also done
similar work for the globular cluster systems of NGC5128 and they
found a population of intermediate age and metal- poor clusters for
the first time.

On the other hand, the $\chi^2_{\rm min}$ SED-fitting of globular
clusters' spectral energy distributions (SEDs) is another efficient
method to determine the parameters on the basis of multi-passband
photometry/imaging data. \citet{rdg03} derive the ages,
metallicities, and reddening of the star clusters associated with
NGC 3310 by the SED-fitting method with the photometry of the
ultraviolet (UV), optical, and near-infrared (NIR) observations
obtained with the HST.
\citet{fan06,ma07,ma09,ma11,ma12,wang10,wang12} have done a series
of SED-fitting works targeting M31 star clusters, based on the
Beijing--Arizona--Taiwan--Connecticut (BATC) multi-color photometry
system, with a 60/90cm Schmidt telescope. They applied the simple
stellar population (SSP) models, \citet[][henceforth BC03]{bc03} and
the Galaxy Evolutionary Synthesis Models \citep[{\sl
  GALEV};][]{lf06,kot09}. To achieve a higher precision, they make use
of multi-band photometry, such as the broad-band $UBVRI$ filters,
$JHK$ bands from the Two Micron All Sky Survey (2MASS), the near-UV
(NUV) and far-UV (FUV) channels of the Galaxy Evolution Explorer ({\sl
  GALEX}), as well as the $ugriz$ bands of the Sloan
Digital Sky Survey (SDSS).

The recent SAGE (Stellar Abundances and Galactic Evolution) survey
\citep[PI: Gang Zhao , see,
e.g.,][]{fan18,zheng18,zheng19a,zheng19b} covers 12,000 deg$^2$ of
the northern sky with eight photometric bands. Its $\rm v_{SAGE}$
filter is self-designed. The project will present one of the largest
catalogs of (a few hundred million) stars with available stellar
atmospheric parameters. The $\rm u_{SC}$, $\rm v_{SAGE}$ photometry
is sensitive to the metallicity [Fe/H] and surface gravity log $g$,
Thus far, the observations of $\rm u_{SC}$, $\rm v_{SAGE}$, $\rm
gri$-bands have almost been completed. The data can be used to
derive the age and metallicity of the stellar populations in
M31/M33, as well as precise interstellar extinctions.

Theoretically, spectroscopy provides much more information than the
corresponding photometry. However, in many cases, the precision of
the spectral flux calibration is much lower than that of photometry,
especially for the blue bands (e.g., $\lambda<4000$ {\AA}) in some
cases, for the limit of observing condition. Thus the  photometry
could be quite useful for improving the precision of the flux
calibration of the spectra. Given that the wavelength coverage of
the spectrum is not wide enough, especially for blue bands, the
photometry seems much more important and can be served as a
complement for the spectroscopy.

In our work, we derive the ages and metallicities of our M33 star
cluster samples with different fitting methods based on the spectra
and photometric data. The spectra were taken from the BFSOC
spectrograph on the Xinglong 2.16-m telescope, while the photometric
data are from the SAGE survey in $\rm u_{SC}$, $\rm v_{SAGE}$ bands,
and the literature in $UBVRI$ and $ugriz$ bands. For comparison,
both the {\sc
  ULySS} models and BC03 models have been applied in the fitting. The
organization of the paper is as follows. In Section \ref{sam.sec},
we describe the selection of sample and introduce the methods
adopted for the fits. In Section \ref{obs.sec} we describe the
observation details for both 2.16-m telescope of Xinglong as well as
that of the SAGE survey; In Section \ref{specfit.sec} the
full-spectrum fitting is described, which provides the best-fitting
results compared to the {\sc ULySS} models and the \citet{bc03}
models; in Section \ref{sedspecfit.sec}, we introduce the fitting
process of the Spectrum-SED fitting, based on $\chi^2_{\rm min}$
fitting with Padova 1994/2000 evolutionary tracks and \citet{chab}/
\citet{salp} IMFS of the BC03 models; different data sets have been
fitted separately; in Section \ref{dis.sec}, we compare the $GALEX$
FUV/NUV data with the SAGE $\rm u_{SC}$ and the Johnson-Cousins
$U$-band, and found some UV excess candidates. Finally, the summary
and concluding remarks are given in Section \ref{sum.sec}.

\section{The Selection of Star Cluster Sample in M33}
\label{sam.sec}
M33 (Triangulum Galaxy) is the third largest
spiral galaxy in our Local Group (LG). The distance of M33 from us
is $847\pm60$ kpc, which corresponds to a distance modulus of
$\rm (m-M)_0 = 24.64 \pm0.15$ mag \citep{gall04}.
The sources were selected from Table~3 of \citet{sara07}, which
contains 451 star cluster candidates in M33. In this work, we select
17 confirmed and luminous clusters ($V<17.5$) as our sample, which
is suitable for spectroscopic observations with a 2-meter class
telescope. These star clusters lack spectroscopic observational
data, especially the metallicity measurements. Thus, it is necessary
to observe the spectra of these sample clusters systematically and
constrain the spectroscopic metallicities and ages in detail.

The observational information of our sample star clusters is listed
in Table~\ref{t1.tab}, which includes the IDs, which are the same as
\citet{sara07}, coordinates, observation dates and exposures. All
the coordinates (R.A. and Dec. in Cols. 2 and 3) are from
\citet{sara07} . The star clusters are sorted by $V$-mag.

\begin{figure}
    \resizebox{\hsize}{!}{\rotatebox{0}{\includegraphics{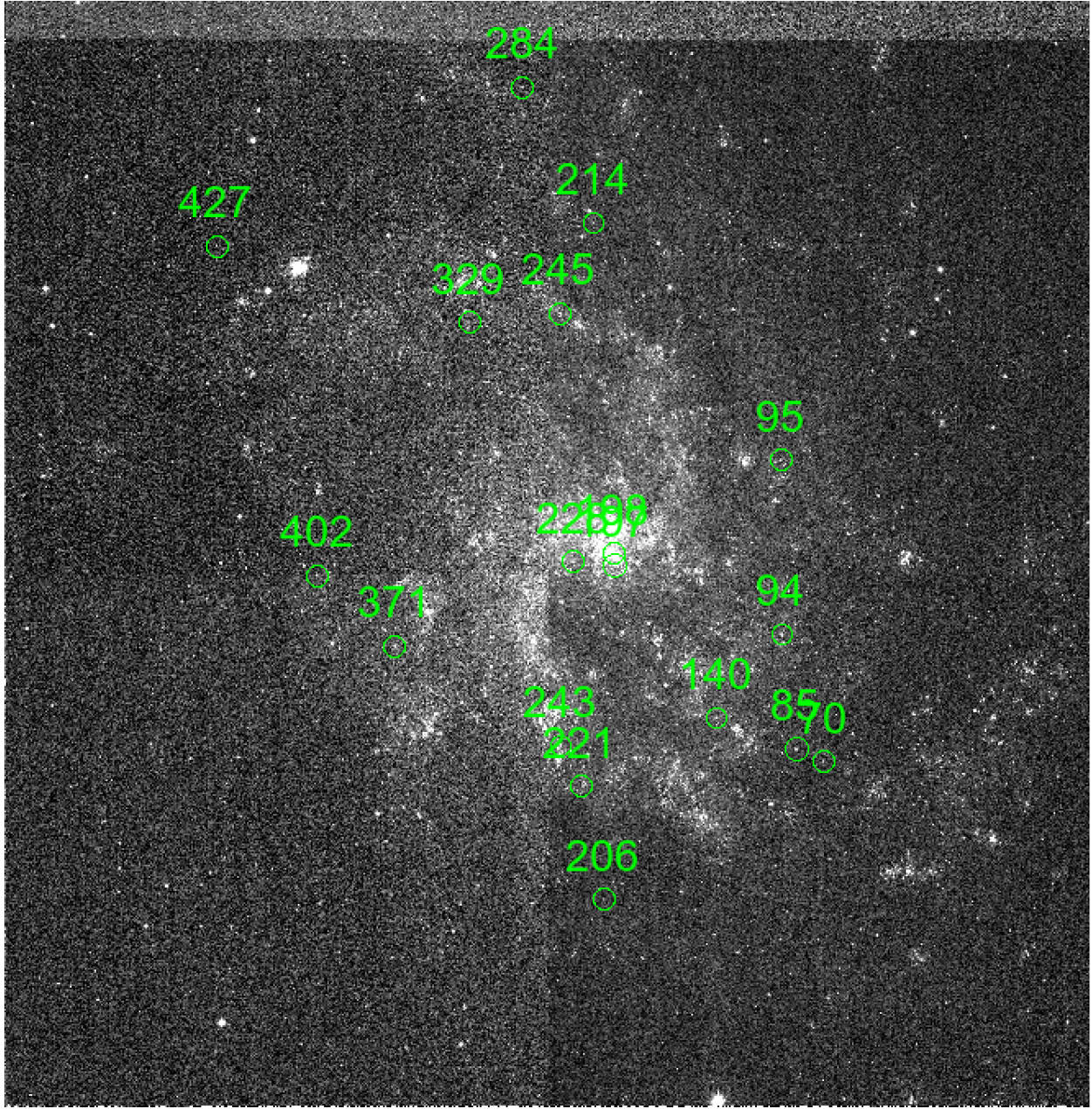}}}
    \caption{The spatial distribution of our M33 star cluster sample.}
  \label{fig1}
\end{figure}

Figure~\ref{fig1} shows the spatial distribution of the sample star
clusters (green circles) in M33. Cluster names are in
Table~\ref{t1.tab}. The images are from SAGE $\rm v_{SAGE}$-band
observations with the Bok 2.3-m (90-inch) telescope of Steward
Observatory, University of Arizona. A four 4k$\times$4k blue
sensitive CCD mosaic is mounted and for each CCD there are four
amplifiers.  The field of view (FoV) is $\sim1$ deg$^2$.

\section{The Spectroscopic Observations and Data Reduction}
\label{obs.sec}

The low-resolution spectroscopic observations were carried out in
2015 with the Xinglong 2.16-m reflector \citep{fan16b} Beijing Faint
Object Spectrograph and Camera (BFOSC) instrument during October 9
to 11, and October 19. The telescope is located in Xinglong
Observatory, National Astronomical Observatories, Chinese Academy of
Sciences (NAOC), in Hebei Province at an altitude of $\sim$900 m.
Most clusters are exposed for 3600 seconds except for SM197 (4200
seconds) due to weather conditions (Table~\ref{t1.tab}). The seeing
was $\sim2''$ to $3"$. We adopted a slit with a width of $1.8"$,
with the grism G4. The first order dispersion is 4.45{\AA}
pixel$^{-1}$ and the wavelength coverage is 3850–7000{\AA}. The
spectral resolution was $R=620$ for slit of $0.6''$ and a central
wavelength of 5007{\AA} \citep{fan16b}. The E2V 55–30–1–348
back-illuminated 1242$\times$1152 pixels$^2$ CCD, AIMO was
installed. The pixel size is 22.5 $\mu$m and pixel scale is
$0''.457$. The gain is 1.08 e$^{-}$ ADU$^{-1}$, with a readout noise
(RN) of 2.54 e$^{-}$. The FOV is $9'.46\times8'.77$ according to the
size of the CCD and the maximum QE is higher than 90\% around 5700
{\AA} of the wavelength.

The data reduction follows the standard procedures with the NOAO
Image Reduction and Analysis Facility ({\sc iraf} v.2.15) software
package. After carefully checking the spectral images by eye, we
perform the bias combinations with {\tt zerocombine} and bias
corrections with {\tt ccdproc}, the flat-field combination,
normalization, and corrections with {\tt flatcombine}, {\tt
response}, and {\tt ccdproc}. Cosmic rays are eliminated with the
package {\tt cosmicrays}. The star cluster spectra and comparison
arc lamp spectra are extracted with {\tt apall}. The wavelength
calibrations are performed with helium/argon-lamp spectra, which
were taken at the beginning and end of each observing night. The
spectral features of the comparison lamps are identified with {\tt
identify}. The wavelength is calibrated with the package {\tt
refspectra}. Then {\tt dispcor} is used for the dispersion
correction and to resample the spectra. We use four Kitt Peak
National Observatory (KPNO) spectral standard stars in
\citet{mass88}  for flux calibrations. The {\tt standard} and {\tt
  sensfunc} packages are used to combine the standard stars and
determine the sensitivity and extinction of the atmosphere. In the
last step, we apply the {\tt calibrate} package to correct the
atmospheric extinction and finish the flux calibration.

\begin{figure}
  \centering
  \includegraphics[angle=0,scale=0.8]{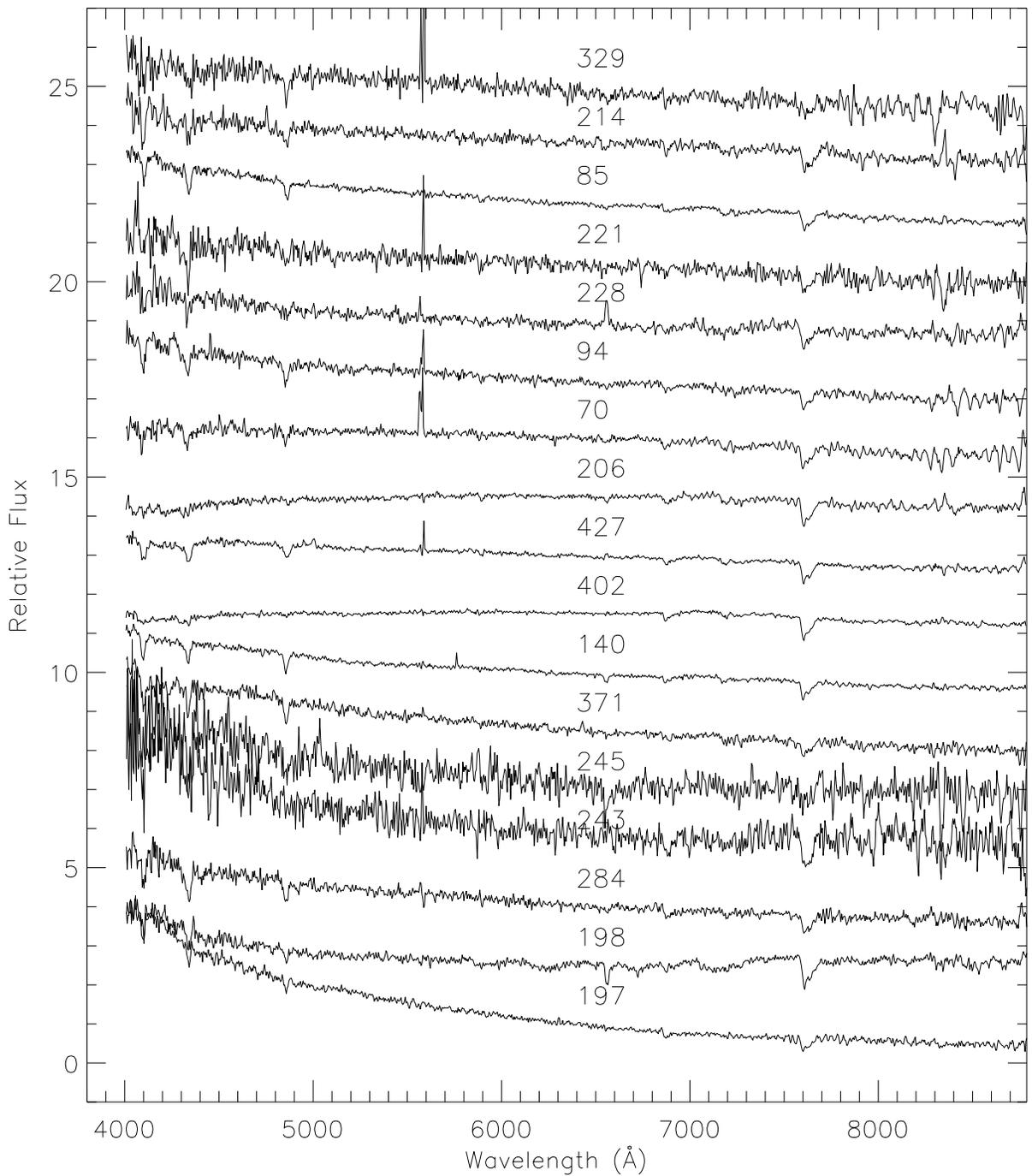}
  \caption{The normalized observed spectra of our star cluster sample
    in M31, which were taken from the 2.16-m telescope in Xinglong
    Observatory, NAOC.}
  \label{fig2}
\end{figure}

We display the normalized, calibrated spectra of our sample star
clusters in Figure~\ref{fig2}, with their names indicated (taken
from \citet{sara07}). Note that the emission lines of night sky [OI]
are at 5577{\AA} and the oxygen absorption lines of Earth's
atmosphere are around 7600 {\AA}. The signal-to-noise ratios (SNRs)
of most clusters are high enough, except for the star clusters SM
243 and 245.

\section{The Full-Spectrum-fit with {\sc ULySS} and BC03 Models}
\label{specfit.sec} In order to compare the fitting results, we also
adopt the {\sc ULySS} \citep{kol09} models for the full spectral
fitting to derive the ages and metallicities of star clusters. The
\citet{vaz} SSP models cover the wavelength ranges of 3540.5{\AA} --
7409.6{\AA} at a full width at half maximum (FWHM) of 2.3{\AA}.  The
models are based on the {\sc Miles} (Medium-resolution INT Library
of Empirical Spectra) spectral library \citep{sb06}. The stellar
initial mass function (IMF) of \citet{salp} is adopted for the
fitting and the solar-scaled theoretical isochrones of \citet{gi00}
have been used. The age range is $10^8$--$1.5\times10^{10}$ yr and the
metallicity is [Fe/H] = $-2.32$ dex ($Z=0.0004$) -- $+0.22$ dex
($Z=0.03$). Furthermore, another independent SSP model, {\sc
  pegase-hr}, which is provided by \citet{leb}, is based on the
empirical spectral library {\sc Elodie} \citep[e.g.,][]{ps01,pr07}.
The wavelength coverage is 3900{\AA} -- 6800{\AA} with a spectral
resolution $R\sim10,000$. In this model, the fitted stellar
atmospheric parameters are effective temperature, $T_{\rm eff}$
(3100--50,000 K), gravity $log~g$ ($-0.25$ dex -- 4.9 dex), and
metallicity $\rm [Fe/H]$ ($-3$ dex -- $+1$ dex). The flux
calibration accuracy is 0.5--2.5\%. We adopt the {\sc pegase-hr} SSP
models with the \citet{salp} IMF. The age ranges from $10^7$ to
$1.5\times10^{10}$ yr, and the metallicity [Fe/H] = $-2.0$ dex
($Z=0.0004$) to $+0.4$ dex ($Z=0.05$).

\begin{figure}
  \centering
  \includegraphics[angle=0,scale=0.8]{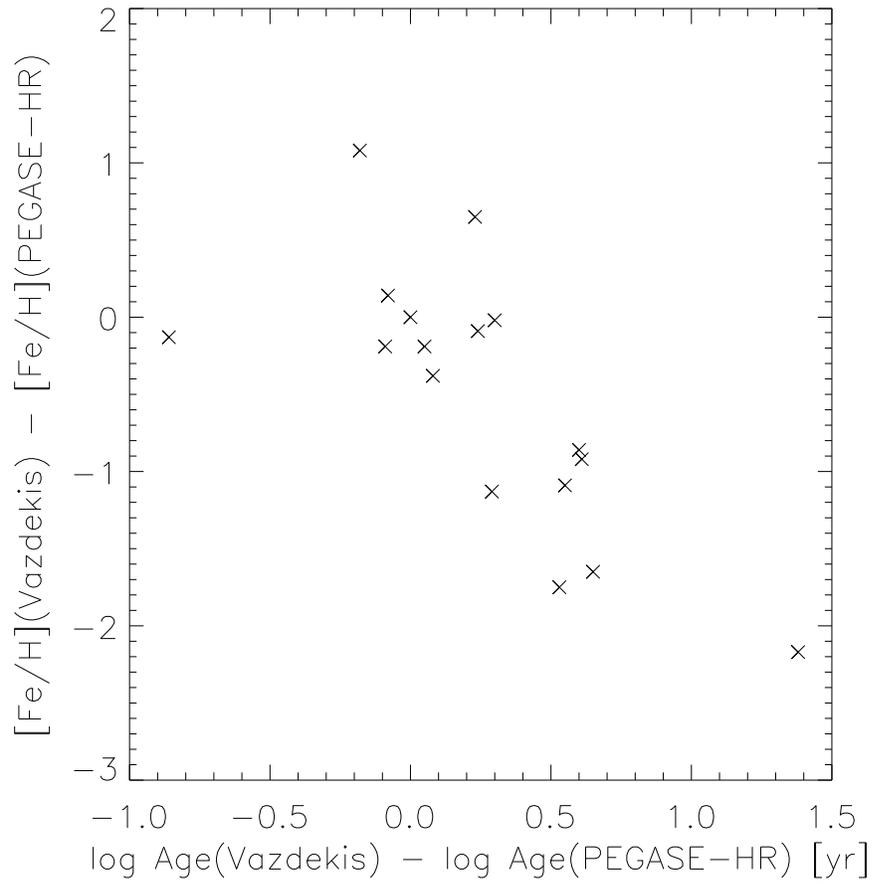}
  \caption{Self-comparisons of full-spectrum fitting results of
    Table~\ref{t2.tab}, ages and metallicities derived with the
    Vazdekis models and {\sc pegase-hr} models, respectively.}
  \label{fig3}
\end{figure}

Table~\ref{t2.tab} lists the ages and metallicities derived from the
full-spectrum fitting with {\sc ULySS} including \citet{vaz} and
{\sc pegase-hr} SSP models. The errors are calculated from
Monte-Carlo
  simulations, which are performed to estimate the biases,
errors and coupling (degeneracies) between the parameters. For the
fitting, a series of random errors are added in the data and then
the resulting the errors. From Table~\ref{t2.tab} we can see that
most of the star clusters are younger than 2 Gyr, except for SM 197,
402 and 206. There is a discrepancy in the age of SM 243, which is
1.48 Gyr from the \citet{vaz} model, but 10.72 Gyr  from {\sc
pegase-hr}. We note that the $\chi^2/dof$ for this source is
relatively large, 2.27 and 2.96 for the two models, which may be the
reason for the discrepancy in age. If we consider it to be the old
population, then 13/17-14/17 (76.5\%-82.3\%) of the sample is the young star
clusters, which agrees with the previous conclusions that the young
star clusters dominate in M33. On the other hand,  the \citet{vaz}
SSP model tends to result in a lower metallicity than {\sc
pegase-hr}, more than (or almost) 1 dex, e.g., SM198, 371, 140, 70,
228, 221 and 85. Meanwhile, it is noted that the age $log~t$ of
these clusters in  \citet{vaz} models is basically (at least)
$\sim0.5-0.6$ dex older than that of  {\sc pegase-hr} models. It may
be due to the different spectral libraries applied 
in the two models: \citet{vaz} SSP models cover wider wavelength
ranges and smaller metallicity lower-limit than  {\sc pegase-hr};
{\sc pegase-hr} SSP models extend the lower limit of ages to $10^7$
yr, which is one order of magnitude smaller than that of
\citet{vaz}. In order to make it clear, we plot
Figure~\ref{fig3} to show the differences in the parameters derived
from the two models. Apparently it is affected by the
age-metallicity degeneracy, which can usually be found in the SED
fitting.

\begin{figure}
  \centering
  \includegraphics[angle=0,scale=0.8]{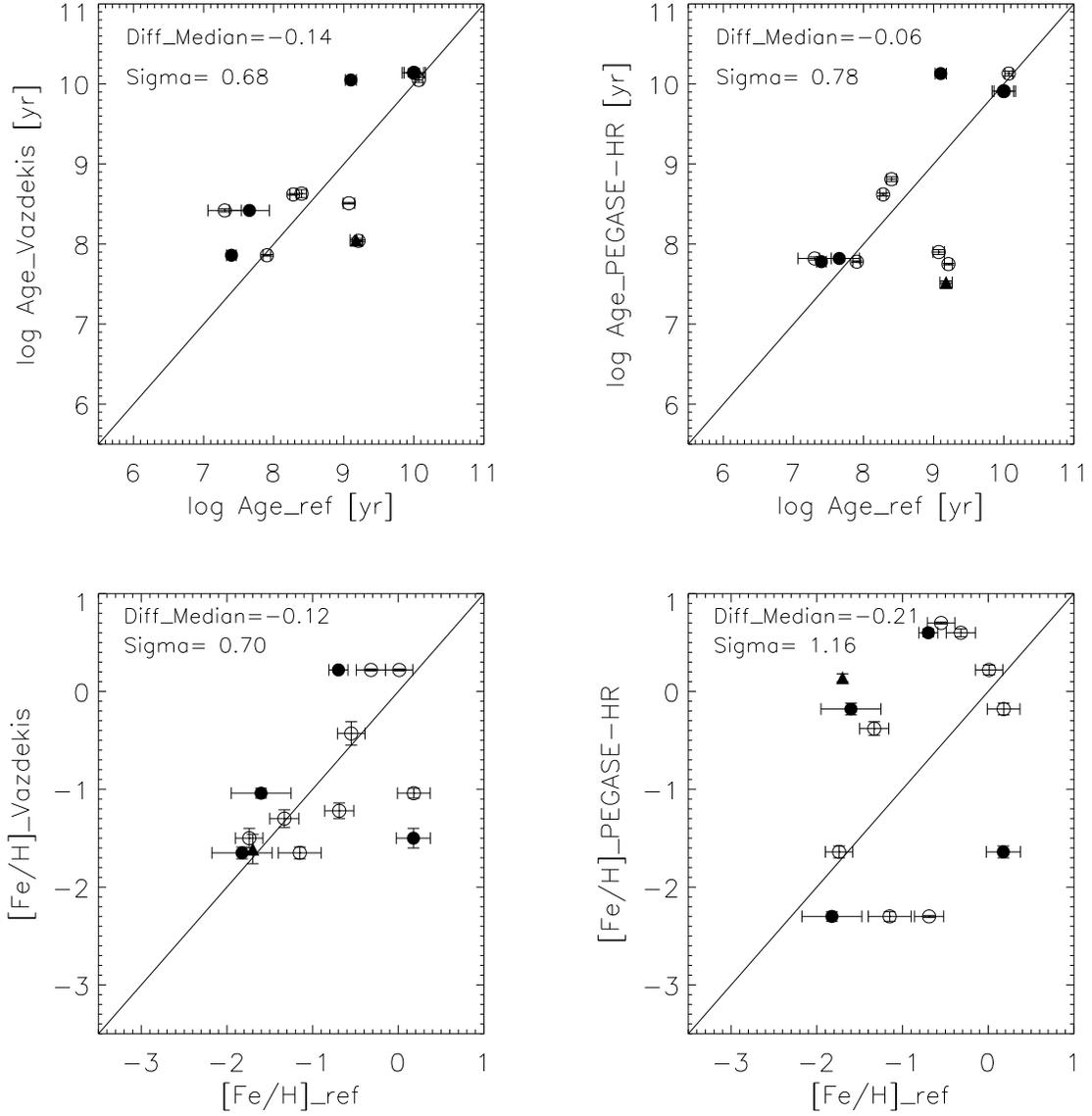}
  \caption{Comparisons of the our full-spectrum fitting results of
    Table~\ref{t2.tab}, ages and metallicities derived with the
    Vazdekis models (left panels) and {\sc pegase-hr} models (right panels) and
    those from \citet{bea15} (open circles), \citet{fan14} (filled circles) and
      \citet{sha10} (filled triangles).}
  \label{fig4}
\end{figure}

Since there are many previous works for determining the ages and
metallicities of M33 star clusters, it is necessary to compare if
there are sources in common. Regarding spectroscopy, for
instance,
  \citet{bea15} provided both age and metallicities based on the
  observations of Gran Telescopio Canarias (GTC) and William
  Herschel Telescope (WHT). \citet{sch91} only provide the velocities
  and did not provide the age and metallicities. \citet{chan02} only
  reported the velocities and ages, but no metallicity
  information. \citet{chan06} did not provide the catalog for age and
  metallicity. \citet{sha10} have only one star cluster in common
  with our work, CBF129 (SM228 in our work), age $t=1.5$ Gyr and
  $\rm [Fe/H]=-1.7$ dex and we have added it to our comparisons.  For the
  photometry, \citet{fan14} used the SED-fitting with the
  photometry in UBVRI/ugriz and JHK bands if available, which is the
  most  comprehensive, updated and homogeneous sample. Thus we only
  adopted the literature from \citet{bea15}, \citet{fan14} and \citet{sha10}.
Figure~\ref{fig4} shows the comparisons between the
results from full-spectrum method of \citet{vaz} and {\sc pegase-hr}
  models and that from \citet{bea15}, \citet{fan14} and
  \citet{sha10}. As shown in the figure, the offsets (median
  difference) are quite small for both the ages ($-0.14$ and $-0.06$),
  with scatters of (0.68 and 0.78 for the \citet{vaz} and {\sc
    pegase-hr} models respectively. The offsets of metallicities $\rm
  [Fe/H]$ and the references are $-0.12$ and $-0.21$ dex, with
  scatters of 0.70 and 1.16. It can be seen that the ages
from both models are comparable, although the offset of the
\citet{vaz} model and the references are smaller, but the scatter is
slightly larger. The largest difference is SM 402,  for which
\citet{fan14} gives 1.26 Gyr, but the result from \citet{bea15} is
11.74 Gyr. In our full-spectrum fitting with {\sc ULySS} model, the
results are also old, 11.22 Gyr in the \citet{vaz} model and 13.49
Gyr in the {\sc pegase-hr} model. Since both the results from
\citet{bea15} and our work are derived from spectroscopy, which
contains more information than the photometry in \citet{fan14},  we
believe our results are more reliable. On the other hand, it is
  known that the models are not sensitive to the ages $>1-2$ Gyr for the
  BC03 SSP models, especially for the SED-fit.

Similarly, we have performed the full-spectrum fitting with the
\citet{bc03} models. The evolutionary stellar population synthesis
models of \citet[][hereafter BC03]{bc03} not only provide spectra
and SEDs for different physical parameters, but also Lick/IDS
absorption-line indices. The models adopt Padova 1994 and Padova
2000 stellar evolutionary tracks, with initial mass functions (IMFs)
of \citet{salp} and \citet{chab}. The wavelength coverage ranges
from 91 {\AA} to 160 $\mu$m. The Padova 1994 model offers six
metallicity options ($Z=0.0001$, 0.0004, 0.004, 0.008, 0.02, and
0.05), and the Padova 2000 model also offers six ($Z=0.0004$, 0.001,
0.004, 0.008, 0.019, and 0.03). In total, there are 221 age steps
from 0 to 20 Gyr. In our work, we adopted both Padova 1994 and
Padova 2000 stellar evolutionary tracks for the purpose of
comparison. However, since there are only six metallicity values in
the model, which seems not enough for the fitting, we interpolate
the metallicities to 61 values. Thus the fitting results can be more
smooth and better distributed.

\begin{figure}
  \centering
  \includegraphics[angle=0,scale=0.8]{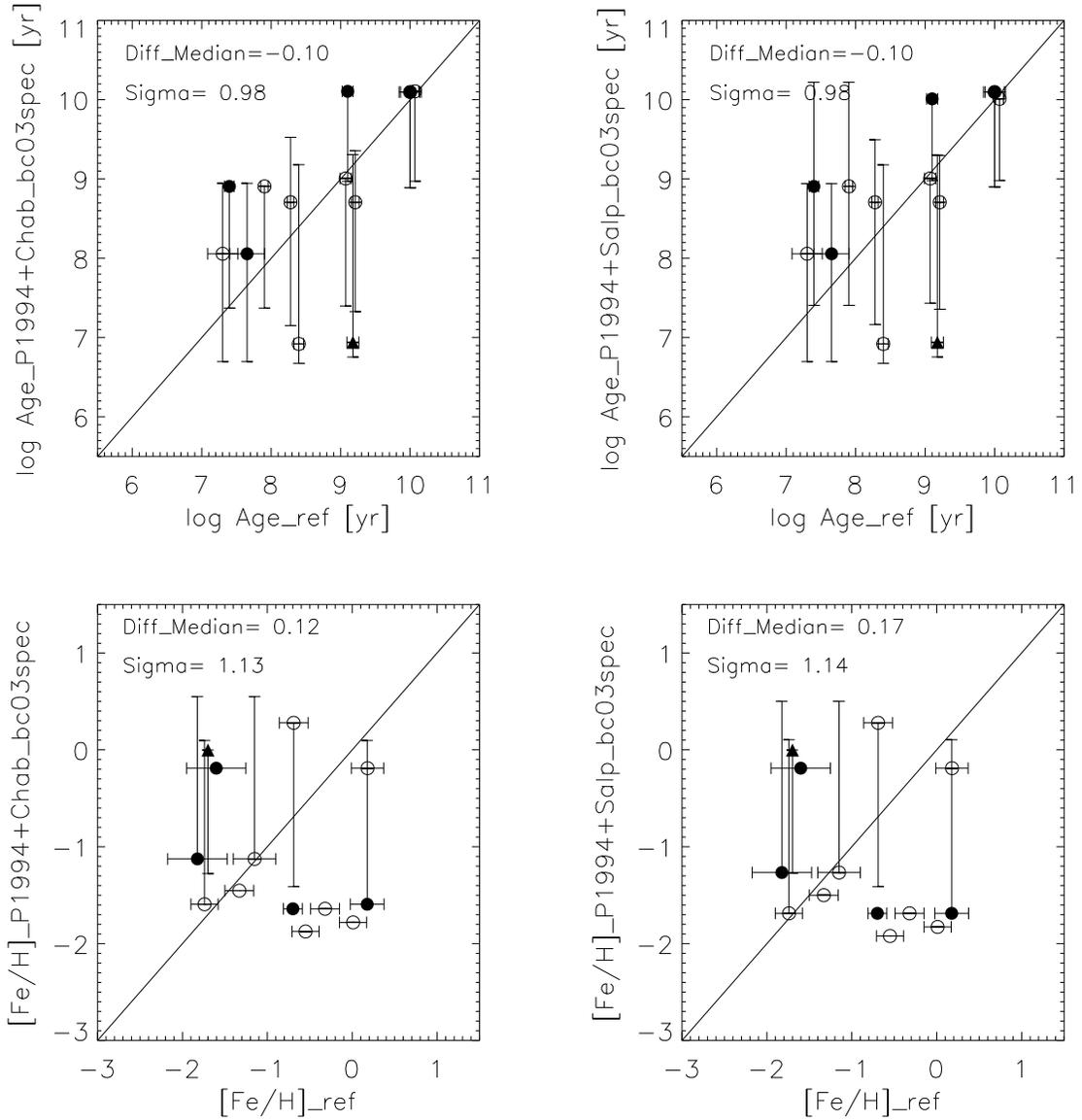}
  \caption{Same as Figure~\ref{fig3}, fitting only with
    spectroscopy, but with P1994 tracks
    + \citet{chab} IMF (left panels) and \citet{salp} IMF  (right panels).
    The open circles are from \citet{bea15}, the filled circles are
      from \citet{fan14} and the filled triangles are from \citet{sha10}.}
  \label{fig5}
\end{figure}

\begin{figure}
  \centering
  \includegraphics[angle=0,scale=0.8]{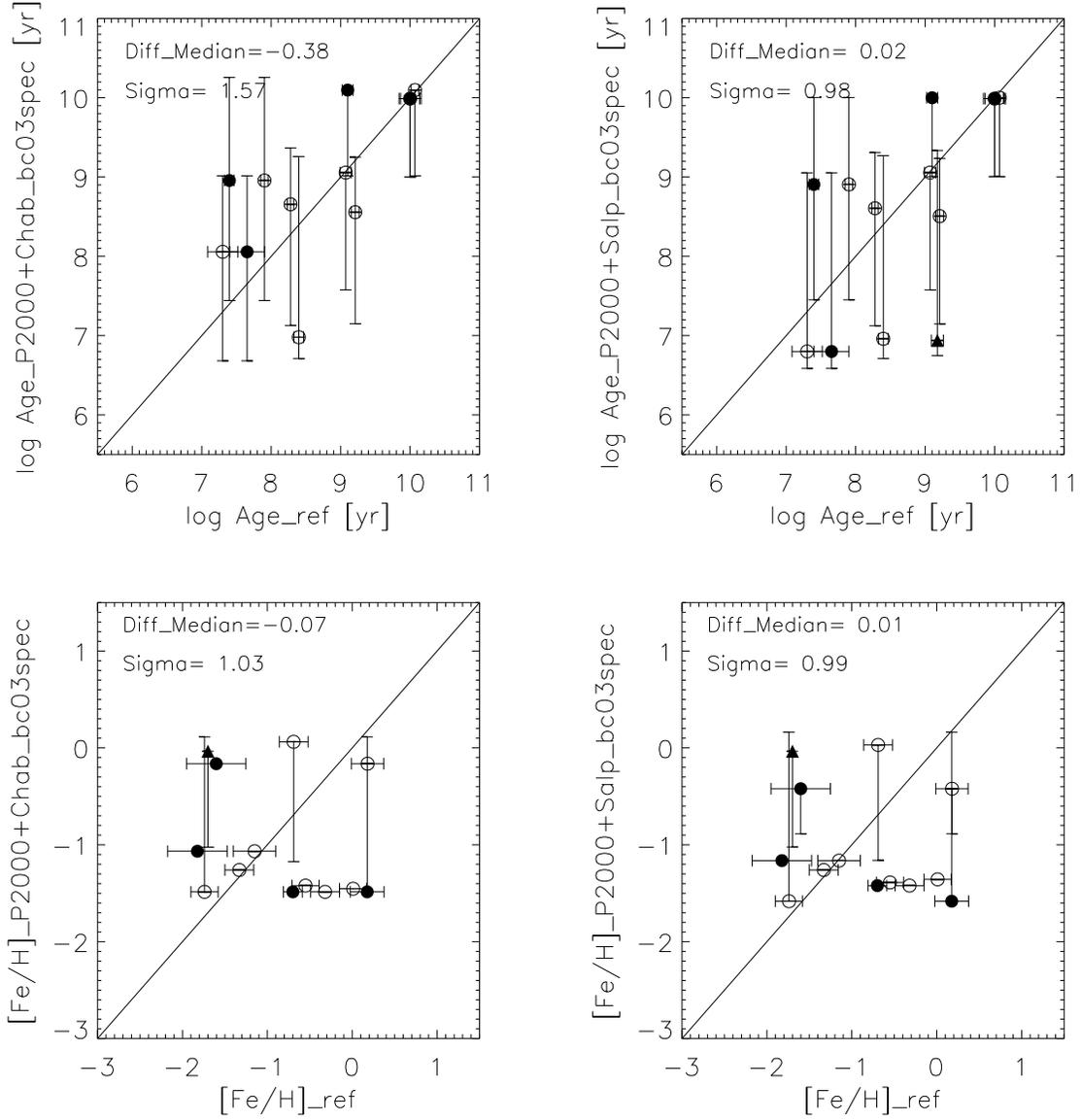}
  \caption{Same as Figure~\ref{fig3}, fitting only with
    spectroscopy. The model applied is the Padova 2000 tracks
    + \citet{chab} IMF (left panels) and \citet{salp} IMF  (right
    panels). The open circles are from \citet{bea15}, the filled circles are
      from \citet{fan14} and the filled triangles are from \citet{sha10}.}
  \label{fig6}
\end{figure}

We show our fitting results in Table~\ref{t3.tab} and
Table~\ref{t4.tab}, which list the ages and metallicities derived
from the full-spectrum fitting with \citet{bc03} SSP models, with
Padova 1994 and Padova 2000 evolutionary tracks respectively. In
each case, we use the \citet{chab} and \citet{salp} IMFs,
separately. It is easy to find that for SM 402 and 206, which are
old in Table~\ref{t2.tab}, they are also old in fitting with BC03
models. However, for SM 197 and 243, the results of BC03 models are
much younger than those from the {\sc ULySS} model.
Figure~\ref{fig5} and Figure~\ref{fig6} are similar comparisons of
the fitting results with the literature of \citet{bea15},
\citet{fan14} and \citet{sha10}. We find that for the ages, the
offset from the Padova 2000 evolutionary track with \citet{salp} IMF
is 0.02, which seems much smaller than that from Padova 1994 track
($-0.10$), except for the comparison with that of models with
  \citet{chab} IMF (offset$=-0.38$). For the metallicity,  
median difference from the Padova 2000 track is slightly smaller
than that from the Padova 1994 track, although the scatters are
comparable. Compared with the {\sc ULySS} model, for the ages, the
median differences of both the Padova 1994 and Padova 2000 tracks
are basically comparable, and the scatters are slightly smaller. For
the metallicity, although the median difference of Padova 1994
tracks is comparable with that of {\sc ULySS} fittings, but for the
Padova 2000 tracks, the median difference of the fitting results are
much smaller, although the scatter is similar.

\section{The SAGE Photometry and Spectrum-SED-fit with Models}
\label{sedspecfit.sec}

\subsection{Fitting with spectroscopy and the SAGE photometry}
\label{sage.sec}

 Since the M33 galaxy is  $847\pm60$ kpc from us, the star
  clusters are almost the point sources (a little bit extended) for the
  ground-based telescopes, either for spectroscopy or for the
  photometry, which guarantee that the data take into account the same
  stellar population/ same region of one star cluster.
As mentioned in Sect.~\ref{intro.sec}, the SAGE
survey\footnote{http://sage.sagenaoc.science/~sagesurvey/}
\citep[PI: Gang Zhao , see,
e.g.,][]{fan18,zheng18,zheng19a,zheng19b} has been operating since
2015 and it covers 12,000 deg$^2$ of the northern sky with
declination $\delta
>-5^{\circ}$, excluding the Galactic disk ($|b|<10^{\circ}$) which is
bright and with high extinction. The survey provides photometry in
eight bands $\rm u_{SC}$,$\rm v_{SAGE}$, $g$, $r$, $i$, $\rm
H{\alpha}_n$, $\rm H{\alpha}_w$ and DDO51. The central wavelengths
of the $\rm u_{SC}$ and $\rm v_{SAGE}$ bands are 3520 and 3950 {\AA}
respectively \citep{fan18}. The wavelength coverage of blue-band
  filter usually includes various metallicity absorption lines, e.g.,
  $\rm u_{SC}$-band. In 
particular, the $\rm v_{SAGE}$ band covers the CaII K line at
$\lambda=3933.44$ {\AA} (between H$\epsilon$ and H$\zeta$), which is
very sensitive to metallicity for FGK stars. The observations of the
$\rm u_{SC}$ and $\rm v_{SAGE}$ bands are almost completed
($\sim88$\% so far). Therefore, we utilized the photometry of the
two bands to derive the ages, metallicities and interstellar
extinctions of our star cluster samples of M33. The observations of
these two bands were carried out by the 90-inch (2.3-m) Bok
telescope of Steward Observatory, University of Arizona. A CCD
mosaic camera, which consists of four 4k$\times$4k CCDs, is mounted
at the prime focus. The field of view is $1\time1$ deg$^2$ and the
pixel size is $0.45''$. The photometry pipeline is based on the {\sc
SExtractor} and MAG$\_$AUTO. For astrometry, the {\sc SCAMP} has
been used. The Position and Proper Motion Extended (PPMX)
\citep{ro08} catalog is adopted in our pipeline as the astrometric
reference. The detailed description of the pipeline of photometry,
astrometry, and flux calibrations could be found in \citet{zheng19a,
zheng19b}. The SNR of 100 corresponds to a limiting magnitude of
$\rm u_{SC}\sim16.5$ mag and $\rm v_{SAGE}\sim15.5$ mag; 5 to $\rm
u_{SC}\sim20$ mag and $\rm v_{SAGE}\sim19.5$ mag. In our photometric
reduction, since star clusters are extended sources, the growth
curve has been calculated to perform the aperture corrections for
the star clusters in both $\rm u_{SC}$, $\rm v_{SAGE}$ bands.

The photometry of star clusters is calibrated in the AB system with
the convolution of the MILES library and our SAGE filters (please
see Table~\ref{t5.tab}). We adopt a value of $E(V-I)=0.06$ mag as
the mean Galactic foreground reddening in the direction of M33
\citep{sara00,san09}. The extinction $A_{\lambda}$ is computed using
equations 6-7 in \citet{ccm89}. The spectra and $\rm u_{SC}$, $\rm 
v_{SAGE}$-band photometry are fitted simultaneously via the equation
as followed,
\begin{equation}
  \chi^2_{\rm min}={\rm
    min}\left[\sum_{i=1}^{n_{Spec}}\left({\frac{m_{\lambda_i}^{\rm
            obs}-m_{\lambda_i}^{\rm mod}(t,\rm [Z/H])}
        {\sigma_{m,i}}}\right)^2+\sum_{j=1}^{n_{Phot}}\left({\frac{M_{\lambda_j}^{\rm
            obs}-M_{\lambda_j}^{\rm mod}(t,\rm [Z/H])}
        {\sigma_{M,j}}}\right)^2W_j^2\right],
  \label{eq1}
\end{equation}
where $m_{\lambda_i}^{\rm obs}$ is the AB magnitude that is
transformed from the dereddened observed spectra;
$m_{\lambda_i}^{\rm
  mod}(t,\rm [Z/H])$ is the $i^{\rm th}$ magnitude provided in the
stellar population model at an age $t$ and metallicity $\rm [Z/H]$;
similarly $M_{\lambda_i}^{\rm obs}$ represents the observed
dereddened magnitude in the $j^{\rm th}$ band; $M_{\lambda_j}^{\rm
mod} (t,\rm [Z/H])$ is the fitted $j^{\rm th}$ magnitude from the
stellar population model at an age $t$, metallicity $\rm [Z/H]$;
$n_{Spec}$ is the interval number of the wavelength/flux of the
spectrum; $n_{Phot}$ is the number of photometric bands, and here it
equals to 2; $W_j$ is the weight for the photometry fitting times of
the $j^{\rm th}$, which is the bandwidth of the SED relative to the
mean wavelength interval of the spectrum in the same wavelength
coverage; $W_j=\delta \lambda_{SED,j}/\overline{\delta
\lambda_{Spec}}$. $W_j$ is important to the second part of the
formula, since it is easy to note that
  the number of photometric bands $n_{Phot}$ is much smaller than spectroscopic
  data points $n_{Spec}$. Thus the two fitting parts for photometry
  and spectroscopy need to be rescaled and relative weight applied.
  In our work, the value of  $W_j$ is $\sim50-80$,
  which is the same order of weight for the total $\chi^2$
  ($g_{PHO}=0.1$) in \citet{werle19}.

We also compute the errors associated with MAG$\_$AUTO with those
related to the flux calibration, as
\begin{equation}
  \sigma_i^{2}=\sigma_{{\rm obs},i}^{2}+\sigma_{{\rm mod},i}^{2},
  \label{eq2}
\end{equation}
where $i$ represents any of the SAGE $\rm u_{SC}$, $\rm v_{SAGE}$
bands. $\sigma_{\rm obs}$ and $\sigma_{\rm mod}$ correspond to the photometric
uncertainties associated with model uncertainties, respectively.

The estimated ages and metallicities with $1-\sigma$ errors of the
M33 star clusters are listed in Tables~\ref{t6.tab} and
~\ref{t7.tab}, which are derived from the BC03 models of Padova 1994
and Padova 2000 evolutionary tracks, respectively. We estimate the
uncertainty associated with a given parameter by fixing the other
parameters to their best values, and vary the parameter of interest.
The error is computed as $1-\sigma~\chi^2_{\rm min}$,  which
means we calculate the
  $1-\sigma$ errors for the parameters (i.e., log~$t$ or Z). When
  the contours of  $\chi^2$ resulting from the parameter gives
  $\chi^2_{0.68}=\chi^2_{\rm min}$+2.3,
  which corresponds to  $1-\sigma$ (significance level = 0.68),
  the contour $\chi^2_{0.68}$ value defines a region of
  confidence in the (log $t$, Z) plane corresponding to the  $1-\sigma$ level of
  significance. A cut along a line of constant metallicity Z is the
  calculation of $\chi^2$ that defines upper and lower values of log $t$,
  corresponding to $\sigma = 1$ for this particular metallicity
  Z. Meanwhile, the errors for metallicity Z are calculated in the same
  way \citep{fan11,fan12,chen16}.  It is found that the fitting results are relatively
younger than those of the previous full-spectrum fittings. In
  Tables~\ref{t6.tab} for the Padova 1994 tracks, it is found that
  10/17 of star clusters are younger than 10 Myr $log~t<7$, while in Table
  ~\ref{t7.tab},  the proportion is much lower, only 4/17 of the
  sample are younger than 10 Myr $log~t<7$ for Padova 2000 tracks.  In
our sample, SM 402 is the only star cluster which is older than 2
Gyr for both models with Padova 1994 and Padova 2000 evolutionary
tracks. It is also noted that SM 206 is relatively older for our
sample in the fitting results, which is consistent with the previous
fittings.

Figures~\ref{fig7}-\ref{fig9} show the spectroscopy and SAGE
photometry fitting with the BC03 models, with the Padova 2000
evolutionary track + \citet{chab} IMF combination.  It can be seen
that most spectra and photometry data pints are fitted well except
for SM198, for which the model is obviously fainter than the observed
spectrum in the red part ($\lambda>7000$ {\AA}). It may be due
to the low SNR in the blue band of the spectrum and also it seems
$\rm u_{SC}$ mag is brighter than the model predicted in the blue
end of the spectrum.

For a consistency check, we compare our fitted ages and
  metallicities to the references. For the spectroscopic study of M33 star clusters,
  \citet{sch91}, \citet{chan02}, \citet{chan06}, \citet{sha10},
  \citet{bea15} can be found. However, most of the works only focus on
  the kinematics and ages except for \citet{bea15} and \citet{sha10}, who
  provide both age and metallicity information which can be used for
  comparison with our work. In fact, \citet{sha10} have only one star
  cluster in common with our work, CBF129 (SM228 in our work), age
  $t=1.5$ Gyr and $\rm [Fe/H]=-1.7$ dex with the full spectrum fitting
  and \citet{vaz99} models. Therefore, we use the spectroscopic
  information from \citet{bea15} and \citet{sha10}. 

As can be seen, the fitting result agrees with the literatures much better for
either ages or metallicities for models of Padova 2000 track
(Figure~\ref{fig11}) than that of Padova 1994 track
(Figure~\ref{fig10}): the median differences are much smaller, although 
the scatters are similar. The reason may be due to that the models of Padova 2000
evolutionary tracks have a new version of stellar
spectral library of \citet{gi00}, which updated equation of state and
low-temperature opacities. Thus the results are more reasonable and reliable
than that of models with Padova 1994 tracks.

However, we also found some outliers (significant differences) in the
comparison. For instance, 
result of SM 70 is $4.0-4.4$ Myr in Table~\ref{t6.tab} but it is
$0.91-1.14$ Gyr in Table~\ref{t7.tab}. For reference, \citet{bea15} provided the
age of 1.18 Gyr, and the results of full-spectrum 
fittings are $\sim1$ Gyr for BC03 model fitting (Tables~\ref{t3.tab} and
\ref{t4.tab}) or several $\times10^8$ yr for {\sc ULySS}
fittings. Thus we considered it due to the updated stellar spectral
library in the models of Padova 2000, which is more reliable.

\subsection{Fitting with spectroscopy and the SAGE photometry, UBVRI and ugriz photometry}
\label{allphot.sec}

In this section, we would like to figure out if adding more
photometry data can improve the fitting results further. Thus, we
have gathered the photometry of our sample star clusters in the $\rm
u_{SC}$ (SAGE), $\rm v_{SAGE}$ (SAGE), $UBVRI$ (Mayall) and $ugriz$
(CFHT) bands. The magnitudes of the $UBVRI$ bands are taken from
\citet{ma13} who carried out the observations by the Mayall 4-m
telescope. They transformed the Vega magnitude system to the AB
system with $\rm m_{AB}-m_{Vega}=$0.79, $-0.09$, 0.02, 0.21 and 0.45
(following:
http://www.astronomy.ohio-state.edu/~martini/usefuldata.html,
Blanton et al. 2007). The $ugriz$ magnitudes were obtained with the
CFHT 3.6-m telescope and in the AB system \citep{san10}. The $\rm
u_{SC}$, $\rm v_{SAGE}$ photometry of the SAGE survey in
Table~\ref{t5.tab} is also in the AB system. The photometry taken
from published work is listed in Table~\ref{t8.tab}. We adopt a
value of $E(V-I)=0.06$ mag as the mean Galactic foreground reddening
in the direction of M33 \citep{sara00,san09}. The extinction
$A_{\lambda}$ is computed using equations 6-7 in \citet{ccm89}. The
spectra and SEDs are fitted simultaneously via the equation as in
the following,
\begin{equation}
  \chi^2_{\rm min}={\rm
    min}\left[\sum_{i=1}^{n_{Spec}}\left({\frac{m_{\lambda_i}^{\rm
            obs}-m_{\lambda_i}^{\rm mod}(t,\rm [Z/H])}
        {\sigma_{m,i}}}\right)^2+\sum_{j=1}^{12}\left({\frac{M_{\lambda_j}^{\rm
            obs}-M_{\lambda_j}^{\rm mod}(t,\rm [Z/H])}
        {\sigma_{M,j}}}\right)^2W_j^2\right],
  \label{eq3}
\end{equation}
where the meanings of all physical quantities are the same as in
eq.~\ref{eq1} but the number of photometric bands is 12 in the
fitting.  Similarly, we compute the errors associated with
MAG$\_$AUTO with those related to the flux calibration, as
\begin{equation}
  \sigma_i^{2}=\sigma_{{\rm obs},i}^{2}+\sigma_{{\rm mod},i}^{2},
  \label{eq4}
\end{equation}
where $i$ represents any of the SAGE $\rm u_{SC}$, $\rm v_{SAGE}$,
$ugriz$ magnitudes, and $UBVRI$ bands. $\sigma_{\rm
  obs}$ and $\sigma_{\rm mod}$ correspond to the photometric
uncertainties associated with the respective model uncertainties.

The estimated ages and metallicities with $1-\sigma$ errors of the
M33 star clusters are listed in Tables~\ref{t9.tab} and
\ref{t10.tab}. We estimate the uncertainty associated with a given
parameter by fixing the other parameters to their best values, and
vary the parameter of interest. The error is computed as
$1-\sigma~\chi^2_{\rm min}$. We found that again, 8/17-9/17 of
  star clusters in our sample for that from the Padova
  1994 models in Tables~\ref{t9.tab} are younger than 10 Myr $log~t<7$,
  while in Table ~\ref{t10.tab},  the proportion is much lower:  only 5/17 of the
  sample are younger than 10 Myr $log~t<7$ for models of Padova 2000
  tracks. The proportions of fitting results are similar to those of in
  Tables~\ref{t6.tab} and \ref{t7.tab}. Particularly, in
  Table~\ref{t10.tab} with the \citet{chab} IMF,  the ages of SM
  206 is $9.51-9.75$ Gyr, which agree with that of  \citet{bea15} (9.98
  Gyr). While for SM 402, it is $6.75-7.74$ Gyr in Table~\ref{t10.tab} with the
  \citet{chab} IMF, but 11.74 Gyr for \citet{bea15}. We consider that
  the age is not very sensitive for the stellar population when $>1-2$
  Gyr in the BC03 SSP models. The fitted 
models and observational spectra and photometry are plotted in
Figures~\ref{fig12}-\ref{fig14}. It is found that most of star
clusters are fitted well except SM198, for which the red part of the
spectral seems much redder than the model, but agree with the
photometry. As for star cluster SM70,  the observed spectrum is not
consistent with its SED or the model spectrum,
especially in the blue part of the spectrum ($\lambda<5500$ {\AA}). The fitting
result is 6.6 Myr, is much younger than the \citet{bea15} value 1.18
Gyr, and that of full-spectrum fitting (1.14 Gyr, see Table~\ref{t4.tab}) /
SAGE+spectrum (1.14 Gyr, see Table~\ref{t7.tab}) with BC03 models. However, the
result is relatively closer to that of the {\sc pegase-hr}  models of {\sc
  ULySS} fittings (79.4 Myr, see Table~\ref{t2.tab}). We have checked
the data and found no 
  problem. Further, we also found the photometry in the similar band,
  $u=18.01$  mag from \citet{san10} are 0.5 mag brighter than our
  photometry $u_{SC}=18.51$ mag, and $U=17.56$ mag from \citet{ma13}
  is even $\sim1$ mag brighter. If we remove the SAGE $u_{SC}$ band
  photometry, only leaving $u$ and $U$ mags in $\lambda<4000$ {\AA}
  for the photometry, the fitting result could become even younger. Thus we think
  our photometry $u_{SC}$ dose not account for the disagreement with literature, and
  probably we may take high S/N spectrum, especially in the blue 
  part ($\lambda<4000$ {\AA}) in the future work.

For a consistency check, we compare our fitted ages and
metallicities to those of \citet{bea15} and \citet{fan14} in
Figures~\ref{fig15} (with Padova 1994 track) and \ref{fig16} (with
Padova 2000 track). Again, the results of the Padova 2000 track are
much better than those of Padova 1994, either for ages or
metallicities, which may be due to the updated library of models.
  As can be seen, in the top panels of Figure~\ref{fig16}, the median
  differences between our fitted ages and the literatures are
  $\delta~log~t=0.12$,  which is slightly larger than that fitted
  with spectroscopy+SAGE photometry in Figure~\ref{fig11}
  ($\delta~log~t=0.04-0.08$). On the other hand, however, it seems that 
   the fitting errors have been reduced in Figures~\ref{fig15} and
  \ref{fig16}, with more bands of photometric 
  data, i.e., spectroscopy and the SAGE photometry, UBVRI and ugriz
  photometry. For the metallicity, we also did not found obvious advantage
  of the fitting in Figures~\ref{fig15} and \ref{fig16},  than that of
  SAGE+spectrum method in Section~\ref{sage.sec}.

\section{Discussion}
\label{dis.sec}

\subsection{The Sample Selection and Results discussion}

Discussion of the sample and results of different methods:

  For our sample, we have selected only 17 star clusters of M33,
  the brightest ones ($V<17.5$) in the galaxy, which are the
  most massive star clusters. Since the distance modulus
  $(m-M)_0=24.64\pm0.15$, the absolute magnitude of our sample star
  clusters are $M_0<-7.1$, for which the mass $>5.8\times10^4$. Thus
  we only focus on the massive star clusters and our sample is not
  effected by the selection effect significantly.

  We have derived the ages and metallicities of our sample star
  clusters, with both {\sc ULySS} \citep{kol09} models, including the
  \citet{vaz} and  {\sc pegase-hr}  SSP models and the BC03
  \citep{bc03} SSP models. It is noted that the ages derived from the
  {\sc ULySS} models lack young star clusters which are $log~
  t<7$ yr in the BC03 models, as the lower limit of ages are $10^8$
  and $10^7$ for \citet{vaz} and {\sc pegase-hr}  SSP models,
  respectively. Thus the results of young star clusters $log~t<7$ yr
  can only be found in that of BC03 models in our work.
  However, for the BC03 models, it is also noted that the number of star
  clusters with age $log~t<7$ yr  from models of Padova 1994 tracks
  are much more than that from the models of Padova 2000 tracks in either
  fitting method 2 (Spectrum +SAGE photometry) or fitting method 3
  (Spectrum + all the photometry), which may be due to the spectral
  library is not updated to that of \citet{gi00}, compared to Padova
  2000 tracks. It is also 
  noted that the number of star clusters with age $log~t<7$ yr are
  comparable for that of the fitting method 2 (Spectrum +SAGE
  photometry) and fitting method 3 (Spectrum + all the
  photometry). After comparing our fitting results with the literature, 
  we found that adding more-band of photometric data besides of SAGE
  photometry in the blue wavelength coverage actually dose not improve
  the fitting results substantially except reducing the fitting errors
  to some extent.

\subsection{The age-metallicity degeneracy}

 It is well-known that the
 age-metallicity degeneracy exists in the SED or spectrum fitting. In order
 to investigate this point, we have done a series of works for
 testing. Figure~\ref{fig17} is the Monte-Carlo simulations for the
 spectrum of one randomly selected star cluster if we add a series of
 errors around 5\% (which is the Gaussian
 distribution) to the spectrum and fit the age and metallicity with
  the {\sc pegase-hr} models (black crosses) and Vazdekis models (red
  crosses). Apparently, we can see the age-metallicity degeneracy,
  especially for the Vazdekis models. Similarly, in Figure~\ref{fig18}
  we have shown the relations between metallicity and ages for the spectrum
  of one randomly selected star cluster with the Padova 2000
  evolutionary tracks and \cite{chab} IMF. We have done the
  ${\chi}^2_{min}$-fit for all the ages provided in the models and fit
  the metallicity. The situations of full-spectrum, SAGE
  photometry+Spectrum (method 2) and all photometry+Spectrum (method
  3) are all plotted and shown. We identify a trend of the
  age-metallicity degeneracy, but it is not significant in all the
  three fitting methods. In fact, we have checked more star clusters,
  and the results are various for different sources. However, the
  age-metallicity relation for this star cluster is quite typical. In
  addition, we also have tried different evolutionary tracks and IMFs, and
  the results are similar, although the results are slightly different.

\subsection{The UV-excess of our sample}

We compare the $GALEX$ FUV, NUV photometry with the SAGE $\rm
u_{SC}$ together with cluster age in Figure~\ref{fig19}. The $GALEX$
photometry \citep{mudd15} was done for the point sources. All the
magnitudes and colors have been dereddened adopting the extinction
law of \citet{ccm89}.

 As indicated in the top left panel of Figure~\ref{fig19}, there is a
  correlation between $NUV-u_{SC}$ vs. cluster age from the left
  columns of Table~\ref{t7.tab}, which are derived from the
  spectroscopy+SAGE photometry fitting, but for the BC03 models with
  Padova 2000 stellar evolutionary tracks, and \cite{chab} IMF.
  Usually young, massive (O/B) stars which are currently formed in the
  young clusters are sources for the emission of UV
  lights. $NUV-u_{SC}$ shows a dependence on age with a correlation
  coefficient of $r=0.44$ (solid line in Figure~\ref{fig19}). Five
  intermediate-age clusters above 400\,Myr (black squares, ID: 245,
  70, 94, 221 and 214) are located further away from the general
  correlation of the sample, by showing bluer color $NUV-u_{SC}$ less
  than 0.2 (defined as the ``UV-excess'' in our work). These five
  clusters (black squares in the top left panel of Figure~\ref{fig19}) have 
  $u_{SC}$ close to young clusters. However, they only have $NUV$
  excess (top right and bottom left panels in Figure~\ref{fig19}), and
  the $FUV-u_{SC}$ colors are normal (bottom right panel in Figure~\ref{fig19}). 
  As the cluster grows old, its massive stars evolved. So that
  $NUV-u_{SC}$ color of the cluster becomes redder, and $u_{SC}$
  magnitude grows fainter at the same time. Young massive stars are
  supposed to be absent in these five clusters. 
  However, they deviate from the monotonic trend of $NUV-u_{SC}$
  vs. $u_{SC}$ (top right and bottom left panels in Figure~\ref{fig19}).  
  Excluding these five clusters, $NUV-u_{SC}$
  dependence on cluster age is enhanced with a correlation coefficient
  of $r=0.76$ (solid line in Figure~\ref{fig19}). 

  Excess emission in UV color has been observed in old populations,
  e.g., discovered in early-type galaxies \citep{deharveng1976, oconnell86};
  Galactic globular clusters: 47 Tuc \citep{oconnell1997}, NGC 6388 and
  NGC 6441 \citep{Rey2007}; and old open clusters: NGC 6791
  \citep{buzzoni2012}. A similar phenomenon has been modeled via
  numerical N-body simulations of \citet{pang16}, which reproduced
  UV-excess in the SED of star clusters up to 
  600\,Myr. Observational and theoretical evidences show that low-mass,
  small-envelope, helium-burning stars in the blue horizontal branch (BHB),
  whose effective temperature reaches above 35000\,K, are promising
  candidates for producing UV excess in old populations \citep{Ree2007,
   Rey2007, buzzoni2012, Bekki2012}. The morphology of BHB depends on
 metallicity. BHB stars are bluer with lower metallicity
 \citep{Yi1999, Yoon2006}. Two intermediate-age clusters (ID: SM 70,
 214) with NUV excess have almost the lowest metallicity ($\rm
 [Fe/H]=-1.51$ and $-1.46$, see mean metallicity of
 Table~\ref{t7.tab} with \citet{chab} and \citet{salp} IMFs) among
 our samples, which is consistent with the theoretical  
 prediction. At the same time, metallicity of three another  
 intermediate-age clusters (ID: SM 245, 94 and 221) are median or higher ($\rm
 [Fe/H]=-0.30, -0.30$ and $-0.63$, see mean metallicity of Table~\ref{t7.tab}
 with \citet{chab} and \citet{salp} IMFs). Including helium
 enrichment, a metal-rich cluster can   
 also produce BHB \citep{Rich1997, Chung2013, Bekki2012}.
 Besides the BHB scenario, \citet{Bekki2012} simulations found that a
 cluster with multiple populations can generate UV-excess when the
 younger generation was helium enhanced from AGBs of the first
 generation. Note that the age and metallicity are degenerated in
 producing blue color. However, with our current data, we cannot
 justify either BHB or multiple population scenarios.

 We found a positive gradient in the $NUV-u_{SC}$ vs. clusters'
 distance to the center of M33 in Figure~\ref{fig20}.  The slope is
 $k=0.23$ and the correlation coefficient $r=0.49$. This color
 gradient is consistent with the age distribution of clusters.
 Younger cluster are located in the inner part of M33 (cross symbols), with bluer
 $NUV-u_{SC}$ color, while older generations are redder and idle
 around the outskirt (squared symbols). This color gradient with the
 slope of $k=0.23$ is also an age gradient, consistent with the upper
 left panel in Figure~\ref{fig19}, implying an inside out formation
 history for M33 galaxy.

To discover more UV-excess among  intermediate-age or old
star clusters in the Local Group, the Chinese Space Station Station
Telescope (CSST) will be an essential equipment in the future,
providing NUV photometry (down to 255 nm) with 10 times higher
spatial resolution $\sim0.15\arcsec$ \citep{cao18,gong19} than
$GALEX$.  Given only five intermediate-age clusters with UV-excess
observed in our small samples, this number will be significantly
increased after CSST 's $NUV$ survey of the Local Group.

\section{Summary and Conclusions}
\label{sum.sec}

In our work, we compared different fitting methods for star clusters
with the stellar population synthesis models based on spectroscopy
and photometry (SED)  data. We choose a sample of 17 star clusters
in M33 which lack previous spectroscopic observations. The spectra
of our sample star clusters were taken with the BFOSC low-resolution
spectrograph on the NAOC Xinglong 2.16-m reflector.

In fact, we applied three different fitting methods: 1)
full-spectrum fitting with  {\sc ULySS}  \citep{kol09} including the
\citet{vaz} and  {\sc
  pegase-hr}  SSP models and \citet{bc03} models;
2) spectroscopy + blue-bands SAGE $\rm u_{SC}$ and $\rm v_{SAGE}$
photometry with \citet{bc03} models; 3) spectroscopy + photometry of
SAGE $\rm u_{SC}$ and $\rm v_{SAGE}$ bands, UBVRI-bands and
ugriz-bands, with \citet{bc03} models. For all the methods, when
using the \citet{bc03} models, the evolutionary tracks of Padova
1994 and Padova 2000 with the IMFs of \citet{chab}  and \citet{salp}
have been applied separately. The $\chi_{\rm min}^2$ technique has
been applied for all the fittings.

We found that 

1. The fitting results of models with Padova 2000
tracks, for which the updated stellar spectral library of \citet{gi00}
is applied, are significantly better than those
of the Padova 1994 track in all the fittings overall,  especially for fitting method 2
(spectroscopy + blue-bands SAGE $\rm u_{SC}$ and $\rm v_{SAGE}$
photometry) and method 3 (spectroscopy + photometry of SAGE $\rm u_{SC}$ and
$\rm v_{SAGE}$ bands, UBVRI-bands and ugriz-band); 

2. Adding the
blue-band photometry (such as SAGE $\rm u_{SC}$ and $\rm v_{SAGE}$
bands) to the spectroscopy fitting can improve the precision of
the fitting results significantly, i.e. the errors have been reduced
significantly.  The median differences between our fitting results
and the literature becomes much smaller, especially for the Padova
2000 tracks model, to $0.04-0.08$ dex for ages $log~t$. Thus the method is
an effective way to combine the spectroscopy and the SEDs. On one
hand, the spectra can provide more information, not only the
continuum, but also the absorption lines. The SAGE $\rm v_{SAGE}$
magnitude is sensitive to metallicity since its bandwidth contains
the CaII K line at $\lambda=$3933.44 {\AA}. On the other hand, the
SED constructed with photometry from the blue bands, e.g., SAGE $\rm
u_{SC}$ and $\rm v_{SAGE}$, is a good complement to the
spectroscopy. The fitting results are consistent with those of
\citet{bea15} and \citet{fan14} in both ages and metallicities,
except for a few outliers. In general, our results agree well with
previous determinations. 

3. Adding photometry in more bands, such as
the UBVRI and ugriz,  did not substantially extend the wavelength
coverage and thus can not provide more information than method 2. 
  It is found that the median difference has not been reduced, but
  slightly larger for ages $log~t$ of the Padova 2000 track model, from
  $0.04-0.08$ to 0.12 dex. Thus the results cannot be improved
significantly. However, we found the fitting errors become much
  smaller, which can be found from Figure~\ref{fig15} and \ref{fig16}.

It is discovered that five candidate star clusters exhibit
UV-excess in FUV and NUV bands. The true magnitude of the UV-excess
is expected to be stronger. The CSST, being the only space telescope
equipped with the NUV band, will validate and discover more
UV-excess star clusters in the Local Group.

Further, as the LAMOST survey \citep{zhao12,cui12} has provides more
  than ten million of stellar spectra already, it is the largest
  spectroscopic dataset in the world currently. In addition, the
  project is still ongoing and the number is increasing. Since  
the limiting magnitudes of $u_{SC}$ and $v_{SAGE}$ bands of SAGE
survey could reach $\sim17-18$ mag with an SNR of $\sim$50, it can
perfectly match the LAMOST magnitude range with high SNR. Thus the
SAGE survey could be quite useful and complementary for the spectrum
fitting of LAMOST spectra. Moreover, it has been demonstrated
that the blue bands are very crucial for distinguishing different
stellar populations.

\acknowledgements We thank the anonymous referee for his/her
thorough review and helpful comments and suggestions, which
significantly contributed to improving the manuscript. This study
is supported by the National Natural Science Foundation of China
(NSFC) under grant Nos. 11988101, 11890694, U1631102; Sino-German
Center Project GZ 1284; National Key Research and Development
Program of China grant Nos. 2019YFA0405502, 2016YFA0400804; and the
Youth Innovation Promotion Association (YIPA), Chinese Academy of
Sciences. X.Y.P. expresses gratitude for support from the Research
Development Fund of Xi'an Jiaotong Liverpool University
(RDF-18-02-32) and the financial support of two grants of National
Natural Science Foundation of China, Nos. 11673032 and 11503015. We
thank Prof. Richard de Grijs for useful discussion and Dr. James
E. Wicker for professional language revision.

\appendix          

\clearpage
\pagestyle{empty}
\begin{deluxetable}{lrrrrc}
  \tablecolumns{5} \tablewidth{0pc} \tablecaption{The observation
    informations of our sample star clusters in M33. The $\rm
    ID_{SM07}$, RA, Dec and $V$-band magnitude are from the Table~3 of \citet{sara07}.
    \label{t1.tab}}
  \tablehead{
    \colhead{$\rm ID_{SM07}$} & \colhead{R.A.} & \colhead{Dec} &
    \colhead{$V$-mag} & \colhead{Date} &
    \colhead{Expose Time} \\
    \colhead{} & \colhead{(J2000)} & \colhead{(J2000)} & \colhead{(mag)} &\colhead{(yyyy/mm/dd)} & \colhead{(sec)} } \startdata
    197 & 01:33:50.85 & +30:38:34.5 & 16.39 & 2015/10/09 &     2400+1800 \\
198 & 01:33:50.90 & +30:38:55.5 & 16.76 & 2015/10/09 &     3600 \\
284 & 01:34:03.12 & +30:52:13.9 & 16.83 & 2015/10/09 &     3600 \\
243 & 01:33:57.87 & +30:33:25.7 & 17.00 & 2015/10/09 &     3600 \\
245 & 01:33:58.01 & +30:45:45.2 & 17.14 & 2015/10/10 &     3600 \\
371 & 01:34:19.89 & +30:36:12.7 & 17.16 & 2015/10/10 &     3600 \\
140 & 01:33:37.24 & +30:34:13.9 & 17.15 & 2015/10/11 &     3600 \\
402 & 01:34:30.20 & +30:38:13.0 & 17.19 & 2015/10/11 &     3600 \\
427 & 01:34:43.70 & +30:47:37.9 & 17.20 & 2015/10/11 &     3600 \\
206 & 01:33:52.20 & +30:29:03.8 & 17.29 & 2015/10/11 &     3600 \\
70  & 01:33:23.10 & +30:33:00.5 & 17.38 & 2015/10/19 &     3600 \\
94  & 01:33:28.70 & +30:36:37.5 & 17.38 & 2015/10/19 &     3600 \\
228 & 01:33:56.18 & +30:38:39.8 & 17.38 & 2015/10/19 &     3600 \\
221 & 01:33:55.00 & +30:32:14.5 & 17.42 & 2015/10/19 &     3600 \\
85  & 01:33:26.75 & +30:33:21.4 & 17.45 & 2015/10/19 &     3600 \\
214 & 01:33:53.69 & +30:48:21.5 & 17.45 & 2015/10/19 &     3600 \\
329 & 01:34:10.09 & +30:45:29.4 & 17.48 & 2015/10/19 &     3600 \\

    \enddata
  \end{deluxetable}

\clearpage
\pagestyle{empty}
\begin{deluxetable}{r|ccc|ccc}
  \tablecolumns{7} \tablewidth{0pc} \tablecaption{The physical
    parameters, ages and metallicities of our sample star clusters
    derived from full-spectrum fits with Vazdekis model and {\sc
      pegase-hr} models of {\sc ULySS}.
    \label{t2.tab}}
  \tablehead{\colhead{} & \multicolumn{3}{c}{Vazdekis model} &\multicolumn{3}{c}{
      {\sc pegase-hr} model} \\
    \cline{2-7}
    \colhead{ID} & \colhead{log $t$} &  \colhead{$\rm
      [Fe/H]$} &  \colhead{$\chi^2_{min}/dof$}  & \colhead{log $t$} &
    \colhead{$\rm [Fe/H]$} &  \colhead{$\chi^2_{min}/dof$}  \\
    \colhead{} & \colhead{(yr)} & \colhead{(dex)} & \colhead{} &
    \colhead{(yr)} & \colhead{(dex)} & \colhead{}}
  \startdata
         197  & $     9.93 \pm    0.07 $ & $  -2.32 \pm  0.01 $ &     0.13 & $    10.02 \pm    0.04 $ & $  -2.13 \pm  0.05 $ &     0.12  \\
       198  & $     7.82 \pm    0.01 $ & $  -2.32 \pm  0.01 $ &     0.36 & $     7.17 \pm    0.02 $ & $  -0.67 \pm  0.13 $ &     0.34  \\
       284  & $     8.56 \pm    0.01 $ & $   0.22 \pm  0.01 $ &     0.43 & $     8.26 \pm    0.03 $ & $   0.24 \pm  0.09 $ &     0.41  \\
       243  & $     9.17 \pm    0.02 $ & $  -2.31 \pm  0.07 $ &     2.27 & $    10.03 \pm    0.07 $ & $  -2.18 \pm  0.03 $ &     2.96  \\
       245  & $     8.63 \pm    0.05 $ & $  -1.22 \pm  0.08 $ &     2.46 & $     8.81 \pm    0.03 $ & $  -2.30 \pm  0.01 $ &     4.74  \\
       371  & $     8.42 \pm    0.02 $ & $  -1.04 \pm  0.05 $ &     0.38 & $     7.82 \pm    0.02 $ & $  -0.18 \pm  0.06 $ &     0.37  \\
       140  & $     8.39 \pm    0.02 $ & $  -0.86 \pm  0.05 $ &     0.08 & $     7.84 \pm    0.02 $ & $   0.23 \pm  0.05 $ &     0.08  \\
       402  & $    10.05 \pm    0.03 $ & $  -1.50 \pm  0.10 $ &     0.12 & $    10.13 \pm    0.03 $ & $  -1.64 \pm  0.06 $ &     0.11  \\
       427  & $     8.56 \pm    0.02 $ & $  -0.16 \pm  0.05 $ &     0.14 & $     8.51 \pm    0.04 $ & $   0.03 \pm  0.08 $ &     0.14  \\
       206  & $    10.14 \pm    0.05 $ & $  -1.65 \pm  0.06 $ &     0.59 & $     9.91 \pm    0.02 $ & $  -2.30 \pm  0.05 $ &     0.55  \\
        70  & $     8.51 \pm    0.01 $ & $  -1.30 \pm  0.09 $ &     0.90 & $     7.90 \pm    0.03 $ & $  -0.38 \pm  0.07 $ &     0.87  \\
        94  & $     8.62 \pm    0.01 $ & $   0.22 \pm  0.01 $ &     0.42 & $     8.62 \pm    0.02 $ & $   0.22 \pm  0.05 $ &     0.36  \\
       228  & $     8.05 \pm    0.02 $ & $  -1.61 \pm  0.15 $ &     0.64 & $     7.52 \pm    0.02 $ & $   0.14 \pm  0.04 $ &     0.65  \\
       221  & $     9.13 \pm    0.01 $ & $  -1.71 \pm  0.05 $ &     1.91 & $     7.75 \pm    0.02 $ & $   0.46 \pm  0.07 $ &     2.03  \\
        85  & $     8.04 \pm    0.01 $ & $  -0.43 \pm  0.12 $ &     0.13 & $     7.75 \pm    0.01 $ & $   0.70 \pm  0.01 $ &     0.13  \\
       214  & $     7.86 \pm    0.01 $ & $   0.22 \pm  0.01 $ &     0.74 & $     7.78 \pm    0.01 $ & $   0.60 \pm  0.04 $ &     0.74  \\
       329  & $     8.16 \pm    0.01 $ & $   0.22 \pm  0.01 $ &     2.07 & $     7.92 \pm    0.02 $ & $   0.31 \pm  0.06 $ &     2.06  \\

  \enddata
\end{deluxetable}

\clearpage
\pagestyle{empty}
\begin{deluxetable}{r|rrr|rrr}
  \tablecolumns{4} \tablewidth{0pc} \tablecaption{Same as in
    Table~\ref{t2.tab} but from full-spectrum fitting with \citet{bc03}
    models and Padova 1994 stellar evolutionary tracks. Both the
    \citet{chab} IMF and \citet{salp} IMF were adopted for the
    fitting separately.
    \label{t3.tab}}
  \tablehead{\colhead{} & \multicolumn{3}{c}{\citet{chab} IMF}
    &\multicolumn{3}{c}{\citet{salp} IMF} \\
    \cline{2-7}
    \colhead{ID} & \colhead{log $t$} &  \colhead{$\rm
      [Fe/H]$} &  \colhead{$\chi^2_{min}/dof$}  & \colhead{log $t$} &  \colhead{$\rm
      [Fe/H]$} &  \colhead{$\chi^2_{min}/dof$}   \\
    \colhead{} & \colhead{(yr)} & \colhead{(dex)} & \colhead{} & \colhead{(yr)} & \colhead{(dex)} & \colhead{}}
  \startdata
  197 & $     6.720_{  0.000}^{+  0.179}  $ & $     0.19_{-  0.00}^{+  0.00} $ &   0.31	 & $     6.720_{  0.000}^{+  0.180}  $ & $     0.19_{-  0.00}^{+  0.00} $ &   0.31 \\
198 & $     6.820_{ -0.083}^{+  0.156}  $ & $     0.56_{-  0.20}^{+  0.00} $ &   0.72	 & $     6.820_{ -0.082}^{+  0.155}  $ & $     0.56_{-  0.20}^{+  0.00} $ &   0.72 \\
284 & $     8.357_{ -1.449}^{+  1.072}  $ & $    -1.27_{-  0.00}^{+  0.00} $ &   0.45	 & $     8.357_{ -1.448}^{+  1.053}  $ & $    -1.27_{-  0.00}^{+  0.00} $ &   0.45 \\
243 & $     6.980_{ -0.331}^{+  2.186}  $ & $     0.37_{-  1.91}^{+  0.00} $ &   0.69	 & $     6.980_{ -0.330}^{+  2.185}  $ & $     0.37_{-  1.92}^{+  0.00} $ &   0.69 \\
245 & $     6.920_{ -0.246}^{+  2.260}  $ & $     0.28_{-  1.69}^{+  0.00} $ &   1.00	 & $     6.920_{ -0.245}^{+  2.259}  $ & $     0.28_{-  1.69}^{+  0.00} $ &   1.00 \\
371 & $     8.057_{ -1.361}^{+  0.890}  $ & $    -0.19_{-  0.00}^{+  0.00} $ &   0.37	 & $     8.057_{ -1.360}^{+  0.885}  $ & $    -0.19_{-  0.00}^{+  0.00} $ &   0.38 \\
140 & $     6.660_{ -0.165}^{+  0.051}  $ & $    -2.01_{-  0.00}^{+  0.38} $ &   0.31	 & $     6.600_{ -0.103}^{+  0.106}  $ & $    -1.92_{-  0.00}^{+  0.42} $ &   0.32 \\
402 & $    10.106_{ -1.134}^{+  0.000}  $ & $    -1.59_{-  0.00}^{+  1.69} $ &   0.38	 & $    10.011_{ -1.029}^{+  0.000}  $ & $    -1.69_{-  0.00}^{+  1.79} $ &   0.37 \\
427 & $     9.057_{ -1.036}^{+  0.000}  $ & $    -1.31_{-  0.00}^{+  0.00} $ &   0.40	 & $     9.057_{ -1.033}^{+  0.000}  $ & $    -1.36_{-  0.00}^{+  0.00} $ &   0.41 \\
206 & $    10.097_{ -1.207}^{+  0.000}  $ & $    -1.13_{-  0.00}^{+  1.67} $ &   0.76	 & $    10.097_{ -1.197}^{+  0.000}  $ & $    -1.27_{-  0.00}^{+  1.77} $ &   0.76 \\
70 & $     9.007_{ -1.609}^{+  0.000}  $ & $    -1.45_{-  0.00}^{+  0.00} $ &   0.98	 & $     9.007_{ -1.573}^{+  0.000}  $ & $    -1.50_{-  0.00}^{+  0.00} $ &   0.98 \\
94 & $     8.707_{ -1.555}^{+  0.819}  $ & $    -1.78_{-  0.00}^{+  0.00} $ &   0.70	 & $     8.707_{ -1.542}^{+  0.789}  $ & $    -1.83_{-  0.00}^{+  0.00} $ &   0.70 \\
228 & $     6.940_{ -0.186}^{+  2.368}  $ & $    -0.00_{-  1.28}^{+  0.00} $ &   0.56	 & $     6.940_{ -0.185}^{+  2.361}  $ & $    -0.00_{-  1.27}^{+  0.00} $ &   0.56 \\
221 & $     8.907_{ -1.779}^{+  0.000}  $ & $    -1.92_{-  0.00}^{+  0.00} $ &   1.83	 & $     8.907_{ -1.768}^{+  0.000}  $ & $    -1.97_{-  0.00}^{+  0.00} $ &   1.83 \\
85 & $     8.707_{ -1.379}^{+  0.651}  $ & $    -1.87_{-  0.00}^{+  0.00} $ &   0.29	 & $     8.707_{ -1.349}^{+  0.591}  $ & $    -1.92_{-  0.00}^{+  0.00} $ &   0.30 \\
214 & $     8.907_{ -1.534}^{+  0.000}  $ & $    -1.64_{-  0.00}^{+  0.00} $ &   0.60	 & $     8.907_{ -1.500}^{+  1.313}  $ & $    -1.69_{-  0.00}^{+  0.00} $ &   0.60 \\
329 & $     9.007_{ -1.659}^{+  0.000}  $ & $    -2.25_{-  0.00}^{+  0.00} $ &   1.11	 & $     8.957_{ -1.618}^{+  0.000}  $ & $    -2.25_{-  0.00}^{+  0.00} $ &   1.12 \\

  \enddata
\end{deluxetable}

\clearpage
\pagestyle{empty}
\begin{deluxetable}{r|rrr|rrr}
  \tablecolumns{4} \tablewidth{0pc} \tablecaption{Same as in
    Table~\ref{t3.tab} but from full-spectrum fitting with Padova 2000
    stellar evolutionary tracks of the \citet{bc03} models. Both the\citet{chab} IMF and
    \citet{salp} IMF were adopted for the fitting separately.
    \label{t4.tab}}
  \tablehead{\colhead{} & \multicolumn{3}{c}{\citet{chab} IMF}
    &\multicolumn{3}{c}{\citet{salp} IMF} \\
    \cline{2-7}
    \colhead{ID} & \colhead{log $t$} &  \colhead{$\rm
      [Fe/H]$} &  \colhead{$\chi^2_{min}/dof$}  & \colhead{log $t$} &  \colhead{$\rm
      [Fe/H]$} &  \colhead{$\chi^2_{min}/dof$}   \\
    \colhead{} & \colhead{(yr)} & \colhead{(dex)} & \colhead{} & \colhead{(yr)} & \colhead{(dex)} & \colhead{}}
  \startdata
  197 & $     6.900_{  0.000}^{+  1.156}  $ & $    -1.00_{-  0.00}^{+  0.00} $ &   0.32	 & $     6.900_{  0.000}^{+  1.157}  $ & $    -1.00_{-  0.00}^{+  0.00} $ &   0.32 \\
198 & $     6.940_{ -0.046}^{+  0.657}  $ & $     0.16_{-  0.35}^{+  0.00} $ &   1.19	 & $     6.940_{ -0.045}^{+  0.656}  $ & $     0.16_{-  0.35}^{+  0.00} $ &   1.24 \\
284 & $     8.307_{ -1.192}^{+  1.171}  $ & $    -1.26_{-  0.00}^{+  0.00} $ &   0.45	 & $     8.307_{ -1.193}^{+  1.082}  $ & $    -1.26_{-  0.00}^{+  0.00} $ &   0.45 \\
243 & $     6.980_{ -0.277}^{+  2.354}  $ & $     0.06_{-  1.25}^{+  0.00} $ &   0.80	 & $     6.960_{ -0.255}^{+  2.196}  $ & $     0.26_{-  1.43}^{+  0.00} $ &   0.79 \\
245 & $     6.980_{ -0.270}^{+  2.279}  $ & $     0.06_{-  1.24}^{+  0.00} $ &   1.03	 & $     6.960_{ -0.249}^{+  2.310}  $ & $     0.03_{-  1.19}^{+  0.00} $ &   1.03 \\
371 & $     8.057_{ -1.375}^{+  0.958}  $ & $    -0.16_{-  0.00}^{+  0.00} $ &   0.37	 & $     6.800_{ -0.214}^{+  2.253}  $ & $    -0.42_{-  0.47}^{+  0.00} $ &   0.37 \\
140 & $     7.544_{ -0.686}^{+  1.584}  $ & $    -0.71_{-  0.00}^{+  0.00} $ &   0.35	 & $     7.544_{ -0.684}^{+  1.582}  $ & $    -0.71_{-  0.00}^{+  0.00} $ &   0.35 \\
402 & $    10.097_{ -1.082}^{+  0.000}  $ & $    -1.49_{-  0.00}^{+  1.60} $ &   0.37	 & $    10.000_{ -0.995}^{+  0.000}  $ & $    -1.58_{-  0.00}^{+  1.75} $ &   0.37 \\
427 & $     9.107_{ -1.102}^{+  0.000}  $ & $    -1.26_{-  0.00}^{+  0.00} $ &   0.40	 & $     9.107_{ -1.094}^{+  0.000}  $ & $    -1.32_{-  0.00}^{+  0.00} $ &   0.41 \\
206 & $     9.989_{ -0.988}^{+  0.000}  $ & $    -1.07_{-  0.00}^{+  0.00} $ &   0.74	 & $     9.989_{ -0.983}^{+  0.000}  $ & $    -1.16_{-  0.00}^{+  0.00} $ &   0.74 \\
70 & $     9.057_{ -1.480}^{+  0.000}  $ & $    -1.26_{-  0.00}^{+  0.00} $ &   0.94	 & $     9.057_{ -1.481}^{+  0.000}  $ & $    -1.26_{-  0.00}^{+  0.00} $ &   0.95 \\
94 & $     8.657_{ -1.528}^{+  0.710}  $ & $    -1.45_{-  0.00}^{+  0.00} $ &   0.71	 & $     8.607_{ -1.483}^{+  0.704}  $ & $    -1.36_{-  0.00}^{+  0.00} $ &   0.72 \\
228 & $     6.940_{ -0.193}^{+  2.402}  $ & $    -0.03_{-  0.99}^{+  0.00} $ &   0.56	 & $     6.940_{ -0.193}^{+  2.396}  $ & $    -0.03_{-  0.99}^{+  0.00} $ &   0.56 \\
221 & $     8.857_{ -1.796}^{+  0.000}  $ & $    -1.65_{-  0.00}^{+  0.00} $ &   1.83	 & $     8.857_{ -1.797}^{+  0.000}  $ & $    -1.65_{-  0.00}^{+  0.00} $ &   1.83 \\
85 & $     8.557_{ -1.408}^{+  0.695}  $ & $    -1.42_{-  0.00}^{+  0.00} $ &   0.31	 & $     8.507_{ -1.360}^{+  0.729}  $ & $    -1.39_{-  0.00}^{+  0.00} $ &   0.31 \\
214 & $     8.957_{ -1.514}^{+  1.299}  $ & $    -1.49_{-  0.00}^{+  0.00} $ &   0.60	 & $     8.907_{ -1.456}^{+  1.096}  $ & $    -1.42_{-  0.00}^{+  0.00} $ &   0.60 \\
329 & $     8.607_{ -2.172}^{+  1.345}  $ & $    -0.74_{-  0.00}^{+  0.00} $ &   1.10	 & $     8.557_{ -2.121}^{+  1.275}  $ & $    -0.68_{-  0.00}^{+  0.00} $ &   1.10 \\

  \enddata
\end{deluxetable}

\clearpage
\pagestyle{empty}
\begin{deluxetable}{rrr}
  \tablecolumns{3} \tablewidth{0pc} \tablecaption{The photometry of
    our sample star clusters. The SM IDs are from \citet{sara07}. The
    $u_{\rm SC}$ and $v_{\rm SAGE}$ are from the
    SAGE survey. The aperture correction have been done and calibrated
    in the AB system.
    \label{t5.tab}}
  \tablehead{\colhead{ID} &  \colhead{$u_{\rm SC}$} &  \colhead{$v_{\rm SAGE}$} \\
    \colhead{} & \colhead{(mag)} & \colhead{(mag)}}
  \startdata
     197& $  16.17\pm   0.01$ & $  16.33\pm   0.01$ \\
   198& $  17.00\pm   0.03$ & $  17.06\pm   0.03$ \\
   284& $  18.30\pm   0.02$ & $  17.49\pm   0.03$ \\
   243& $  17.29\pm   0.01$ & $  17.31\pm   0.02$ \\
   245& $  18.25\pm   0.02$ & $  17.76\pm   0.03$ \\
   371& $  17.83\pm   0.01$ & $  17.39\pm   0.03$ \\
   140& $  18.06\pm   0.02$ & $  17.55\pm   0.02$ \\
   402& $  18.76\pm   0.03$ & $  18.27\pm   0.05$ \\
   427& $  18.95\pm   0.04$ & $  18.19\pm   0.05$ \\
   206& $  19.46\pm   0.04$ & $  19.07\pm   0.07$ \\
    70& $  18.51\pm   0.02$ & $  18.11\pm   0.03$ \\
    94& $  19.03\pm   0.04$ & $  18.33\pm   0.04$ \\
   228& $  18.54\pm   0.03$ & $  18.36\pm   0.05$ \\
   221& $  19.69\pm   0.07$ & $  19.56\pm   0.13$ \\
    85& $  18.19\pm   0.02$ & $  17.83\pm   0.03$ \\
   214& $  18.57\pm   0.02$ & $  18.08\pm   0.04$ \\
   329& $  19.23\pm   0.05$ & $  18.30\pm   0.05$ \\

 \enddata
\end{deluxetable}

\clearpage
\pagestyle{empty}
\begin{deluxetable}{r|rrr|rrr}
  \tablecolumns{4} \tablewidth{0pc} \tablecaption{Same as in
    Table~\ref{t3.tab} but from the spectroscopy+ SAGE photometry fitting. BC03
    models with Padova 1994 stellar evolutionary tracks, and the
    \citet{chab} IMF and \citet{salp} IMF were adopted.
    \label{t6.tab}}
  \tablehead{\colhead{} & \multicolumn{3}{c}{\citet{chab} IMF}
    &\multicolumn{3}{c}{\citet{salp} IMF} \\
    \cline{2-7}
    \colhead{ID} & \colhead{log $t$} &  \colhead{$\rm
      [Fe/H]$} &  \colhead{$\chi^2_{min}/dof$}  & \colhead{log $t$} &  \colhead{$\rm
      [Fe/H]$} &  \colhead{$\chi^2_{min}/dof$}   \\
    \colhead{} & \colhead{(yr)} & \colhead{(dex)} & \colhead{} & \colhead{(yr)} & \colhead{(dex)} & \colhead{}}
  \startdata
  197 & $     6.660_{ -0.184}^{+  0.077}  $ & $    -1.59_{-  0.27}^{+  0.00} $ &   0.56	 & $     6.660_{ -0.184}^{+  0.078}  $ & $    -1.59_{-  0.27}^{+  0.00} $ &   0.56 \\
198 & $     6.820_{ -0.064}^{+  0.103}  $ & $     0.56_{-  0.14}^{+  0.00} $ &   1.57	 & $     6.820_{ -0.063}^{+  0.102}  $ & $     0.56_{-  0.14}^{+  0.00} $ &   1.62 \\
284 & $     8.657_{ -0.563}^{+  0.403}  $ & $    -0.56_{-  1.67}^{+  0.00} $ &   0.80	 & $     8.657_{ -0.563}^{+  0.402}  $ & $    -0.56_{-  0.00}^{+  0.00} $ &   0.81 \\
243 & $     6.980_{ -0.241}^{+  0.967}  $ & $     0.19_{-  1.52}^{+  0.36} $ &   0.84	 & $     6.980_{ -0.240}^{+  0.967}  $ & $     0.19_{-  1.52}^{+  0.37} $ &   0.85 \\
245 & $     6.580_{ -0.081}^{+  0.111}  $ & $    -2.25_{-  0.00}^{+  0.34} $ &   1.59	 & $     6.580_{ -0.081}^{+  0.110}  $ & $    -2.25_{-  0.00}^{+  0.33} $ &   1.59 \\
371 & $     8.157_{ -0.309}^{+  0.865}  $ & $    -2.25_{-  0.00}^{+  0.00} $ &   0.55	 & $     8.157_{ -0.315}^{+  0.861}  $ & $    -2.25_{-  0.00}^{+  0.00} $ &   0.55 \\
140 & $     6.540_{ -0.042}^{+  0.158}  $ & $    -2.25_{-  0.00}^{+  0.29} $ &   0.46	 & $     6.560_{ -0.061}^{+  0.131}  $ & $    -2.20_{-  0.00}^{+  0.29} $ &   0.47 \\
402 & $     9.942_{ -0.848}^{+  0.000}  $ & $    -1.55_{-  0.00}^{+  0.98} $ &   0.51	 & $     9.903_{ -0.805}^{+  0.000}  $ & $    -1.59_{-  0.00}^{+  1.01} $ &   0.51 \\
427 & $     9.007_{ -0.561}^{+  0.362}  $ & $    -0.47_{-  1.56}^{+  0.78} $ &   0.53	 & $     9.007_{ -0.571}^{+  0.382}  $ & $    -0.52_{-  1.55}^{+  0.83} $ &   0.53 \\
206 & $     9.107_{ -0.302}^{+  0.539}  $ & $     0.47_{-  1.38}^{+  0.00} $ &   0.91	 & $     9.057_{ -0.285}^{+  0.452}  $ & $     0.56_{-  0.92}^{+  0.00} $ &   0.92 \\
70 & $     6.640_{ -0.141}^{+  0.056}  $ & $    -2.25_{-  0.00}^{+  0.29} $ &   1.06	 & $     6.600_{ -0.100}^{+  0.085}  $ & $    -2.16_{-  0.00}^{+  0.28} $ &   1.06 \\
94 & $     8.907_{ -0.857}^{+  0.538}  $ & $    -1.78_{-  0.00}^{+  1.57} $ &   1.03	 & $     8.857_{ -0.818}^{+  0.421}  $ & $    -1.69_{-  0.00}^{+  1.63} $ &   1.05 \\
228 & $     6.640_{ -0.142}^{+  0.061}  $ & $    -2.01_{-  0.00}^{+  0.29} $ &   0.61	 & $     6.620_{ -0.121}^{+  0.075}  $ & $    -1.97_{-  0.00}^{+  0.38} $ &   0.61 \\
221 & $     6.580_{ -0.084}^{+  0.124}  $ & $    -2.25_{-  0.00}^{+  0.51} $ &   2.03	 & $     6.580_{ -0.084}^{+  0.124}  $ & $    -2.25_{-  0.00}^{+  0.50} $ &   2.03 \\
85 & $     6.820_{ -0.056}^{+  0.159}  $ & $    -0.38_{-  0.25}^{+  0.38} $ &   0.39	 & $     6.820_{ -0.057}^{+  0.156}  $ & $    -0.33_{-  0.29}^{+  0.33} $ &   0.41 \\
214 & $     6.580_{ -0.081}^{+  0.109}  $ & $    -2.20_{-  0.00}^{+  0.32} $ &   0.68	 & $     6.580_{ -0.081}^{+  0.108}  $ & $    -2.20_{-  0.00}^{+  0.31} $ &   0.68 \\
329 & $     8.657_{ -0.616}^{+  0.399}  $ & $    -0.19_{-  0.00}^{+  0.00} $ &   1.22	 & $     8.657_{ -0.615}^{+  0.398}  $ & $    -0.19_{-  0.00}^{+  0.00} $ &   1.22 \\

  \enddata
\end{deluxetable}

\clearpage
\pagestyle{empty}
\begin{deluxetable}{r|rrr|rrr}
  \tablecolumns{4} \tablewidth{0pc} \tablecaption{Same as in
    Table~\ref{t6.tab} from spectroscopy+ SAGE photometry fitting, but
    for the BC03 models with Padova 2000 stellar evolutionary tracks. The
    \citet{chab} IMF and \citet{salp} IMF were adopted separately.
    \label{t7.tab}}
  \tablehead{\colhead{} & \multicolumn{3}{c}{\citet{chab} IMF}
    &\multicolumn{3}{c}{\citet{salp} IMF} \\
    \cline{2-7}
    \colhead{ID} & \colhead{log $t$} &  \colhead{$\rm
      [Fe/H]$} &  \colhead{$\chi^2_{min}/dof$}  & \colhead{log $t$} &  \colhead{$\rm
      [Fe/H]$} &  \colhead{$\chi^2_{min}/dof$}   \\
    \colhead{} & \colhead{(yr)} & \colhead{(dex)} & \colhead{} & \colhead{(yr)} & \colhead{(dex)} & \colhead{}}
  \startdata
  197 & $     6.740_{ -0.230}^{+  0.053}  $ & $    -0.29_{-  0.00}^{+  0.00} $ &   0.38	 & $     6.680_{ -0.207}^{+  0.819}  $ & $    -1.23_{-  0.00}^{+  0.00} $ &   0.53 \\
198 & $     6.940_{ -0.037}^{+  0.564}  $ & $     0.13_{-  0.32}^{+  0.00} $ &   1.63	 & $     6.940_{ -0.035}^{+  0.649}  $ & $     0.13_{-  0.33}^{+  0.00} $ &   1.82 \\
284 & $     8.657_{ -1.362}^{+  0.235}  $ & $     0.26_{-  0.00}^{+  0.00} $ &   0.65	 & $     8.707_{ -0.582}^{+  0.399}  $ & $    -0.55_{-  0.00}^{+  0.00} $ &   0.80 \\
243 & $     7.000_{ -0.256}^{+  1.774}  $ & $     0.16_{-  1.23}^{+  0.00} $ &   0.85	 & $     6.980_{ -0.231}^{+  0.943}  $ & $     0.16_{-  1.20}^{+  0.00} $ &   0.86 \\
245 & $     8.757_{ -1.579}^{+  0.329}  $ & $    -0.20_{-  0.00}^{+  0.00} $ &   1.43	 & $     8.357_{ -0.598}^{+  0.627}  $ & $    -0.39_{-  0.00}^{+  0.00} $ &   1.93 \\
371 & $     6.800_{ -0.060}^{+  2.126}  $ & $    -0.39_{-  0.29}^{+  0.00} $ &   0.39	 & $     7.857_{ -0.877}^{+  0.584}  $ & $     0.03_{-  0.00}^{+  0.00} $ &   0.52 \\
140 & $     8.657_{ -1.370}^{+  0.457}  $ & $    -0.78_{-  0.00}^{+  0.00} $ &   0.45	 & $     8.307_{ -0.640}^{+  0.768}  $ & $    -1.23_{-  0.00}^{+  0.00} $ &   0.80 \\
402 & $     9.989_{ -0.889}^{+  0.000}  $ & $    -1.32_{-  0.00}^{+  1.13} $ &   0.38	 & $     9.929_{ -0.774}^{+  0.000}  $ & $    -1.61_{-  0.00}^{+  1.08} $ &   0.50 \\
427 & $     9.007_{ -0.480}^{+  0.333}  $ & $    -0.23_{-  0.00}^{+  0.00} $ &   0.46	 & $     9.007_{ -0.440}^{+  0.137}  $ & $    -0.29_{-  0.00}^{+  0.00} $ &   0.48 \\
206 & $     9.157_{ -0.291}^{+  1.109}  $ & $     0.29_{-  1.25}^{+  0.00} $ &   0.80	 & $     9.157_{ -0.268}^{+  0.709}  $ & $     0.26_{-  1.14}^{+  0.00} $ &   0.87 \\
70 & $     9.057_{ -1.044}^{+  0.694}  $ & $    -1.36_{-  0.00}^{+  1.36} $ &   0.95	 & $     8.957_{ -0.935}^{+  0.388}  $ & $    -1.65_{-  0.00}^{+  1.18} $ &   2.03 \\
94 & $     8.607_{ -1.717}^{+  0.279}  $ & $     0.22_{-  0.00}^{+  0.00} $ &   0.89	 & $     8.607_{ -0.833}^{+  0.504}  $ & $    -0.81_{-  0.00}^{+  0.00} $ &   0.96 \\
228 & $     6.940_{ -0.179}^{+  2.058}  $ & $    -0.03_{-  0.91}^{+  0.00} $ &   0.67	 & $     7.220_{ -0.366}^{+  1.059}  $ & $     0.13_{-  0.59}^{+  0.00} $ &   0.73 \\
221 & $     8.707_{ -1.971}^{+  0.467}  $ & $     0.26_{-  0.00}^{+  0.00} $ &   1.94	 & $     8.907_{ -1.301}^{+  0.969}  $ & $    -1.52_{-  0.00}^{+  0.00} $ &   2.14 \\
85 & $     8.557_{ -1.202}^{+  0.554}  $ & $    -1.16_{-  0.00}^{+  0.00} $ &   0.31	 & $     6.820_{ -0.056}^{+  0.157}  $ & $    -0.36_{-  0.27}^{+  0.33} $ &   0.40 \\
214 & $     8.957_{ -1.279}^{+  0.606}  $ & $    -1.26_{-  0.00}^{+  1.40} $ &   0.61	 & $     8.907_{ -0.883}^{+  0.408}  $ & $    -1.65_{-  0.00}^{+  1.31} $ &   1.20 \\
329 & $     8.657_{ -1.795}^{+  0.339}  $ & $     0.26_{-  0.00}^{+  0.00} $ &   1.14	 & $     8.607_{ -0.575}^{+  0.307}  $ & $     0.29_{-  0.00}^{+  0.00} $ &   1.20 \\

  \enddata
\end{deluxetable}

\clearpage
\pagestyle{empty}
\begin{deluxetable}{rrrrrrrrrrr}
  \rotate
  \tablecolumns{11} \tablewidth{0pc} \tablecaption{The photometry of
    our sample star clusters from references. The $UBVRI$ photometry
    are from \citet{ma13}, which are in the Johnson-Cousins
    system. The $ugriz$ photometry are from \citet{san10} which are in
    the AB system.
    \label{t8.tab}}
  \tablehead{\colhead{ID} &
    \colhead{$U$} &  \colhead{$B$} & \colhead{$V$} &  \colhead{$R$}  &
    \colhead{$I$} & \colhead{$u$} & \colhead{$g$} & \colhead{$r$} &
    \colhead{$i$} & \colhead{$z$} \\
    \colhead{} & \colhead{(mag)} &
    \colhead{(mag)} & \colhead{(mag)} & \colhead{(mag)} &
    \colhead{(mag)} & \colhead{(mag)} &
    \colhead{(mag)} & \colhead{(mag)} & \colhead{(mag)} & \colhead{(mag)}}
  \startdata
     197& $  15.54$ & $  16.41$ & $  16.38$ & $  16.51$ & $  16.60$ & $  99.99$ & $  99.99$ & $  99.99$ & $  99.99$ & $  99.99$ \\
   198& $  16.37$ & $  17.09$ & $  16.78$ & $  16.38$ & $  15.65$ & $  16.81$ & $  16.81$ & $  16.63$ & $  16.24$ & $  15.77$ \\
   284& $  17.04$ & $  17.17$ & $  16.82$ & $  16.66$ & $  16.39$ & $  99.99$ & $  99.99$ & $  99.99$ & $  99.99$ & $  99.99$ \\
   243& $  16.52$ & $  17.20$ & $  17.08$ & $  16.97$ & $  16.72$ & $  17.07$ & $  17.10$ & $  17.20$ & $  17.15$ & $  17.05$ \\
   245& $  17.24$ & $  17.50$ & $  17.17$ & $  16.94$ & $  16.64$ & $  17.90$ & $  17.31$ & $  17.26$ & $  17.14$ & $  16.98$ \\
   371& $  17.04$ & $  17.39$ & $  17.12$ & $  16.98$ & $  16.76$ & $  17.72$ & $  17.22$ & $  17.22$ & $  17.20$ & $  17.08$ \\
   140& $  17.12$ & $  17.42$ & $  17.12$ & $  16.94$ & $  16.68$ & $  17.38$ & $  17.09$ & $  17.05$ & $  16.95$ & $  16.82$ \\
   402& $  17.85$ & $  17.88$ & $  17.13$ & $  16.62$ & $  16.11$ & $  18.62$ & $  17.57$ & $  17.06$ & $  16.72$ & $  16.44$ \\
   427& $  17.78$ & $  17.75$ & $  17.25$ & $  16.97$ & $  16.58$ & $  99.99$ & $  99.99$ & $  99.99$ & $  99.99$ & $  99.99$ \\
   206& $  18.30$ & $  18.13$ & $  17.29$ & $  16.77$ & $  16.28$ & $  18.83$ & $  17.78$ & $  17.18$ & $  16.82$ & $  16.59$ \\
    70& $  17.56$ & $  17.89$ & $  17.59$ & $  17.41$ & $  17.09$ & $  18.01$ & $  17.63$ & $  17.65$ & $  17.55$ & $  17.47$ \\
    94& $  17.92$ & $  18.05$ & $  17.71$ & $  17.50$ & $  17.20$ & $  18.45$ & $  17.89$ & $  17.85$ & $  17.70$ & $  17.58$ \\
   228& $  17.93$ & $  18.47$ & $  18.17$ & $  17.90$ & $  17.34$ & $  18.32$ & $  18.18$ & $  18.07$ & $  17.72$ & $  17.30$ \\
   221& $  18.87$ & $  18.94$ & $  18.59$ & $  18.44$ & $  18.25$ & $  99.99$ & $  99.99$ & $  99.99$ & $  99.99$ & $  99.99$ \\
    85& $  17.39$ & $  17.84$ & $  17.57$ & $  17.40$ & $  17.11$ & $  17.79$ & $  17.49$ & $  17.50$ & $  17.39$ & $  17.31$ \\
   214& $  17.56$ & $  17.89$ & $  17.55$ & $  17.34$ & $  17.00$ & $  18.23$ & $  17.69$ & $  17.63$ & $  17.50$ & $  17.34$ \\
   329& $  17.81$ & $  17.90$ & $  17.48$ & $  17.34$ & $  16.94$ & $  18.70$ & $  17.95$ & $  18.00$ & $  17.98$ & $  17.90$ \\

 \enddata
\end{deluxetable}

\clearpage
\pagestyle{empty}
\begin{deluxetable}{r|rrr|rrr}
  \tablecolumns{4} \tablewidth{0pc} \tablecaption{Same as in
    Table~\ref{t6.tab} but from the spectroscopy+ SAGE, UVBRI, ugriz
    photometry fitting. BC03 models with Padova 1994 stellar
    evolutionary tracks, and the \citet{chab} IMF and \citet{salp} IMF
    have been adopted separately for the fitting.
    \label{t9.tab}}
  \tablehead{\colhead{} & \multicolumn{3}{c}{\citet{chab} IMF}
    &\multicolumn{3}{c}{\citet{salp} IMF} \\
    \cline{2-7}
    \colhead{ID} & \colhead{log $t$} &  \colhead{$\rm
      [Fe/H]$} &  \colhead{$\chi^2_{min}/dof$}  & \colhead{log $t$} &  \colhead{$\rm
      [Fe/H]$} &  \colhead{$\chi^2_{min}/dof$}   \\
    \colhead{} & \colhead{(yr)} & \colhead{(dex)} & \colhead{} & \colhead{(yr)} & \colhead{(dex)} & \colhead{}}
  \startdata
  197 & $     6.700_{ -0.072}^{+  0.043}  $ & $    -0.14_{-  1.74}^{+  0.57} $ &   1.19	 & $     6.700_{ -0.075}^{+  0.044}  $ & $    -0.14_{-  1.75}^{+  0.58} $ &   1.19 \\
198 & $     6.940_{ -0.017}^{+  0.146}  $ & $     0.19_{-  0.19}^{+  0.24} $ &   3.17	 & $     6.940_{ -0.017}^{+  0.146}  $ & $     0.19_{-  0.19}^{+  0.24} $ &   3.38 \\
284 & $     8.307_{ -0.356}^{+  0.278}  $ & $    -0.42_{-  1.75}^{+  0.00} $ &   2.12	 & $     8.307_{ -0.356}^{+  0.276}  $ & $    -0.42_{-  1.77}^{+  0.00} $ &   2.13 \\
243 & $     6.860_{ -0.110}^{+  0.061}  $ & $     0.04_{-  0.25}^{+  0.32} $ &   1.81	 & $     6.980_{ -0.109}^{+  0.443}  $ & $    -0.24_{-  1.11}^{+  0.25} $ &   1.82 \\
245 & $     6.520_{ -0.021}^{+  0.065}  $ & $    -2.25_{-  0.00}^{+  0.19} $ &   2.09	 & $     6.540_{ -0.040}^{+  0.142}  $ & $    -2.20_{-  0.00}^{+  0.19} $ &   2.07 \\
371 & $     7.740_{ -0.346}^{+  0.305}  $ & $    -1.13_{-  0.57}^{+  0.00} $ &   0.86	 & $     7.740_{ -0.345}^{+  0.305}  $ & $    -1.13_{-  0.58}^{+  0.00} $ &   0.86 \\
140 & $     7.602_{ -0.208}^{+  0.526}  $ & $    -0.94_{-  0.45}^{+  1.38} $ &   2.73	 & $     7.602_{ -0.210}^{+  0.524}  $ & $    -0.89_{-  0.50}^{+  1.33} $ &   2.74 \\
402 & $     9.916_{ -0.690}^{+  0.000}  $ & $    -1.73_{-  0.00}^{+  0.71} $ &   0.87	 & $     9.889_{ -0.661}^{+  0.000}  $ & $    -1.83_{-  0.00}^{+  0.76} $ &   0.86 \\
427 & $     9.057_{ -0.403}^{+  0.465}  $ & $    -1.55_{-  0.00}^{+  0.84} $ &   1.64	 & $     9.057_{ -0.389}^{+  0.409}  $ & $    -1.59_{-  0.00}^{+  0.87} $ &   1.67 \\
206 & $    10.000_{ -0.619}^{+  0.000}  $ & $    -1.13_{-  0.88}^{+  0.52} $ &   1.57	 & $     9.954_{ -0.598}^{+  0.000}  $ & $    -1.17_{-  0.95}^{+  0.55} $ &   1.62 \\
70 & $     6.820_{ -0.036}^{+  0.094}  $ & $    -0.33_{-  0.18}^{+  0.19} $ &   2.52	 & $     6.580_{ -0.065}^{+  0.084}  $ & $    -2.01_{-  0.15}^{+  0.18} $ &   2.57 \\
94 & $     6.520_{ -0.021}^{+  0.074}  $ & $    -2.25_{-  0.00}^{+  0.16} $ &   2.08	 & $     6.540_{ -0.040}^{+  0.141}  $ & $    -2.25_{-  0.00}^{+  0.21} $ &   2.11 \\
228 & $     7.240_{ -0.323}^{+  0.432}  $ & $     0.19_{-  0.25}^{+  0.23} $ &   2.41	 & $     7.240_{ -0.323}^{+  0.433}  $ & $     0.19_{-  0.25}^{+  0.23} $ &   2.51 \\
221 & $     8.207_{ -0.450}^{+  0.340}  $ & $    -0.28_{-  0.00}^{+  0.00} $ &   2.38	 & $     8.207_{ -0.449}^{+  0.339}  $ & $    -0.28_{-  0.00}^{+  0.00} $ &   2.38 \\
85 & $     6.840_{ -0.066}^{+  0.122}  $ & $    -0.33_{-  0.23}^{+  0.27} $ &   1.55	 & $     6.840_{ -0.065}^{+  0.121}  $ & $    -0.33_{-  0.23}^{+  0.27} $ &   1.56 \\
214 & $     6.580_{ -0.063}^{+  0.082}  $ & $    -2.06_{-  0.17}^{+  0.17} $ &   1.34	 & $     6.580_{ -0.061}^{+  0.080}  $ & $    -2.06_{-  0.19}^{+  0.16} $ &   1.33 \\
329 & $     6.540_{ -0.040}^{+  0.125}  $ & $    -2.25_{-  0.00}^{+  0.18} $ &   7.28	 & $     8.257_{ -0.369}^{+  0.390}  $ & $    -1.13_{-  0.00}^{+  0.00} $ &   7.31 \\

  \enddata
\end{deluxetable}

\clearpage
\pagestyle{empty}
\begin{deluxetable}{r|rrr|rrr}
  \tablecolumns{4} \tablewidth{0pc} \tablecaption{Same as in
    Table~\ref{t9.tab} from the spectroscopy+ SAGE, UVBRI, ugriz
    photometry fitting, but for the Padova 2000 stellar evolutionary
    tracks. The \citet{chab} IMF and \citet{salp} IMF were
    adopted separately for the fitting.
    \label{t10.tab}}
  \tablehead{\colhead{} & \multicolumn{3}{c}{\citet{chab} IMF}
    &\multicolumn{3}{c}{\citet{salp} IMF} \\
    \cline{2-7}
    \colhead{ID} & \colhead{log $t$} &  \colhead{$\rm
      [Fe/H]$} &  \colhead{$\chi^2_{min}/dof$}  & \colhead{log $t$} &  \colhead{$\rm
      [Fe/H]$} &  \colhead{$\chi^2_{min}/dof$}   \\
    \colhead{} & \colhead{(yr)} & \colhead{(dex)} & \colhead{} & \colhead{(yr)} & \colhead{(dex)} & \colhead{}}
  \startdata
  197 & $     6.560_{ -0.100}^{+  0.193}  $ & $    -0.49_{-  0.18}^{+  0.16} $ &   1.20	 & $     6.560_{ -0.100}^{+  0.193}  $ & $    -0.49_{-  0.19}^{+  0.16} $ &   1.20 \\
198 & $     6.940_{ -0.016}^{+  0.132}  $ & $     0.13_{-  0.18}^{+  0.00} $ &   3.34	 & $     6.940_{ -0.016}^{+  0.132}  $ & $     0.13_{-  0.18}^{+  0.00} $ &   3.55 \\
284 & $     8.307_{ -0.338}^{+  0.294}  $ & $    -0.36_{-  0.00}^{+  0.00} $ &   2.05	 & $     8.307_{ -0.338}^{+  0.292}  $ & $    -0.36_{-  0.00}^{+  0.00} $ &   2.06 \\
243 & $     6.860_{ -0.115}^{+  0.060}  $ & $    -0.00_{-  0.22}^{+  0.00} $ &   1.79	 & $     6.860_{ -0.114}^{+  0.062}  $ & $    -0.00_{-  0.23}^{+  0.00} $ &   1.79 \\
245 & $     8.007_{ -0.418}^{+  0.280}  $ & $     0.13_{-  1.68}^{+  0.00} $ &   2.45	 & $     8.007_{ -0.417}^{+  0.281}  $ & $     0.13_{-  1.69}^{+  0.00} $ &   2.46 \\
371 & $     7.757_{ -0.266}^{+  0.336}  $ & $    -0.00_{-  1.62}^{+  0.00} $ &   0.77	 & $     7.757_{ -0.268}^{+  0.336}  $ & $    -0.00_{-  1.63}^{+  0.00} $ &   0.77 \\
140 & $     7.591_{ -0.198}^{+  0.526}  $ & $    -0.74_{-  0.37}^{+  0.00} $ &   2.73	 & $     7.591_{ -0.196}^{+  0.526}  $ & $    -0.74_{-  0.37}^{+  0.00} $ &   2.73 \\
402 & $     9.889_{ -0.654}^{+  0.000}  $ & $    -1.65_{-  0.00}^{+  0.73} $ &   0.83	 & $     9.829_{ -0.608}^{+  0.000}  $ & $    -1.65_{-  0.00}^{+  0.74} $ &   0.84 \\
427 & $     9.057_{ -0.449}^{+  0.489}  $ & $    -1.49_{-  0.00}^{+  1.10} $ &   1.74	 & $     9.057_{ -0.424}^{+  0.458}  $ & $    -1.55_{-  0.00}^{+  1.16} $ &   1.75 \\
206 & $     9.989_{ -0.664}^{+  0.000}  $ & $    -1.03_{-  0.00}^{+  0.51} $ &   1.52	 & $     9.978_{ -0.655}^{+  0.000}  $ & $    -1.10_{-  0.00}^{+  0.54} $ &   1.57 \\
70 & $     6.820_{ -0.036}^{+  0.094}  $ & $    -0.32_{-  0.19}^{+  0.17} $ &   2.53	 & $     6.820_{ -0.035}^{+  0.092}  $ & $    -0.32_{-  0.18}^{+  0.17} $ &   2.63 \\
94 & $     8.157_{ -0.364}^{+  0.285}  $ & $    -0.29_{-  0.00}^{+  0.00} $ &   2.36	 & $     8.157_{ -0.362}^{+  0.284}  $ & $    -0.29_{-  0.00}^{+  0.00} $ &   2.37 \\
228 & $     7.240_{ -0.325}^{+  0.336}  $ & $     0.13_{-  0.22}^{+  0.00} $ &   2.40	 & $     7.240_{ -0.324}^{+  0.335}  $ & $     0.13_{-  0.22}^{+  0.00} $ &   2.49 \\
221 & $     8.207_{ -0.465}^{+  0.388}  $ & $    -0.32_{-  0.00}^{+  0.00} $ &   2.36	 & $     8.207_{ -0.463}^{+  0.386}  $ & $    -0.32_{-  0.00}^{+  0.00} $ &   2.36 \\
85 & $     6.840_{ -0.066}^{+  0.122}  $ & $    -0.29_{-  0.27}^{+  0.22} $ &   1.54	 & $     6.820_{ -0.045}^{+  0.141}  $ & $    -0.26_{-  0.30}^{+  0.18} $ &   1.54 \\
214 & $     7.628_{ -0.195}^{+  0.641}  $ & $    -0.71_{-  0.38}^{+  0.45} $ &   1.75	 & $     7.628_{ -0.194}^{+  0.644}  $ & $    -0.71_{-  0.38}^{+  0.45} $ &   1.79 \\
329 & $     8.157_{ -0.383}^{+  0.591}  $ & $    -1.20_{-  0.00}^{+  0.00} $ &   7.24	 & $     8.157_{ -0.381}^{+  0.578}  $ & $    -1.20_{-  0.00}^{+  0.00} $ &   7.26 \\

  \enddata
\end{deluxetable}

\clearpage

\begin{figure}
\begin{subfigure}
  \centering
  \includegraphics[width=.5\linewidth]{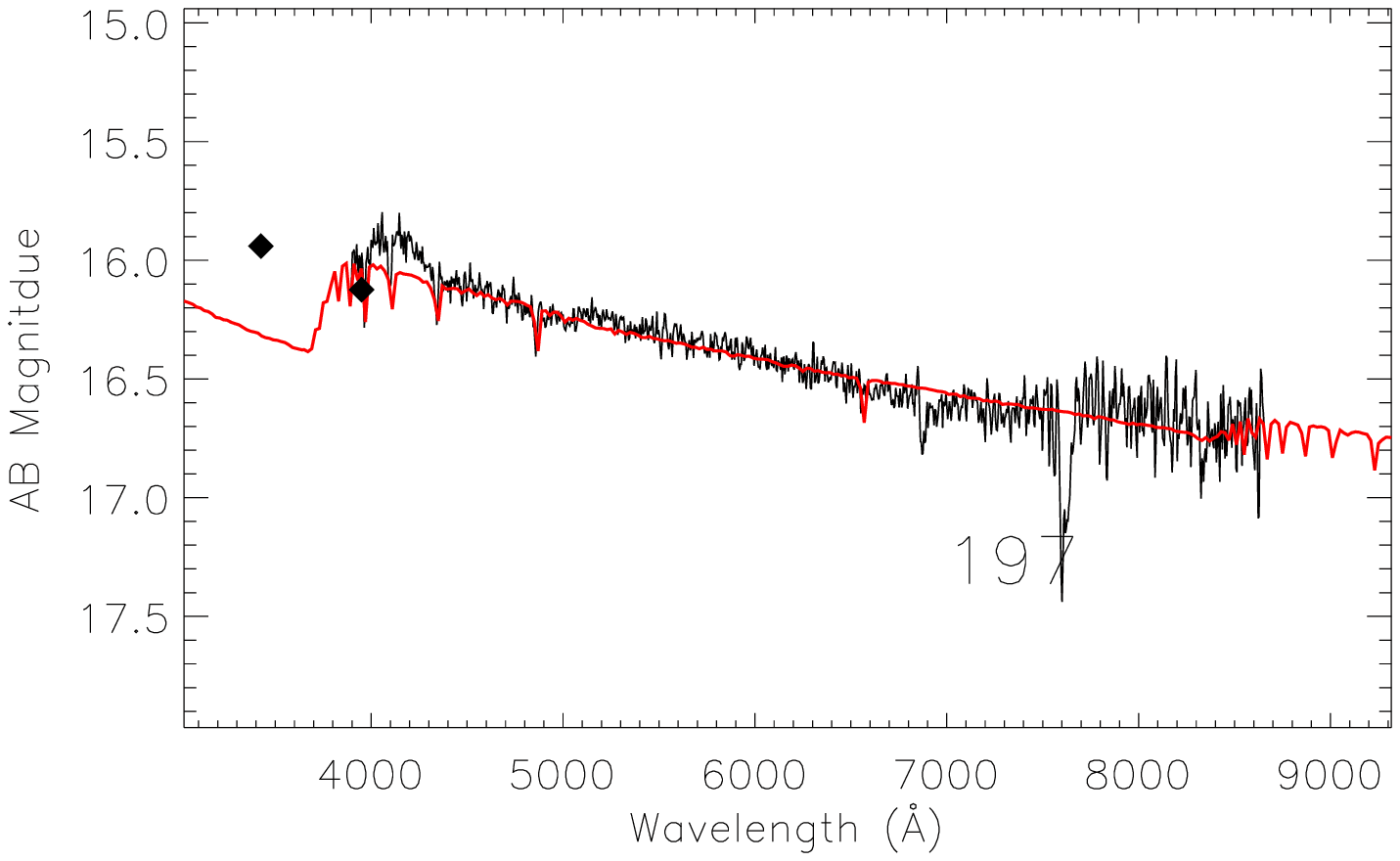}
  \includegraphics[width=.5\linewidth]{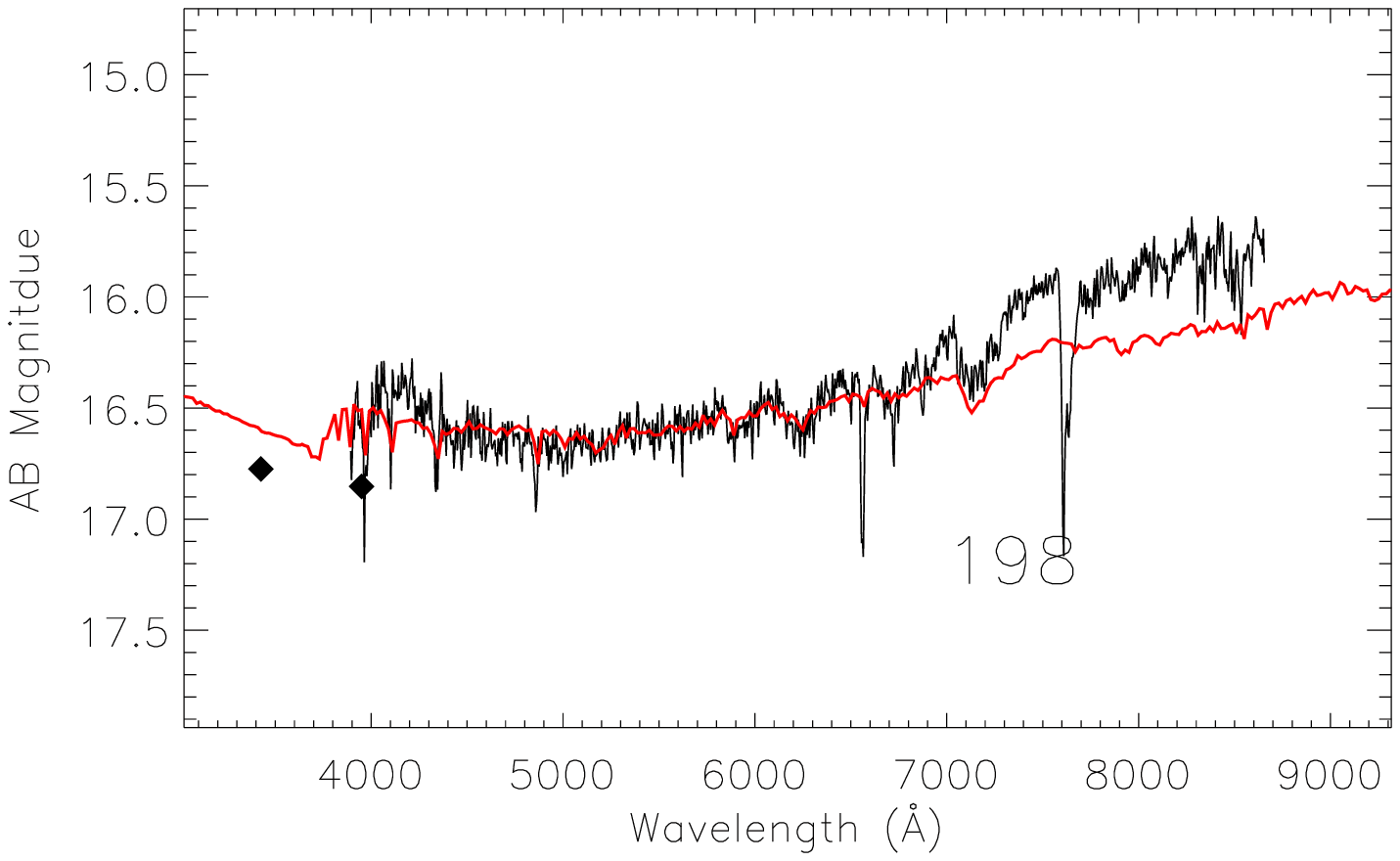}
\end{subfigure}%
\begin{subfigure}
  \centering
  \includegraphics[width=.5\linewidth]{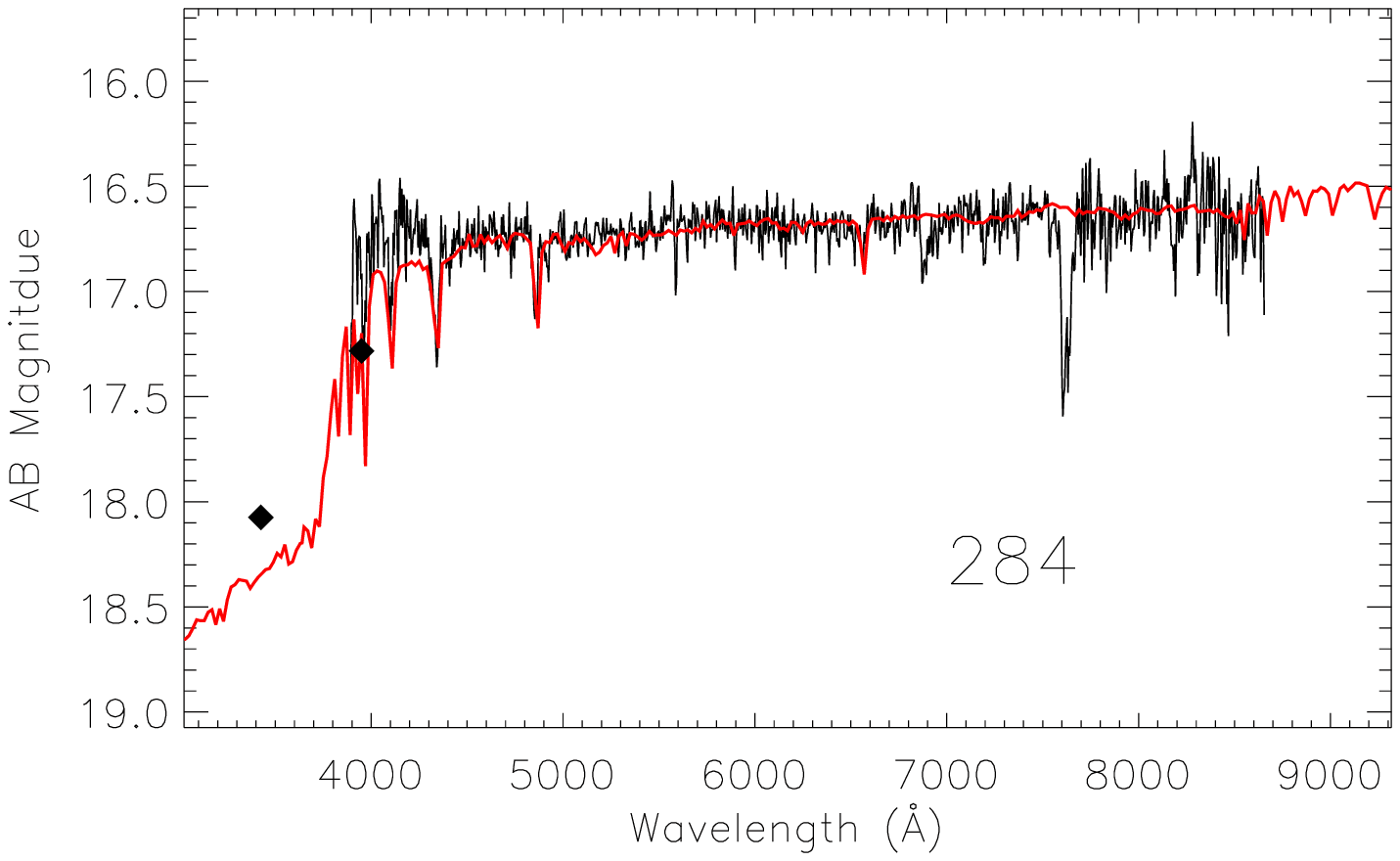}
  \includegraphics[width=.5\linewidth]{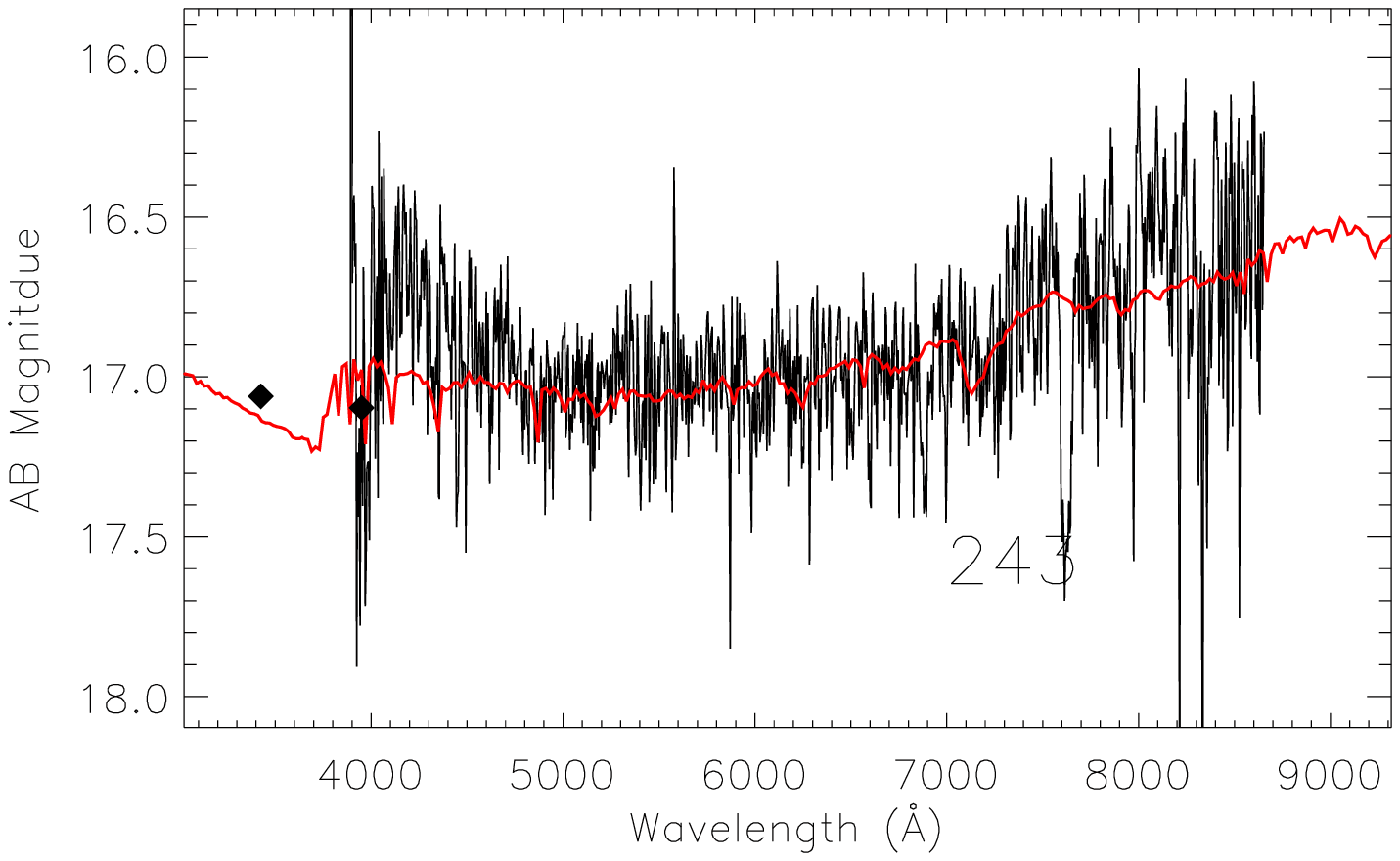}
\end{subfigure}%
\begin{subfigure}
  \centering
  \includegraphics[width=.5\linewidth]{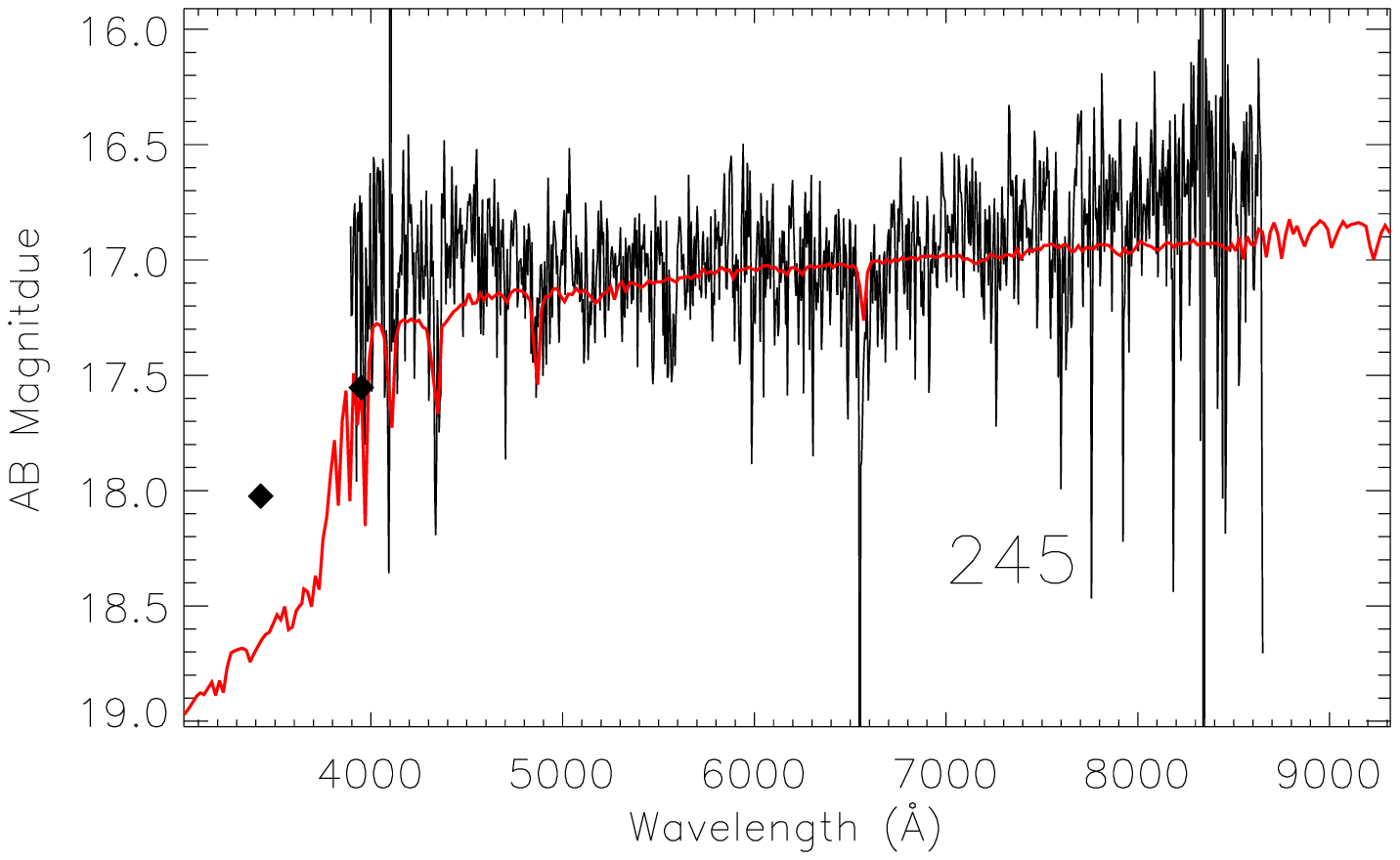}
  \includegraphics[width=.5\linewidth]{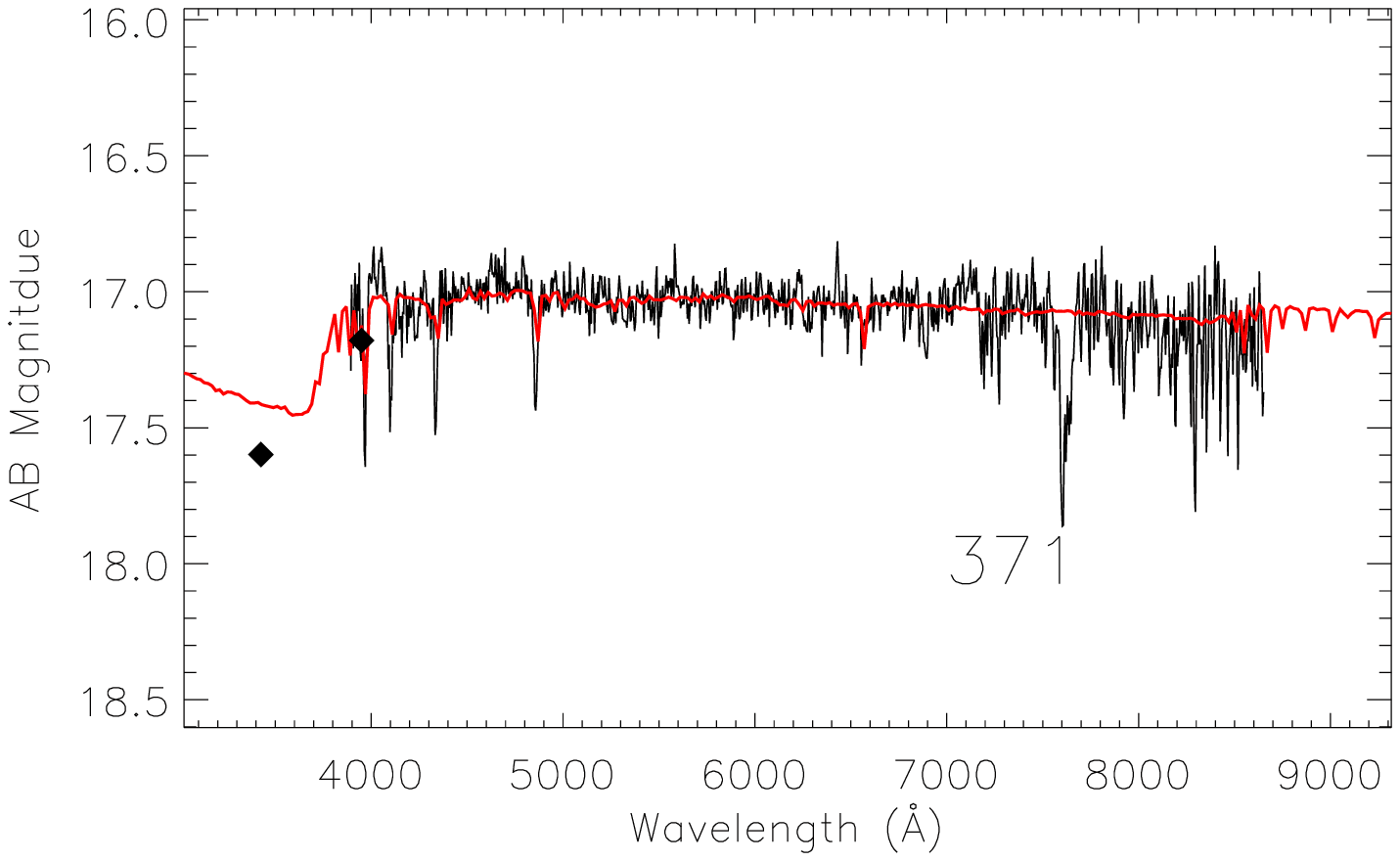}
\end{subfigure}
\begin{subfigure}
  \centering
  \includegraphics[width=.5\linewidth]{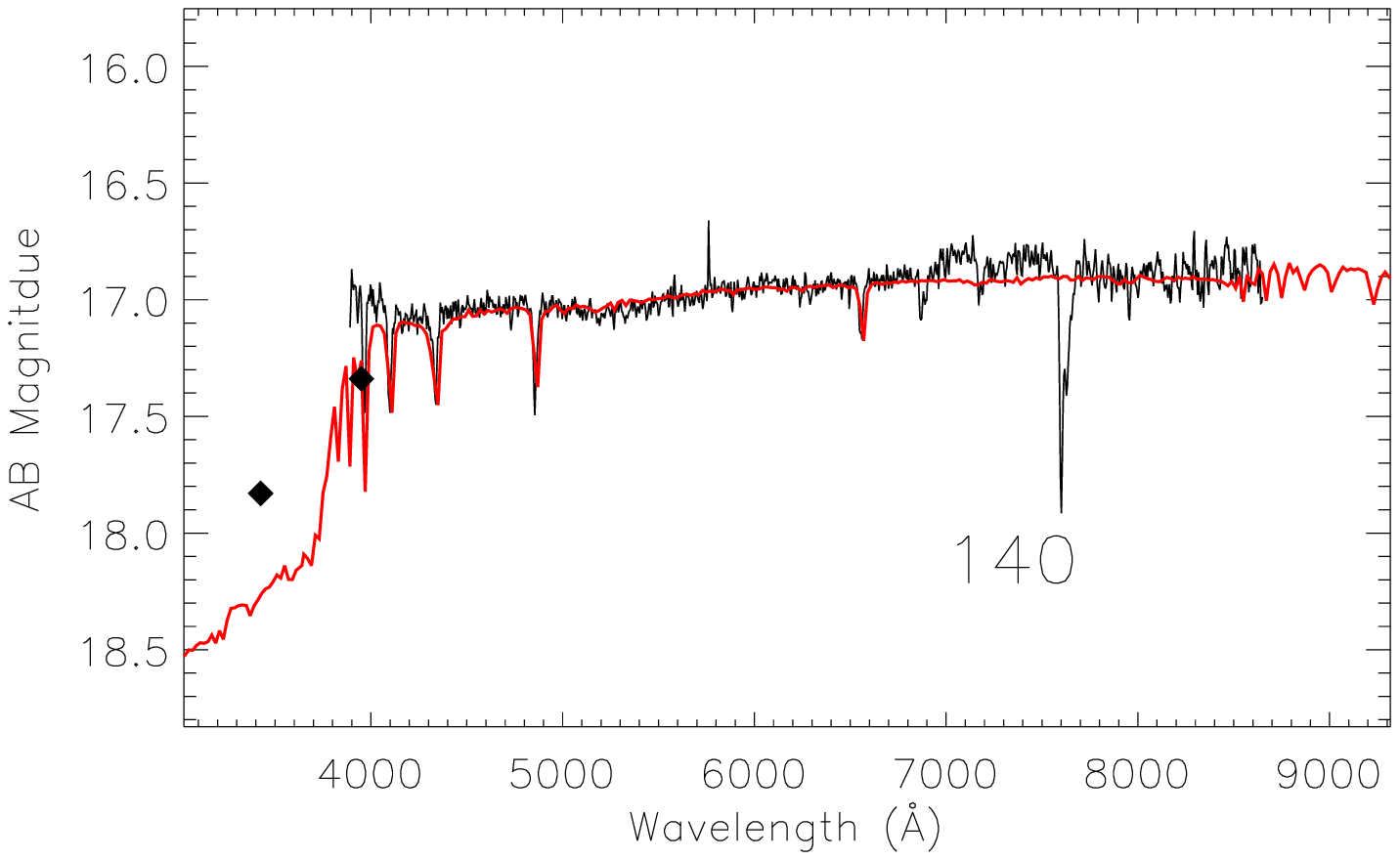}
  \includegraphics[width=.5\linewidth]{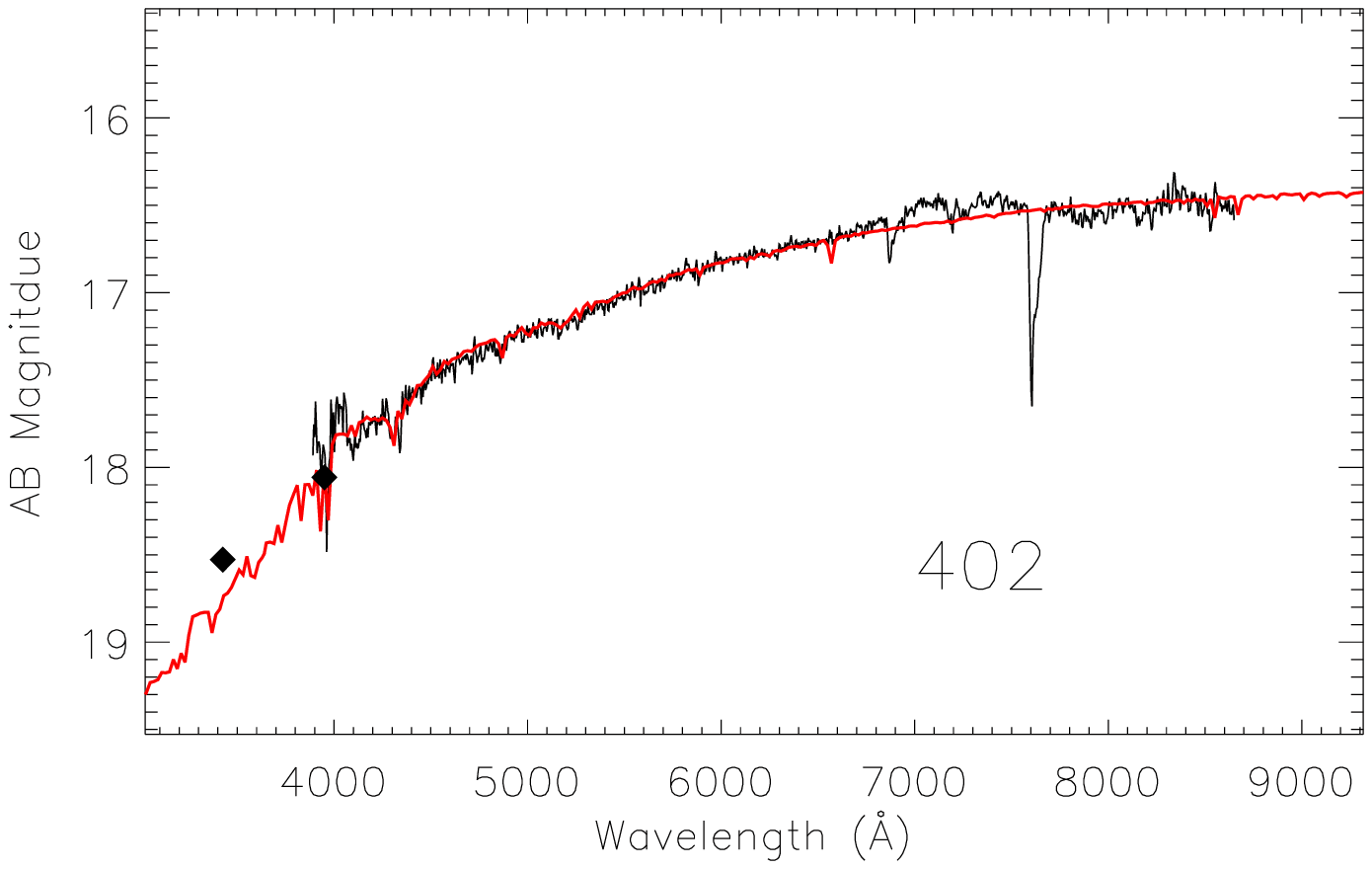}
\end{subfigure}
\caption{SEDs of our sample star clusters in M33. Fitting with spectroscopy
  and the only photometry from SAGE $\rm u_{SC}$ and $\rm v_{SAGE}$. The BC03 models
  with \citet{chab} IMF and Padova 2000 tracks are applied for the fitting.}
\label{fig7}
\end{figure}

\begin{figure}
\begin{subfigure}
  \centering
  \includegraphics[width=.5\linewidth]{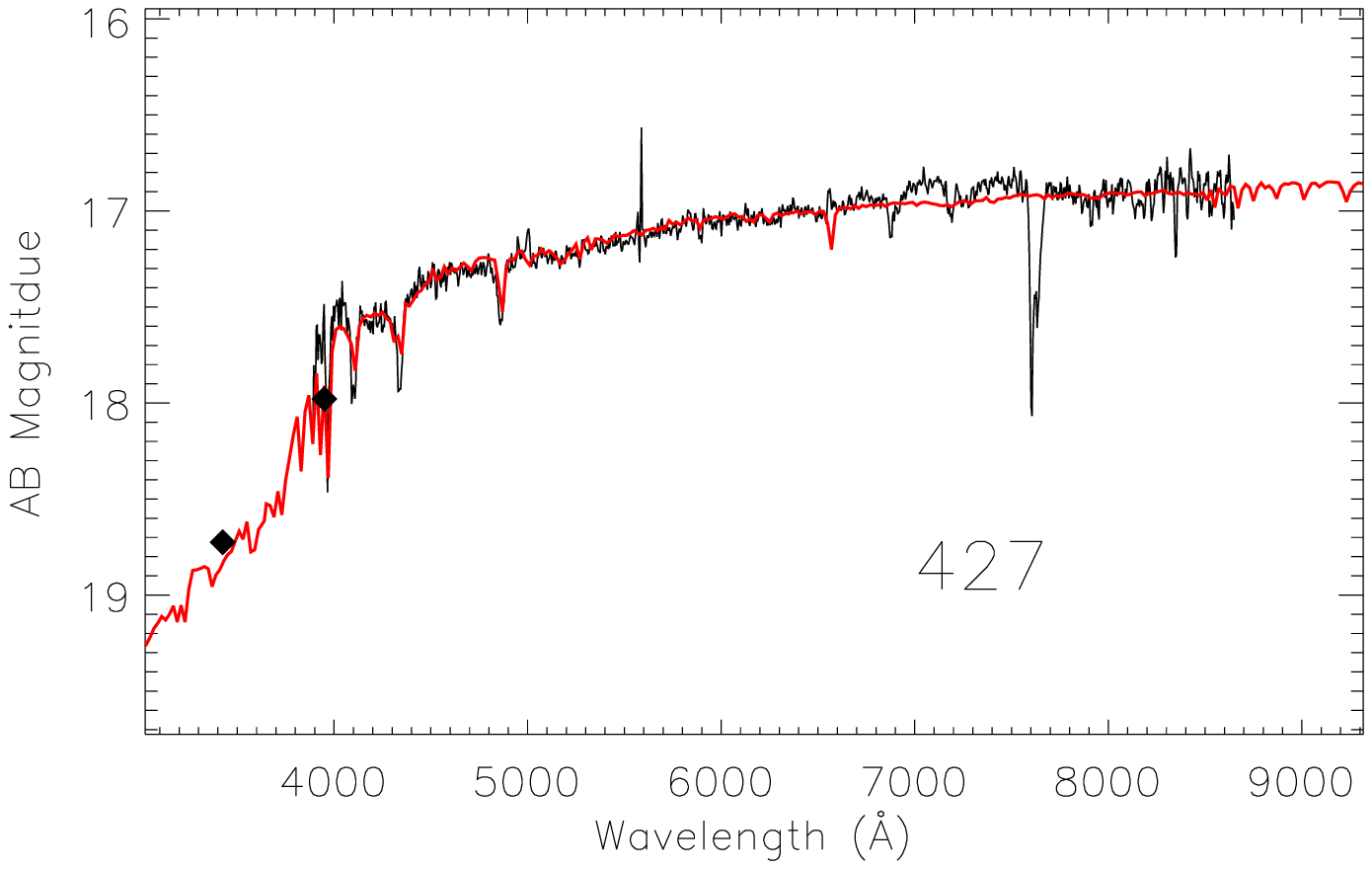}
  \includegraphics[width=.5\linewidth]{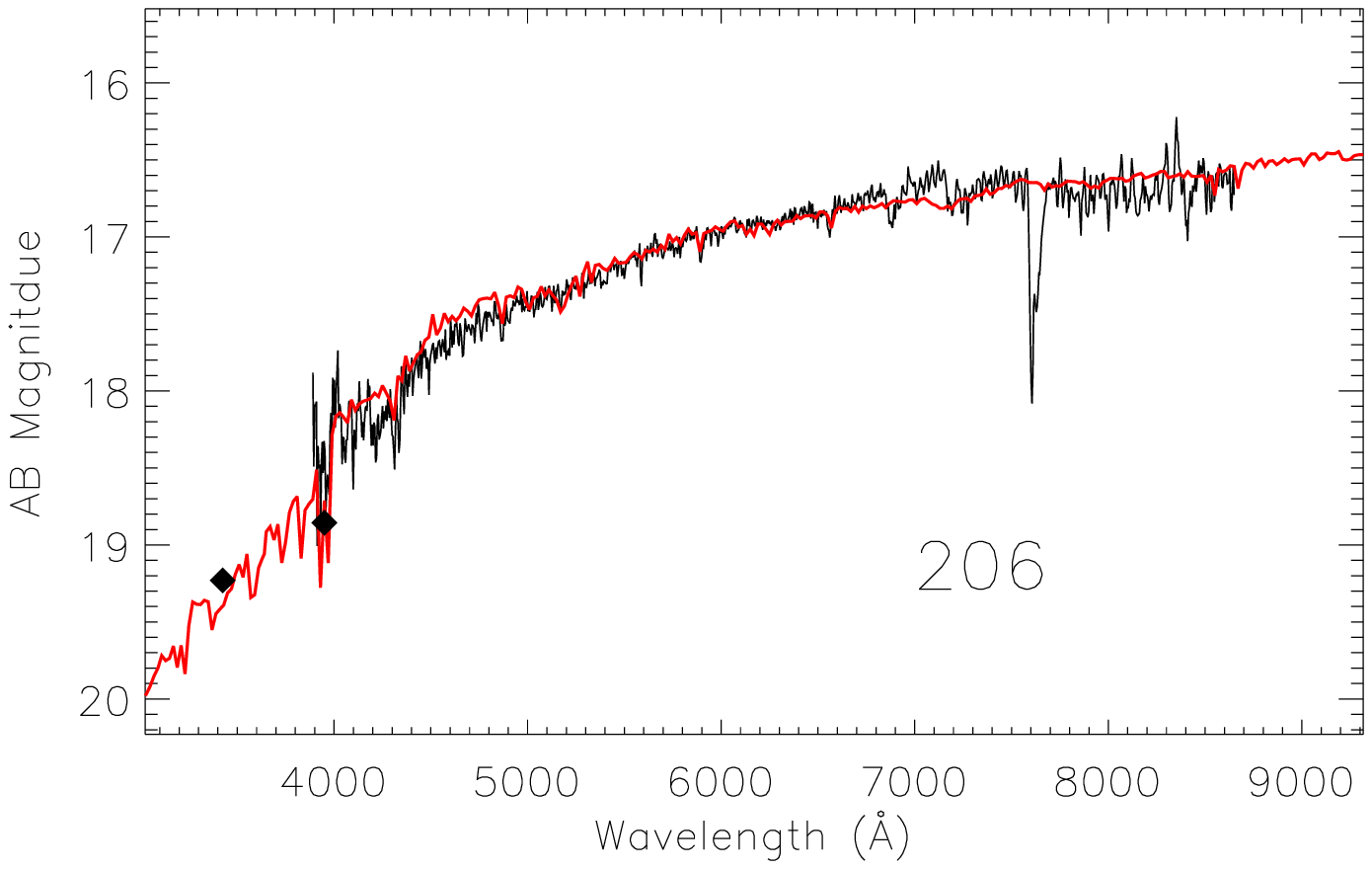}
\end{subfigure}
\begin{subfigure}
  \centering
  \includegraphics[width=.5\linewidth]{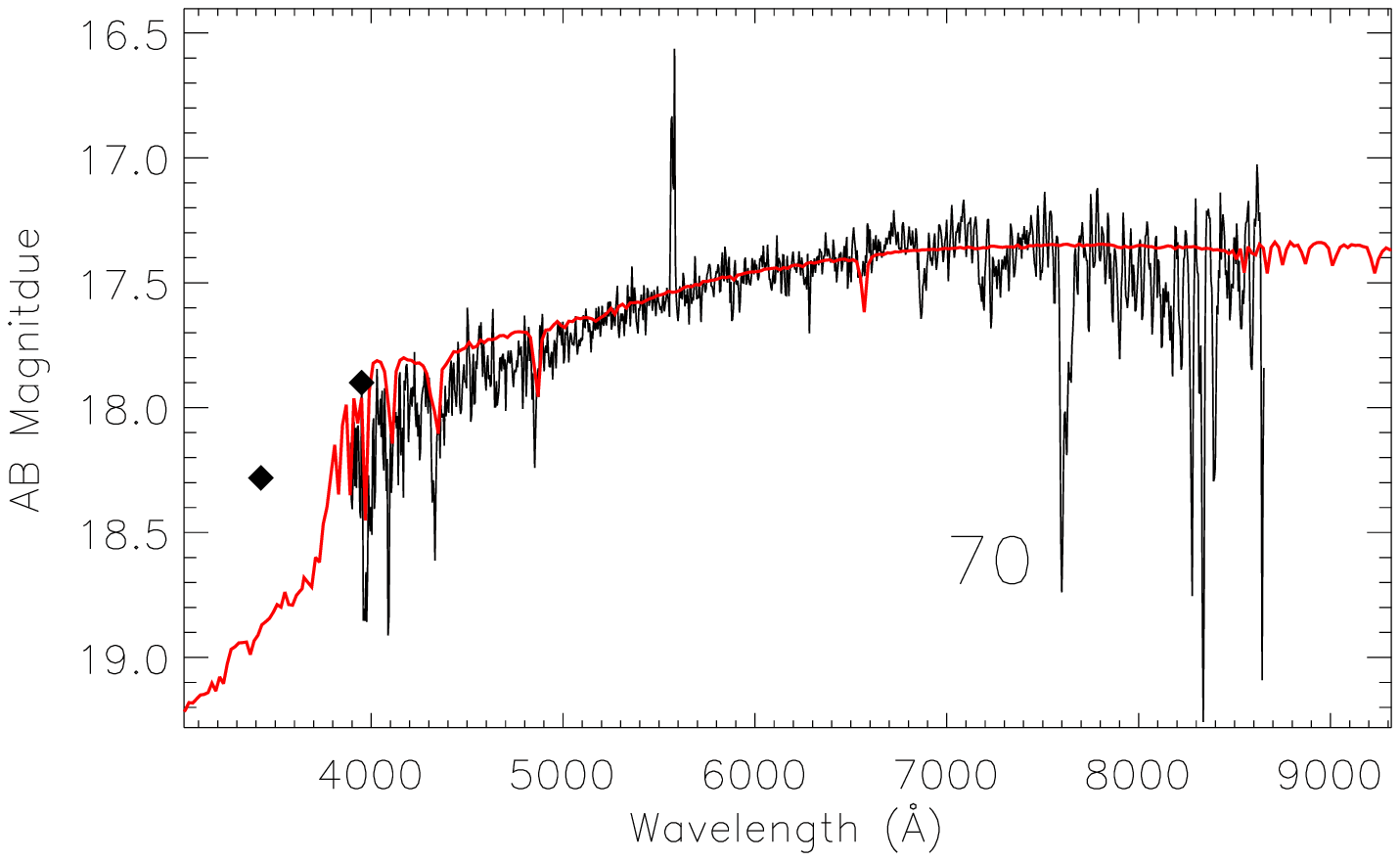}
  \includegraphics[width=.5\linewidth]{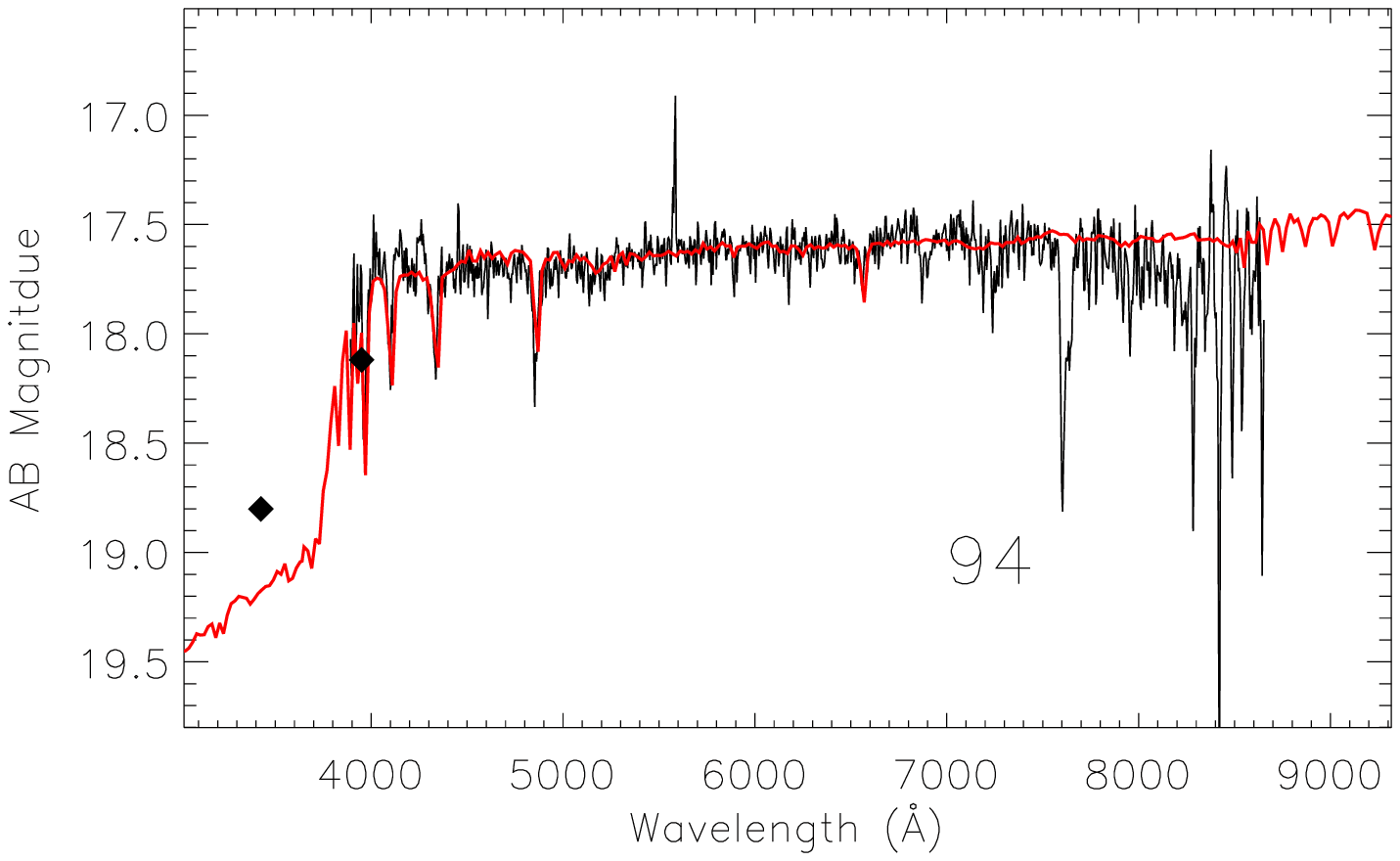}
\end{subfigure}
\begin{subfigure}
  \centering
  \includegraphics[width=.5\linewidth]{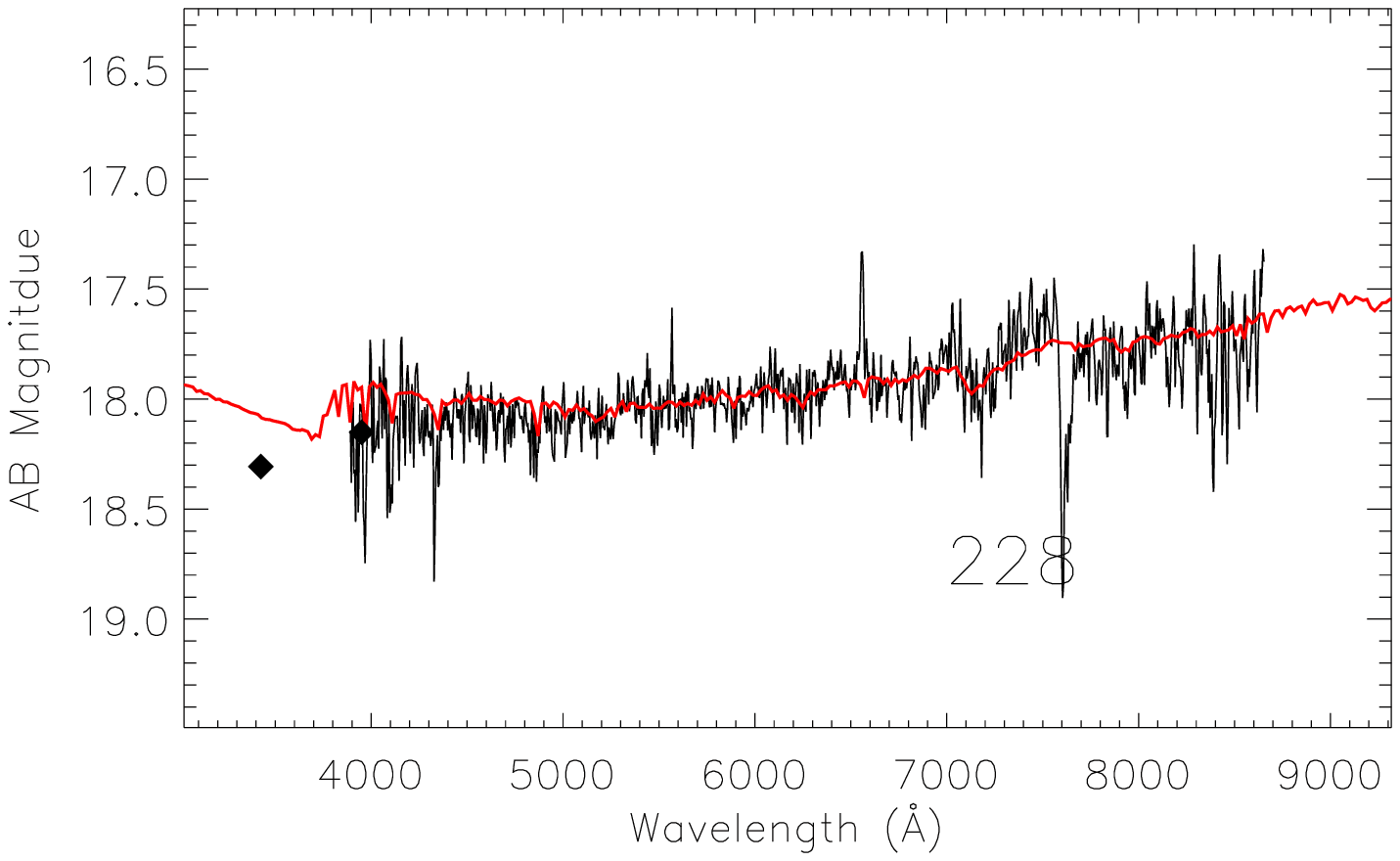}
  \includegraphics[width=.5\linewidth]{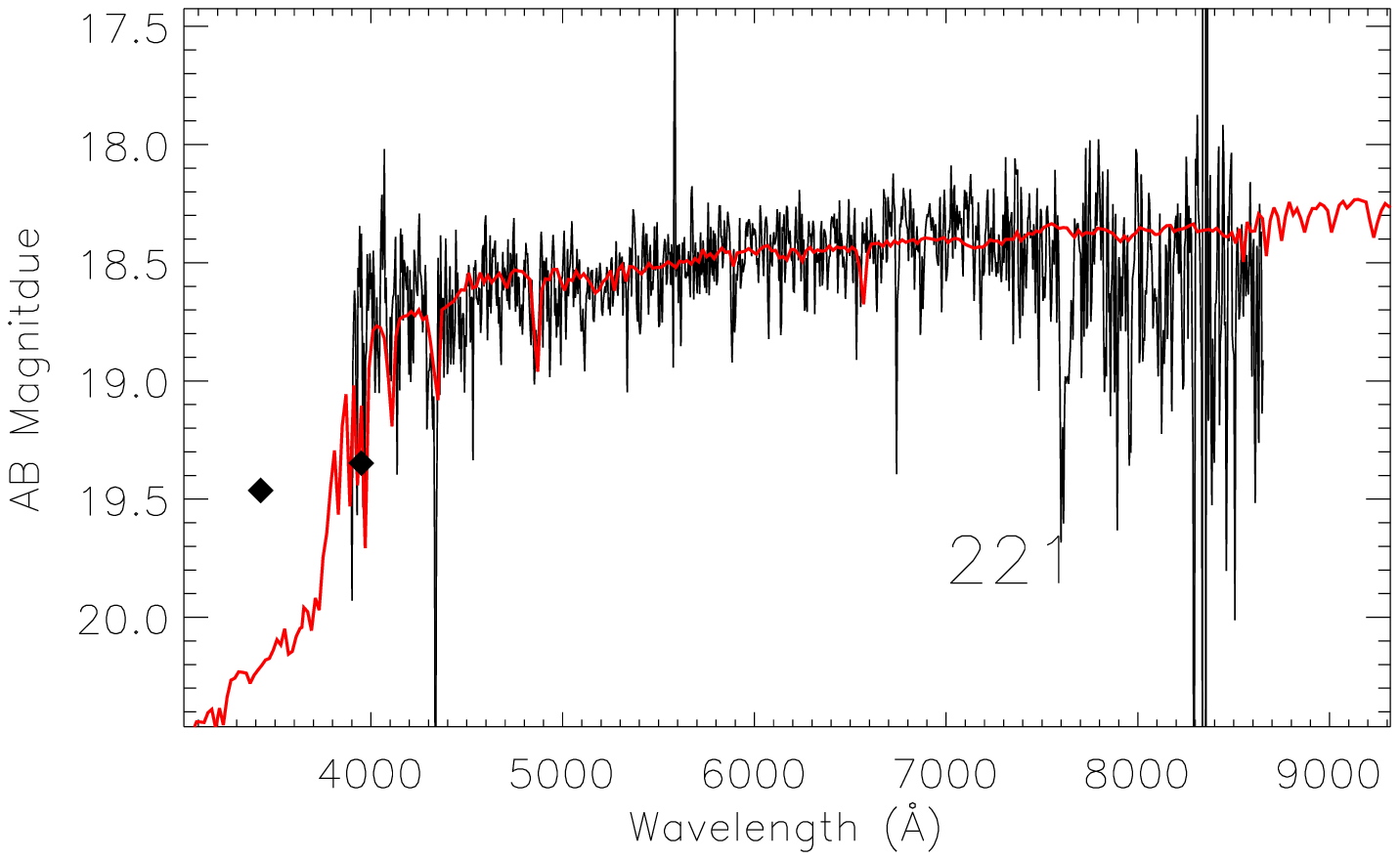}
\end{subfigure}
\begin{subfigure}
  \centering
  \includegraphics[width=.5\linewidth]{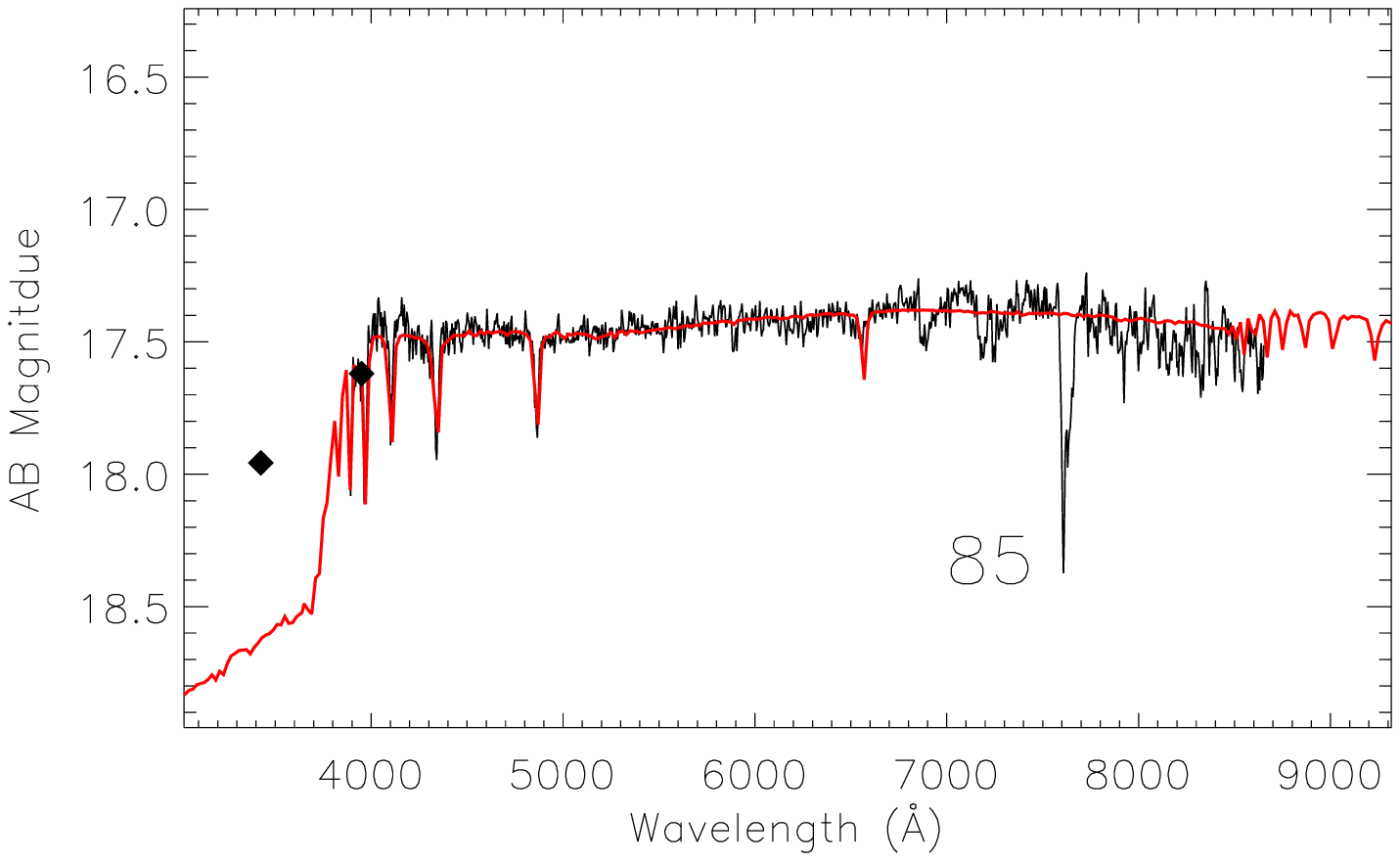}
  \includegraphics[width=.5\linewidth]{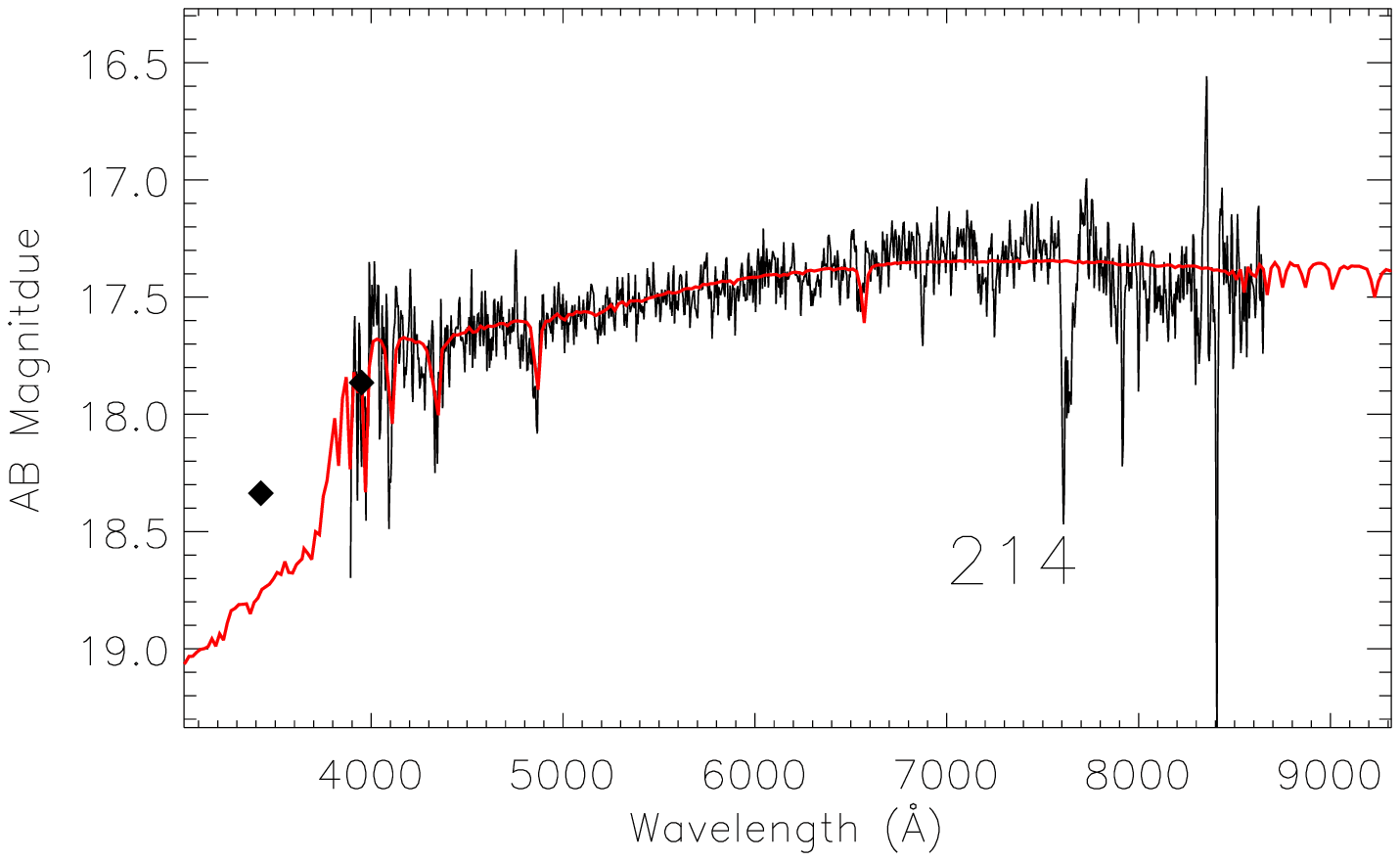}
\end{subfigure}
\caption{-Continued.}
\label{fig8}
\end{figure}
\begin{figure}
\begin{subfigure}
  \centering
  \includegraphics[width=.5\linewidth]{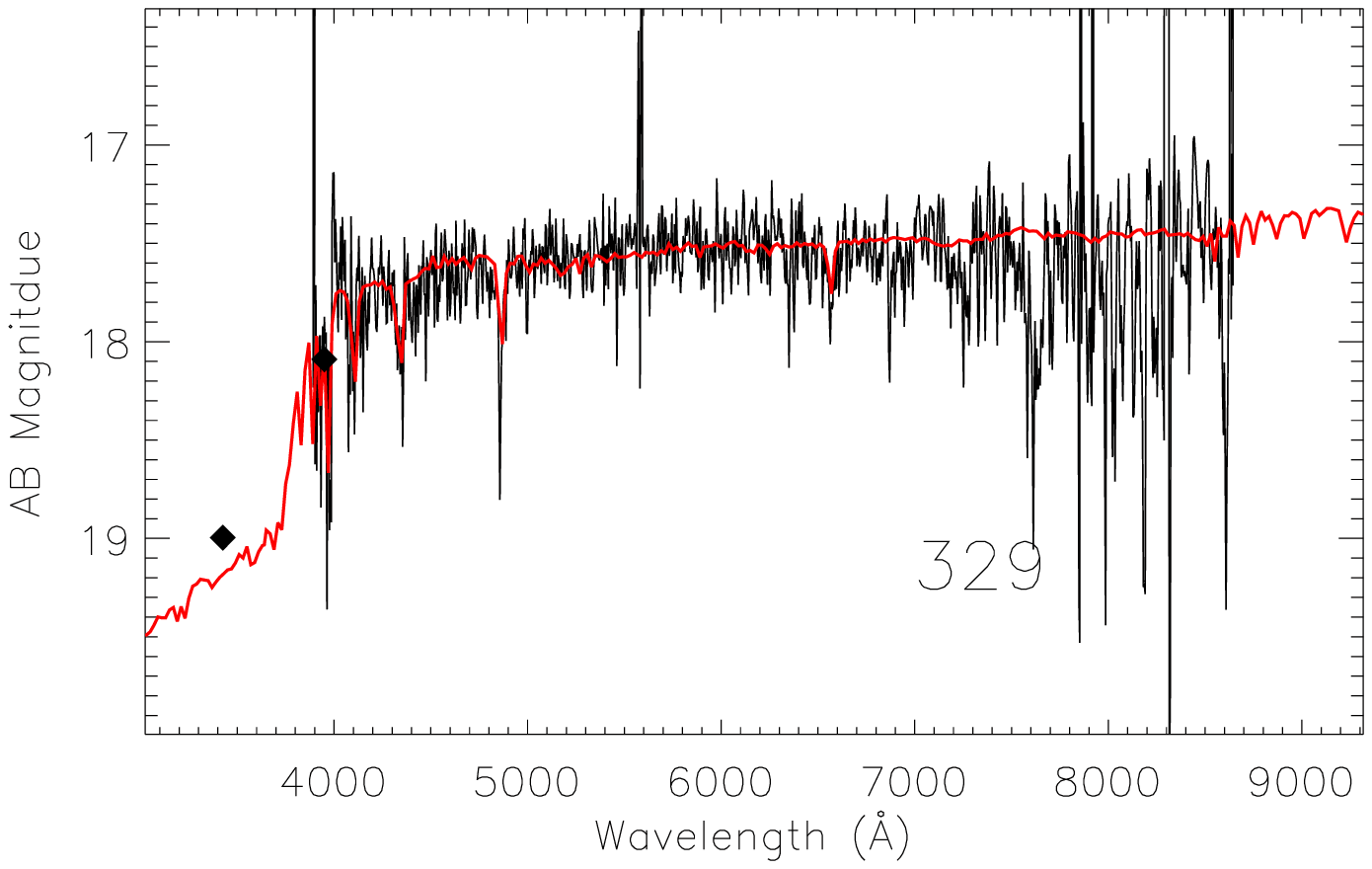}
\end{subfigure}
\caption{-Continued.}
\label{fig9}
\end{figure}

\begin{figure}
  \centering
  \includegraphics[angle=0,scale=0.8]{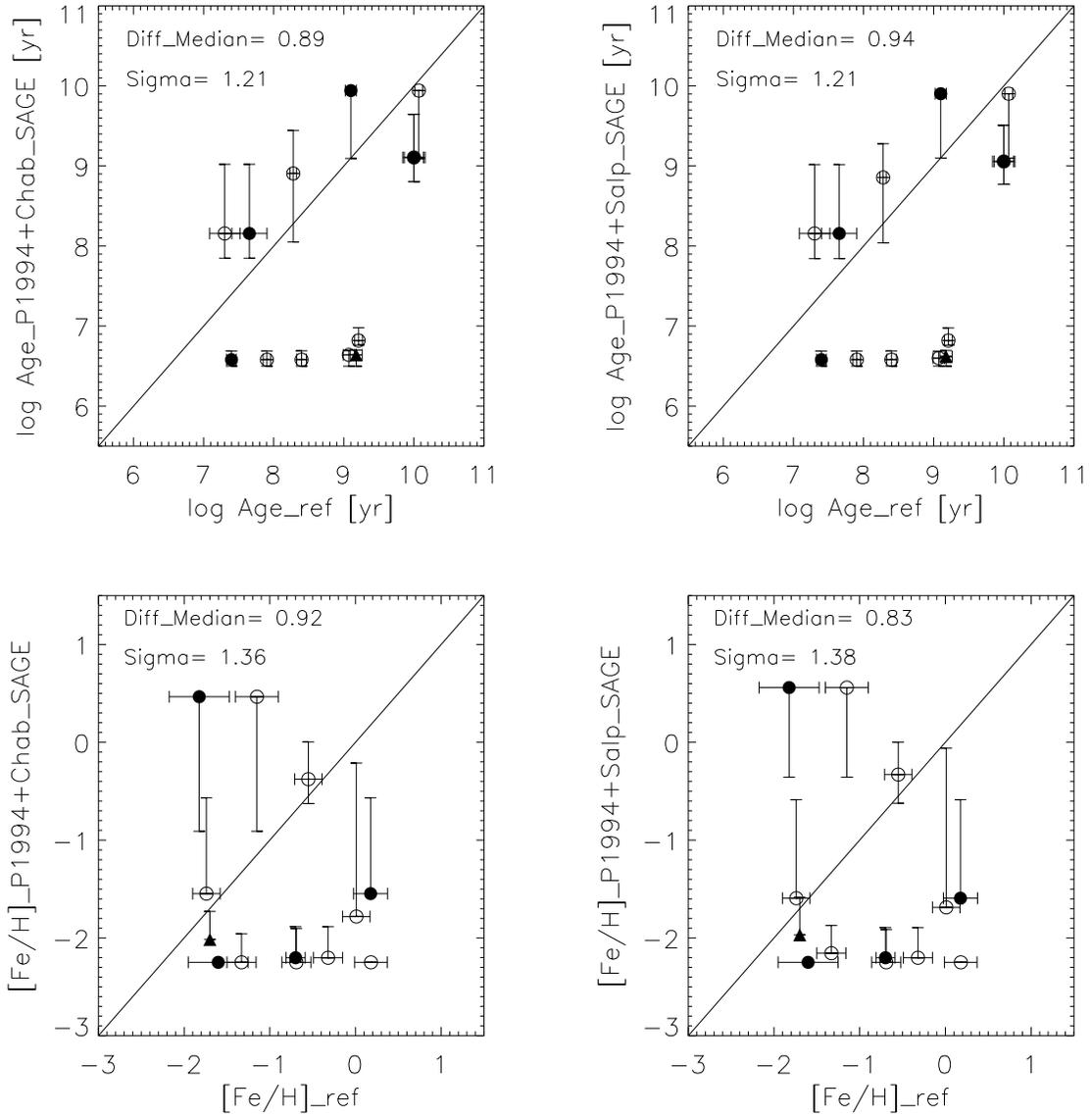}
  \caption{Same as Figure~\ref{fig7} but only using the SAGE $\rm
    u_{SC}$ and $\rm v_{SAGE}$ photometry.  The model applied is the Padova 1994
    + \citet{chab} IMF (left panels) and \citet{salp} IMF  (right panels) and
    those from \citet{bea15} (open circles), the filled circles are
      from \citet{fan14} and the filled triangles are from \citet{sha10}.}
  \label{fig10}
\end{figure}

\begin{figure}
  \centering
  \includegraphics[angle=0,scale=0.8]{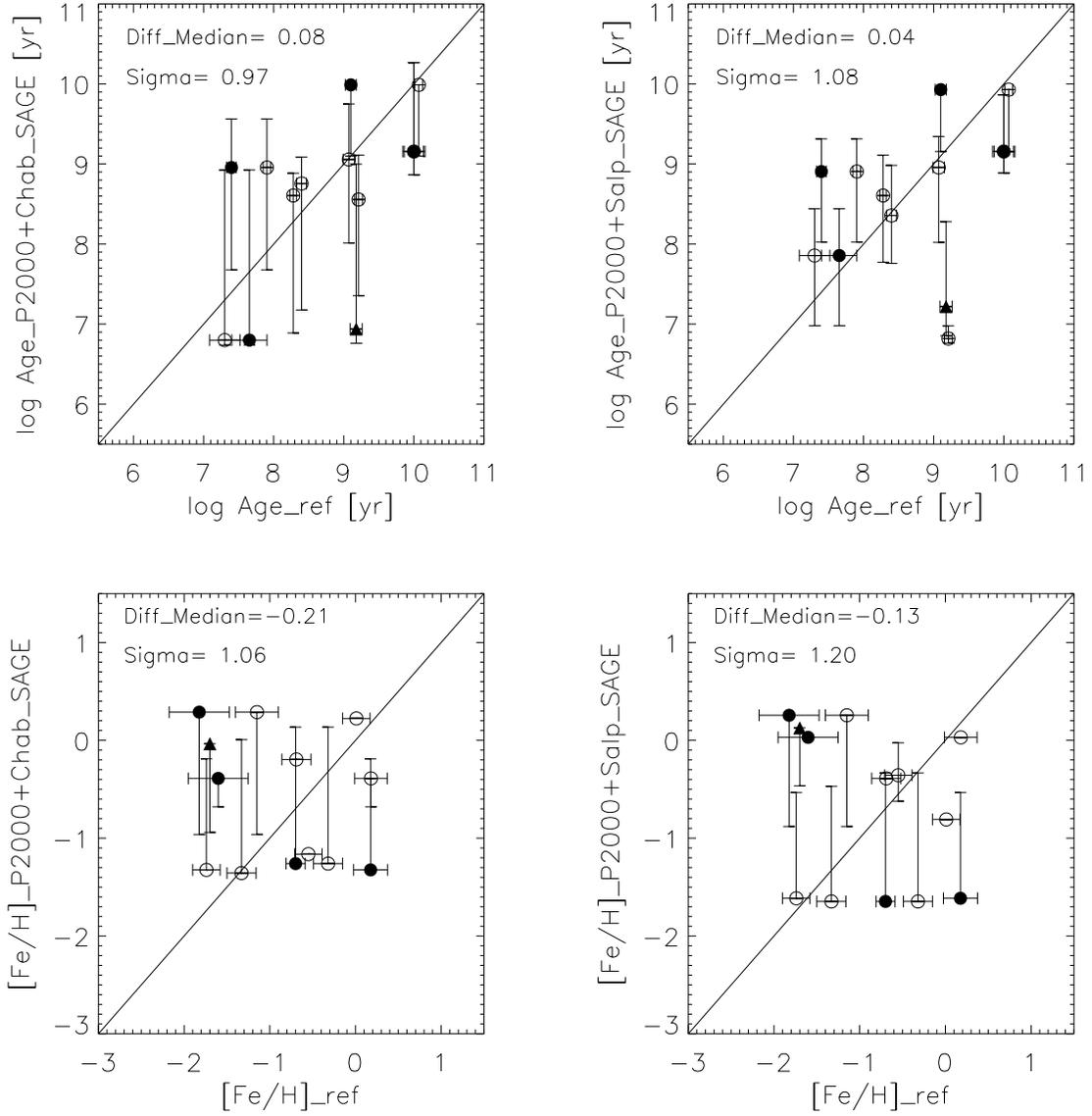}
  \caption{Same as Figure~\ref{fig7} but only using the SAGE $\rm
    u_{SC}$ and $\rm v_{SAGE}$ photometry. The model applied is the
    Padova 2000
    + \citet{chab} IMF (left panels) and \citet{salp} IMF  (right panels) and
    those from \citet{bea15} (open circles),  the filled circles are
      from \citet{fan14} and the filled triangles are from \citet{sha10}.}
  \label{fig11}
\end{figure}

\begin{figure}
\begin{subfigure}
  \centering
  \includegraphics[width=.49\linewidth]{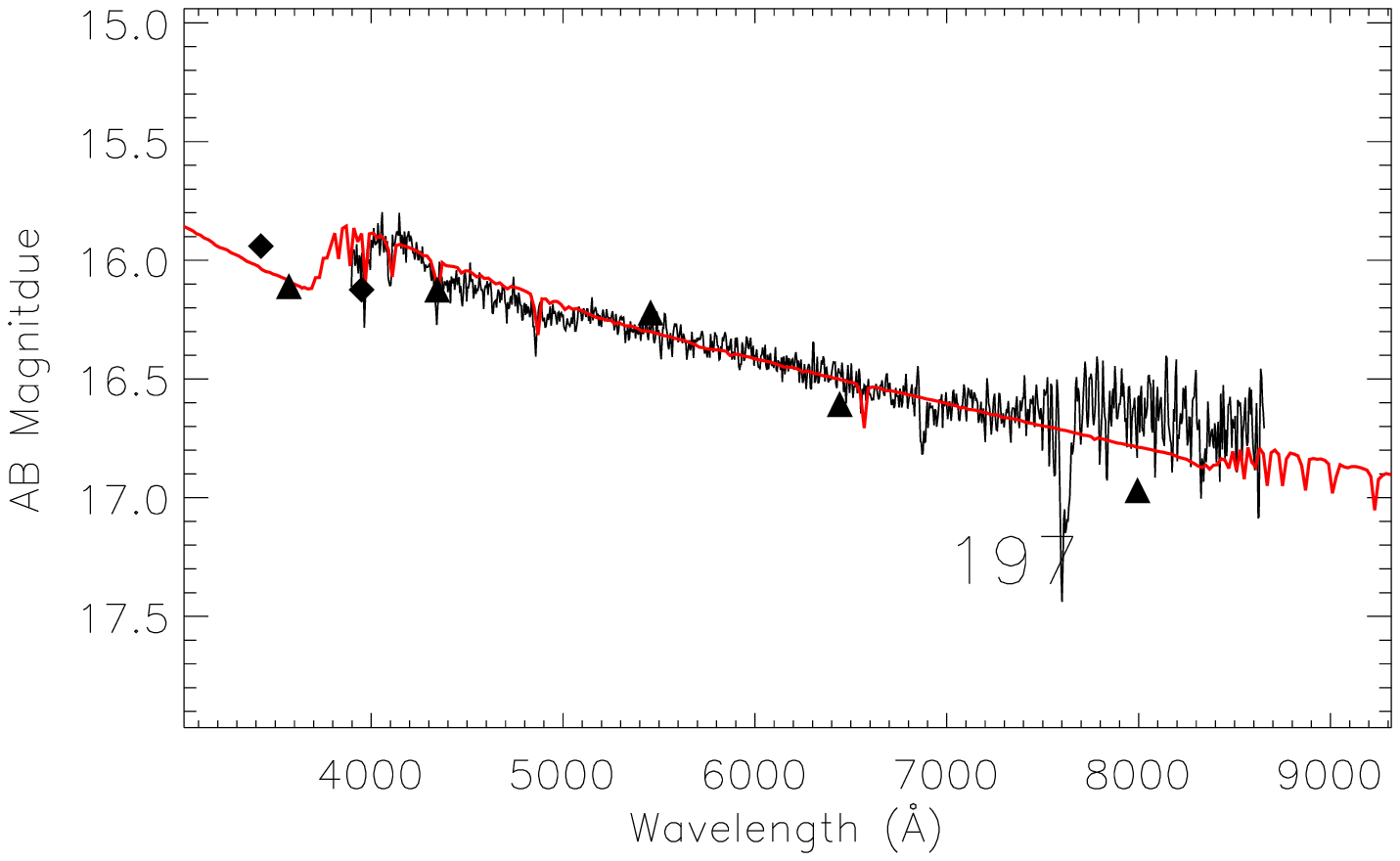}
  \includegraphics[width=.49\linewidth]{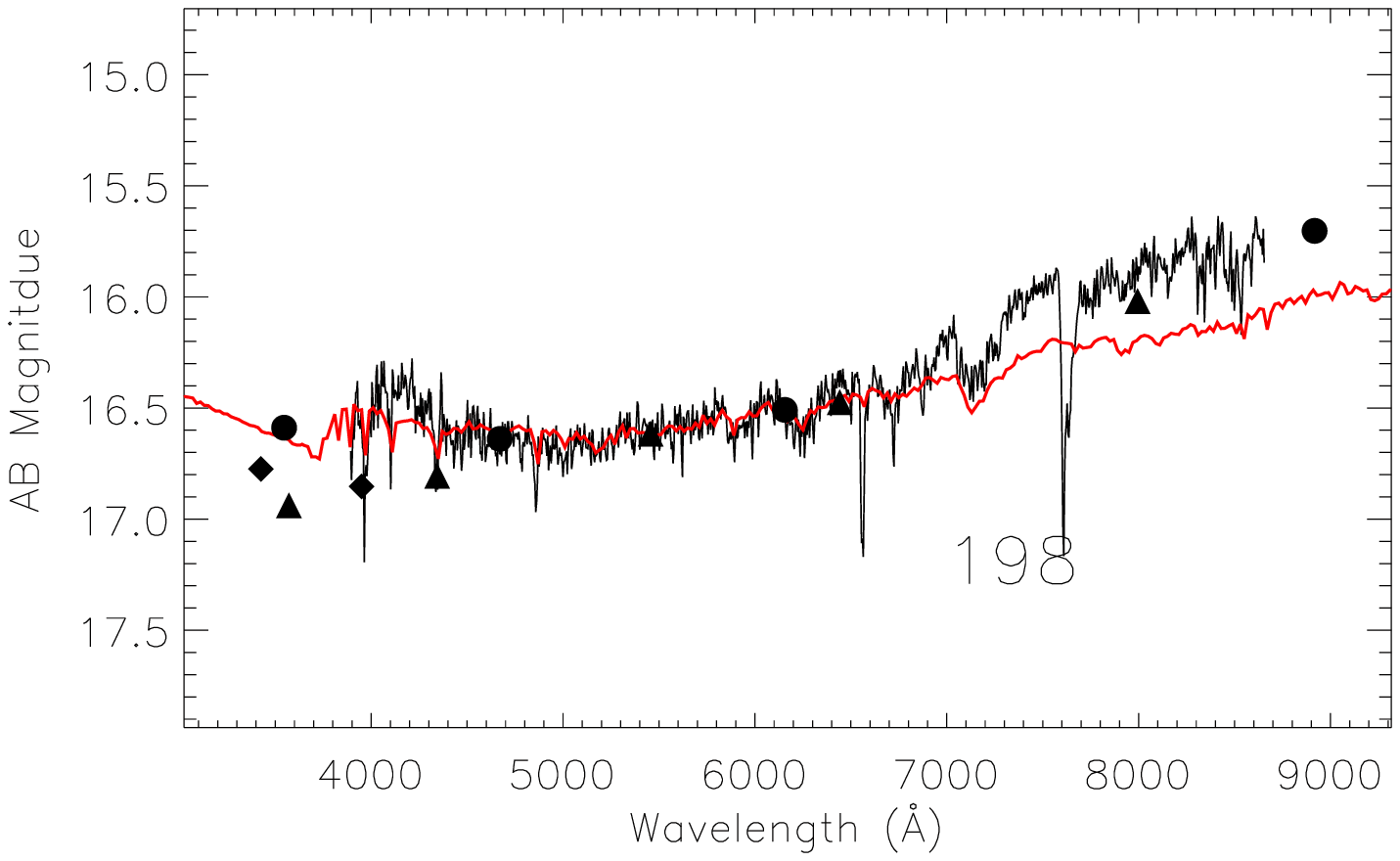}
\end{subfigure}%
\begin{subfigure}
  \centering
  \includegraphics[width=.49\linewidth]{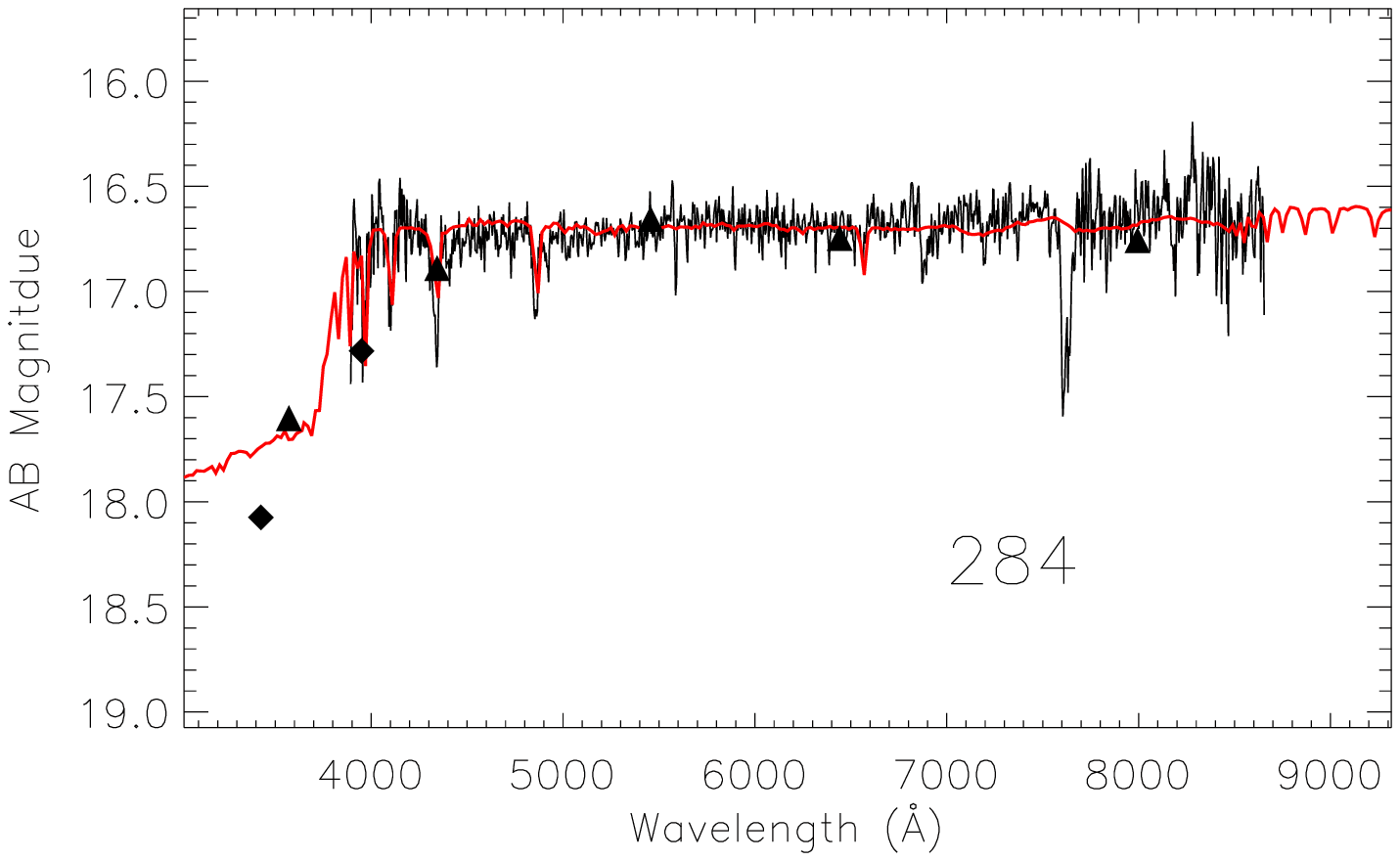}
  \includegraphics[width=.49\linewidth]{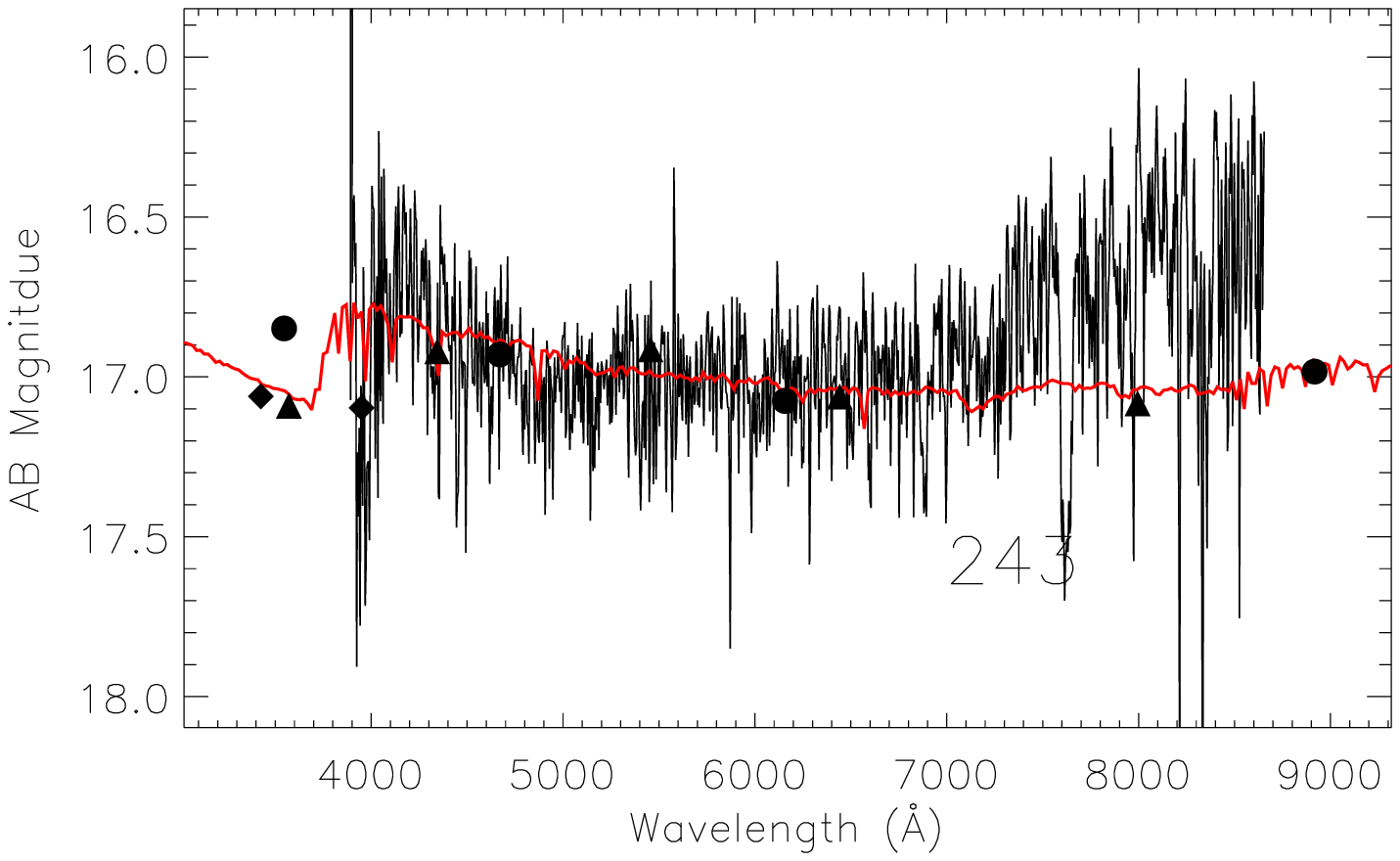}
\end{subfigure}%
\begin{subfigure}
  \centering
  \includegraphics[width=.49\linewidth]{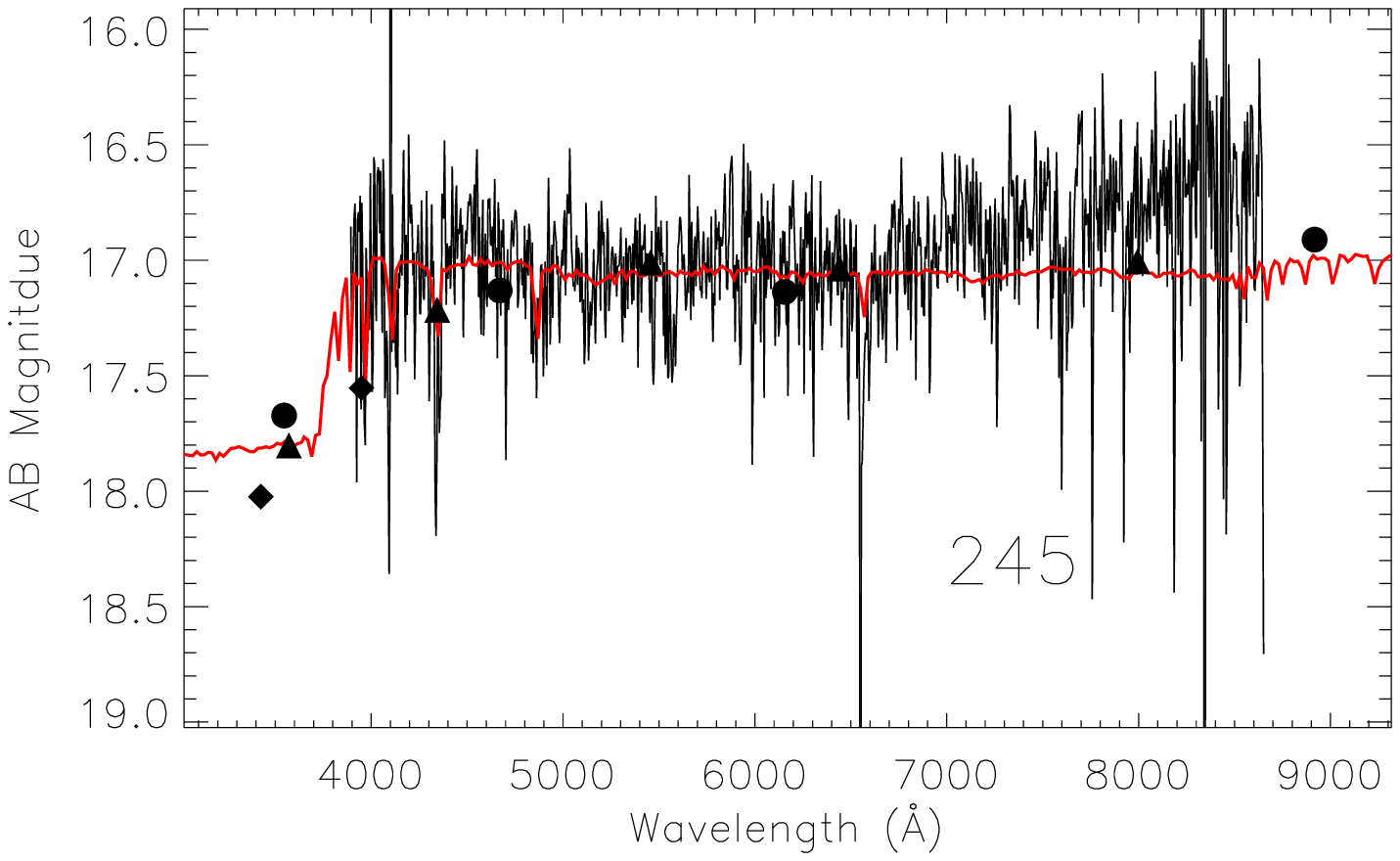}
  \includegraphics[width=.49\linewidth]{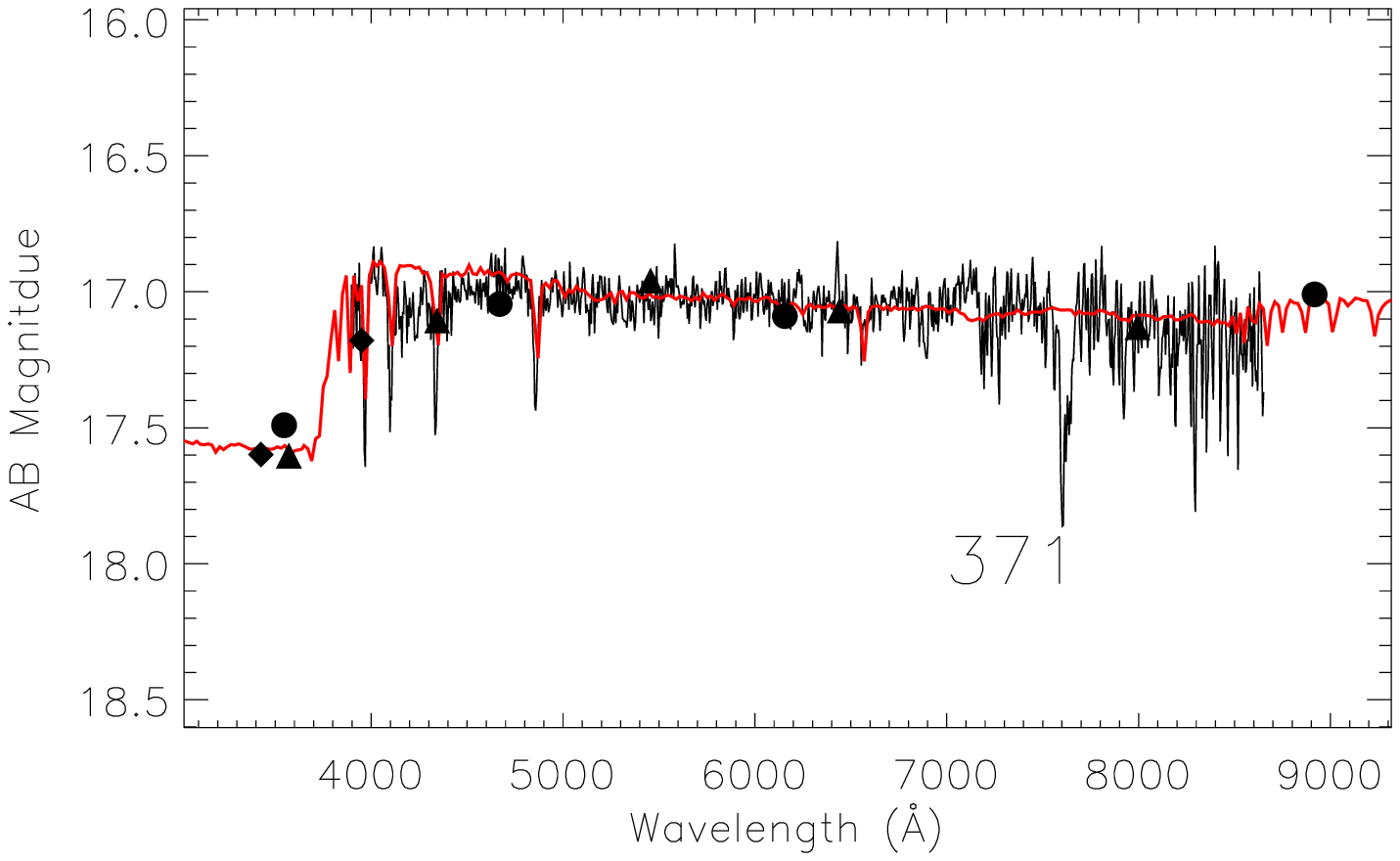}
\end{subfigure}
\begin{subfigure}
  \centering
  \includegraphics[width=.49\linewidth]{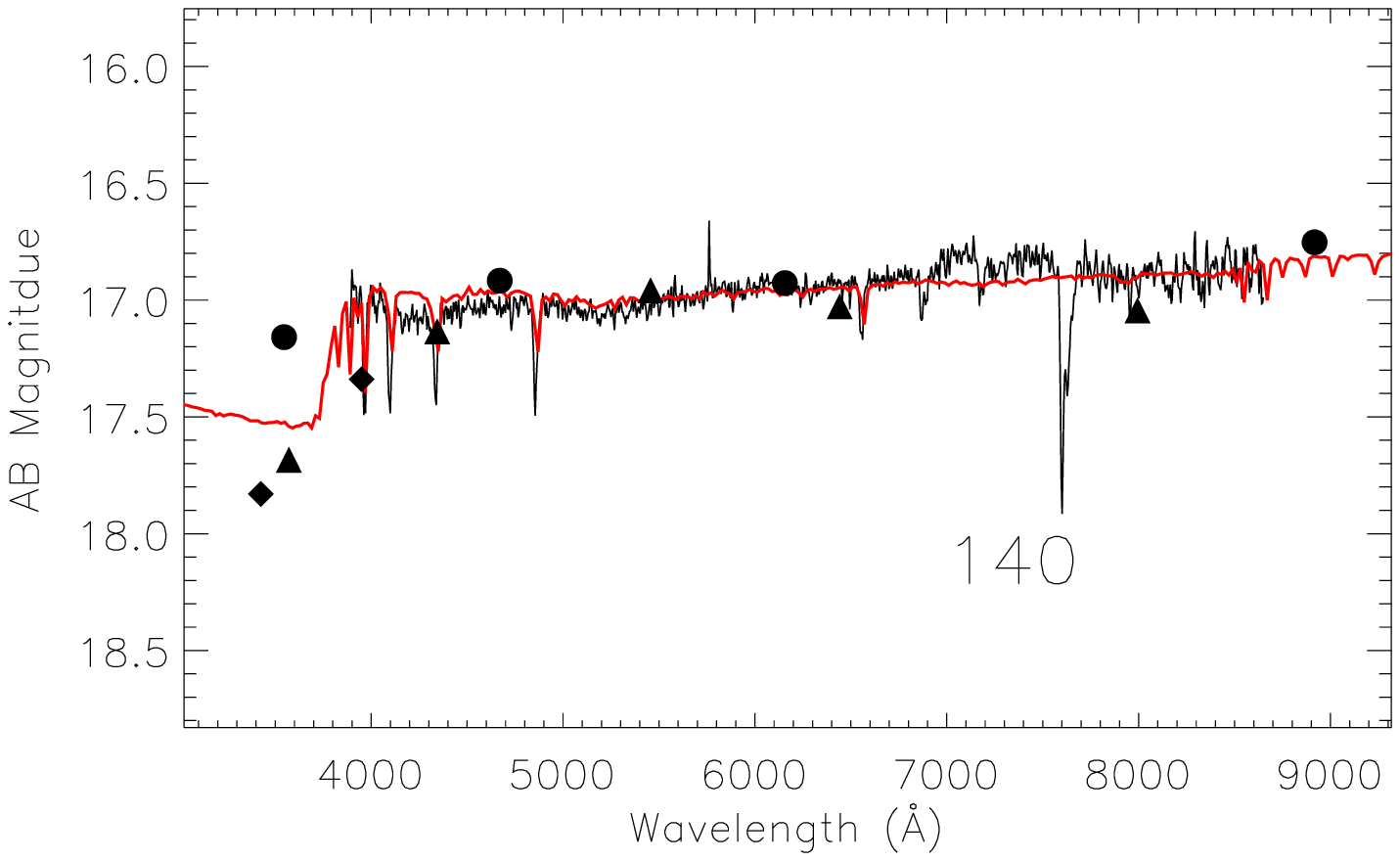}
  \includegraphics[width=.49\linewidth]{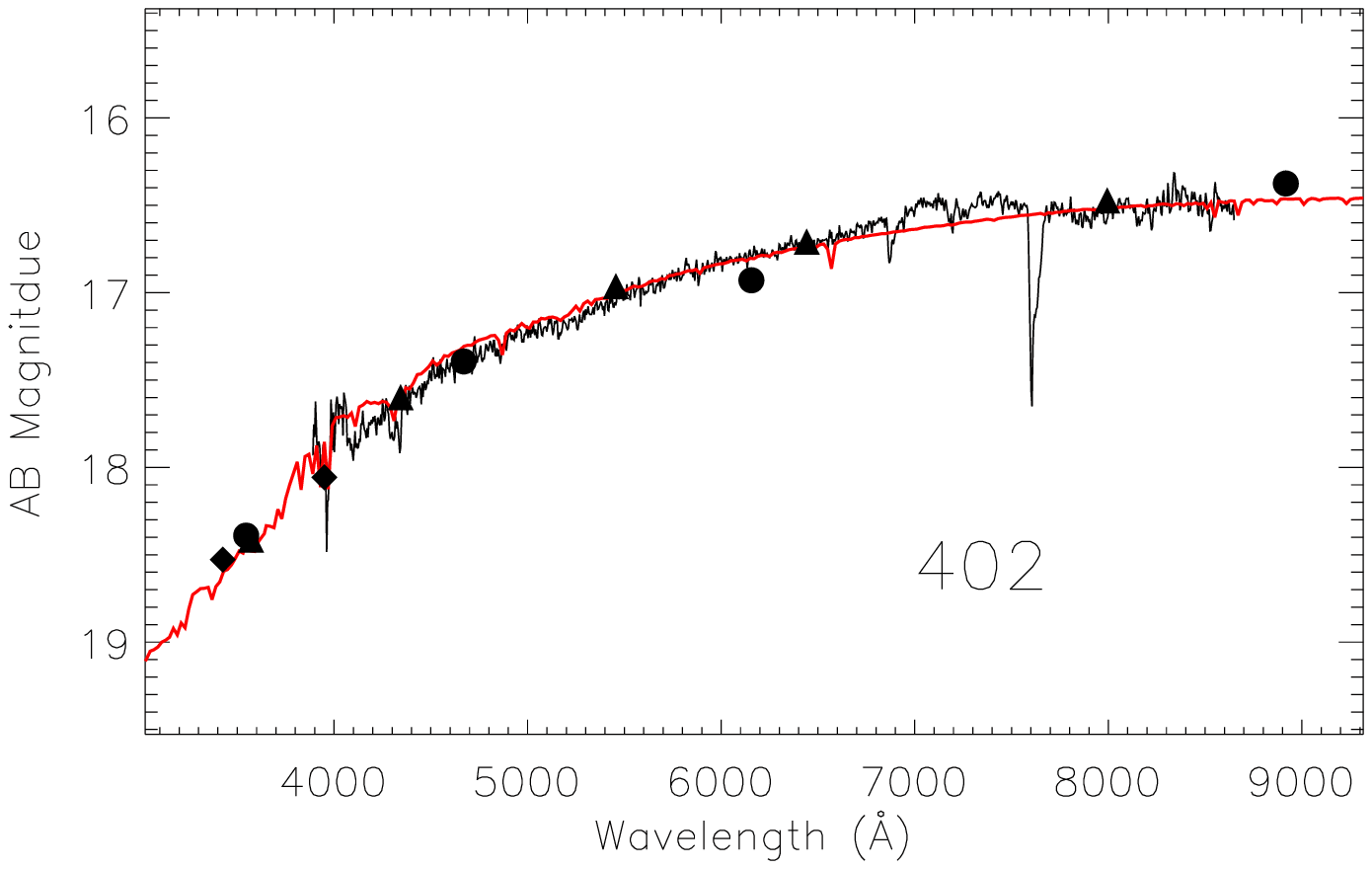}
\end{subfigure}
\caption{Same as in Figure~\ref{fig6} but for spectroscopy and photometry
  from SAGE $\rm u_{SC}$ and $\rm v_{SAGE}$, UBVRI and ugriz.  The \citet{bc03} models
  with \citet{chab} IMF and Padova 2000 tracks are applied for the
  fitting. The filled diamonds represent SAGE data; triangles represent UBVRI;
  filled circles represent ugriz.}
\label{fig12}
\end{figure}

\begin{figure}
\begin{subfigure}
  \centering
 \includegraphics[width=.5\linewidth]{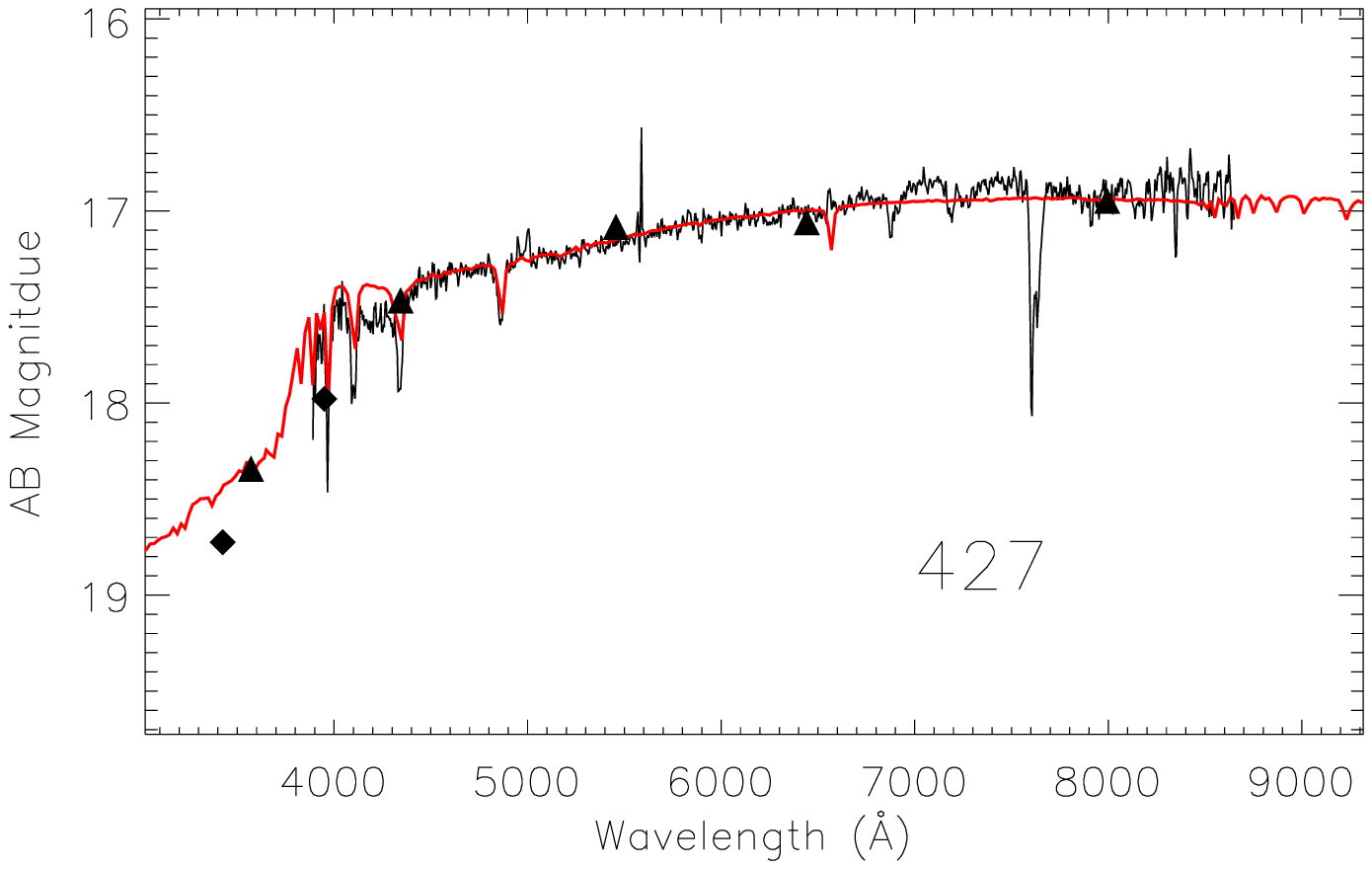}
  \includegraphics[width=.5\linewidth]{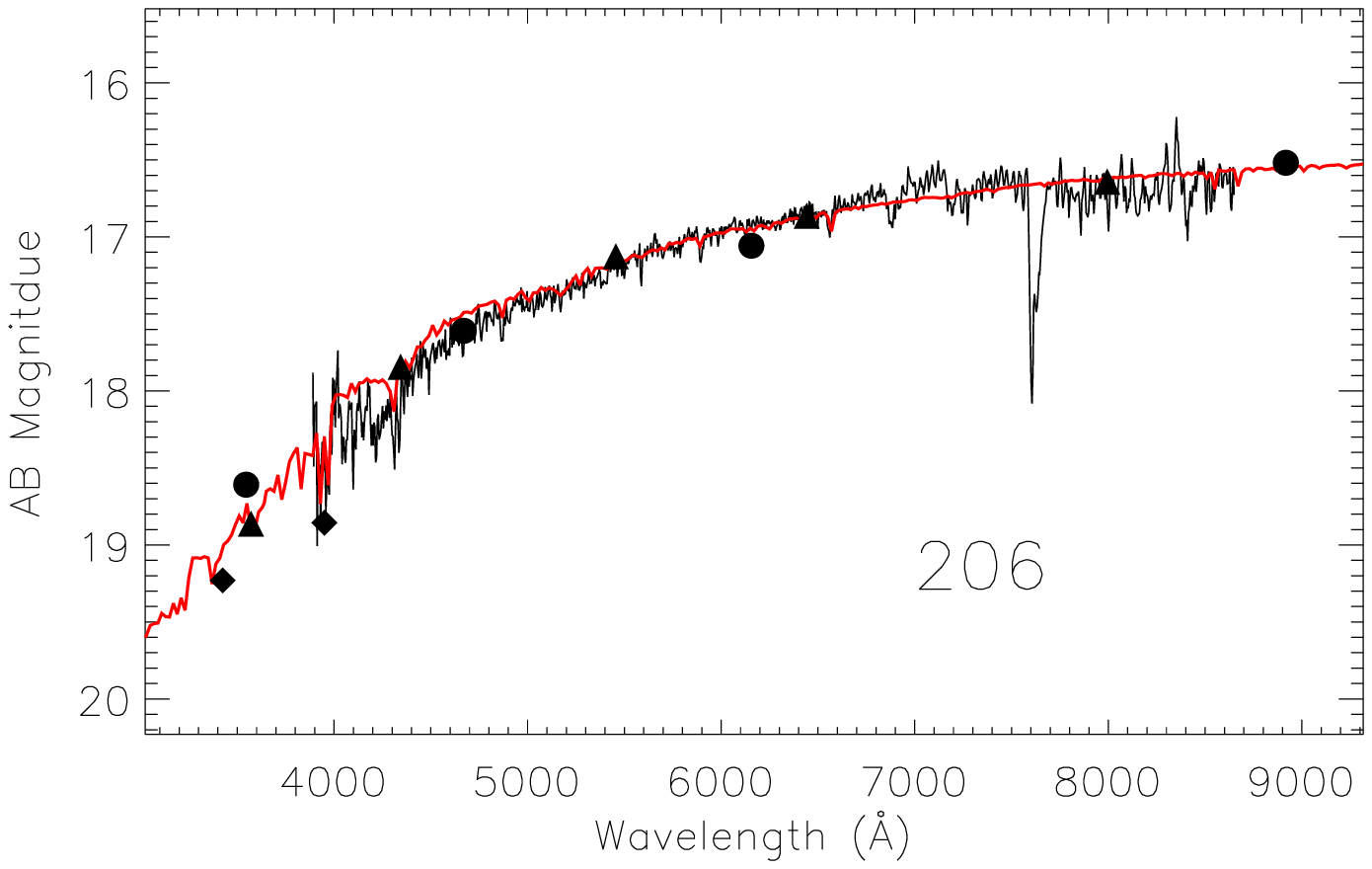}
\end{subfigure}
\begin{subfigure}
  \centering
 \includegraphics[width=.5\linewidth]{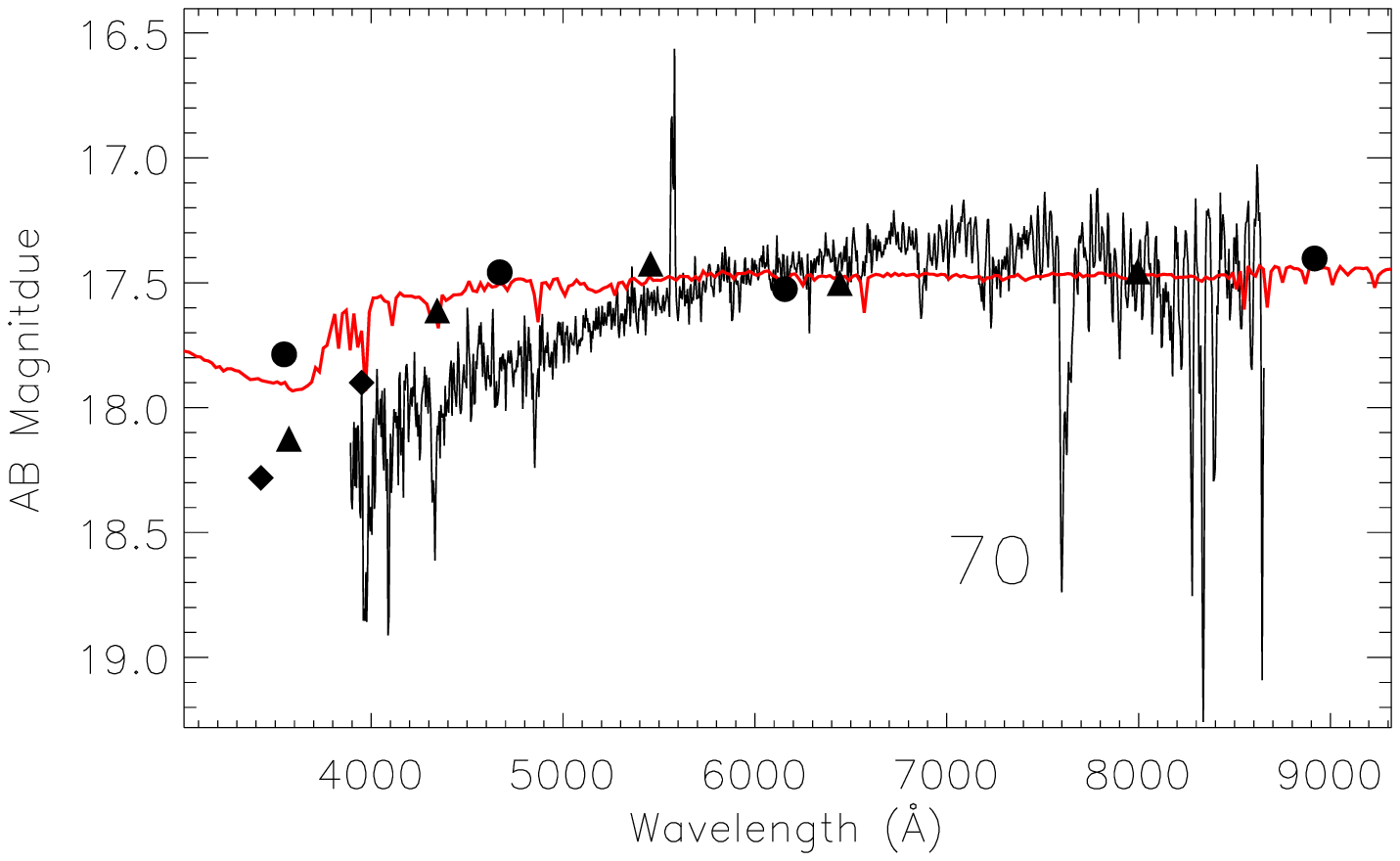}
  \includegraphics[width=.5\linewidth]{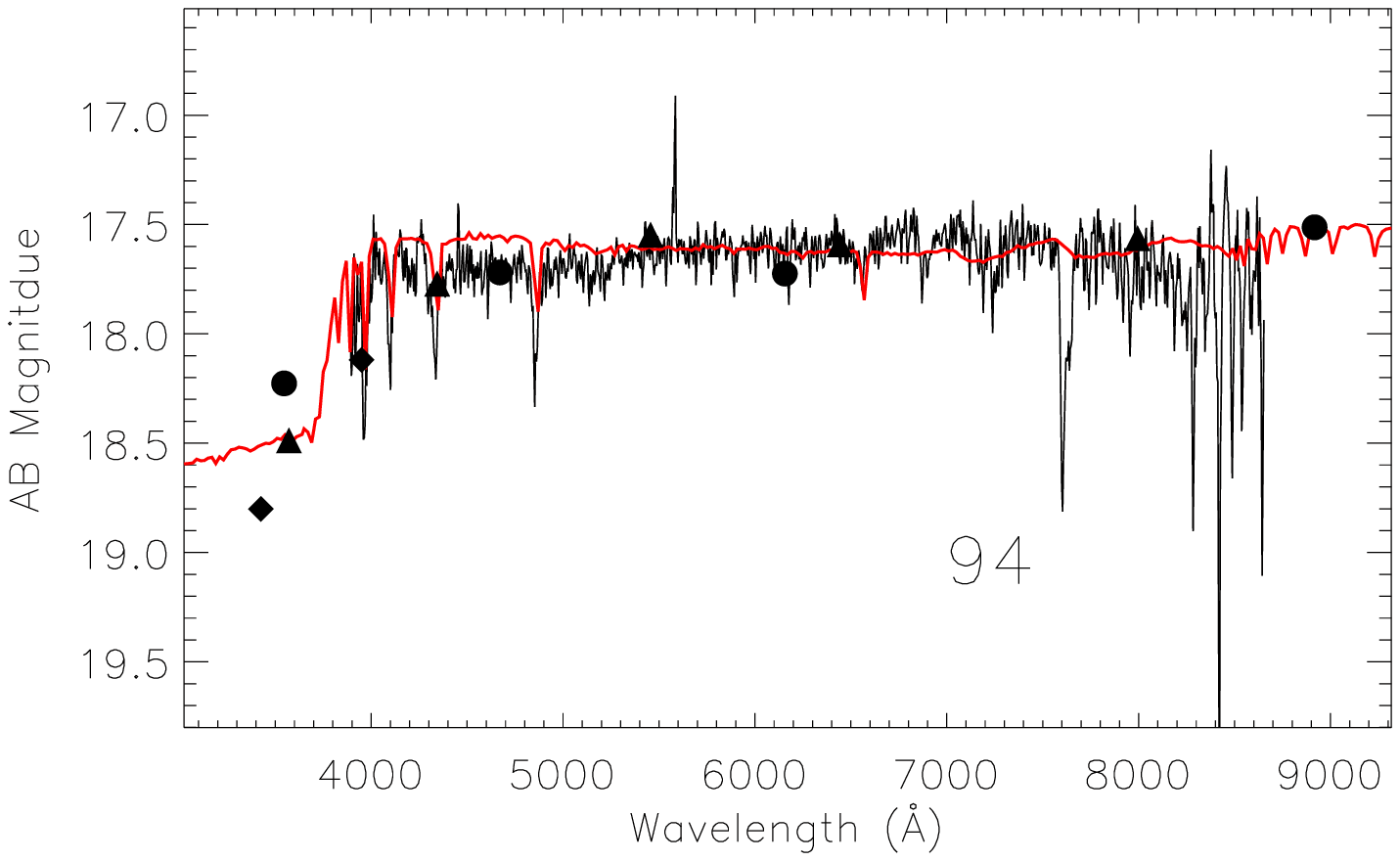}
\end{subfigure}
\begin{subfigure}
  \centering
 \includegraphics[width=.5\linewidth]{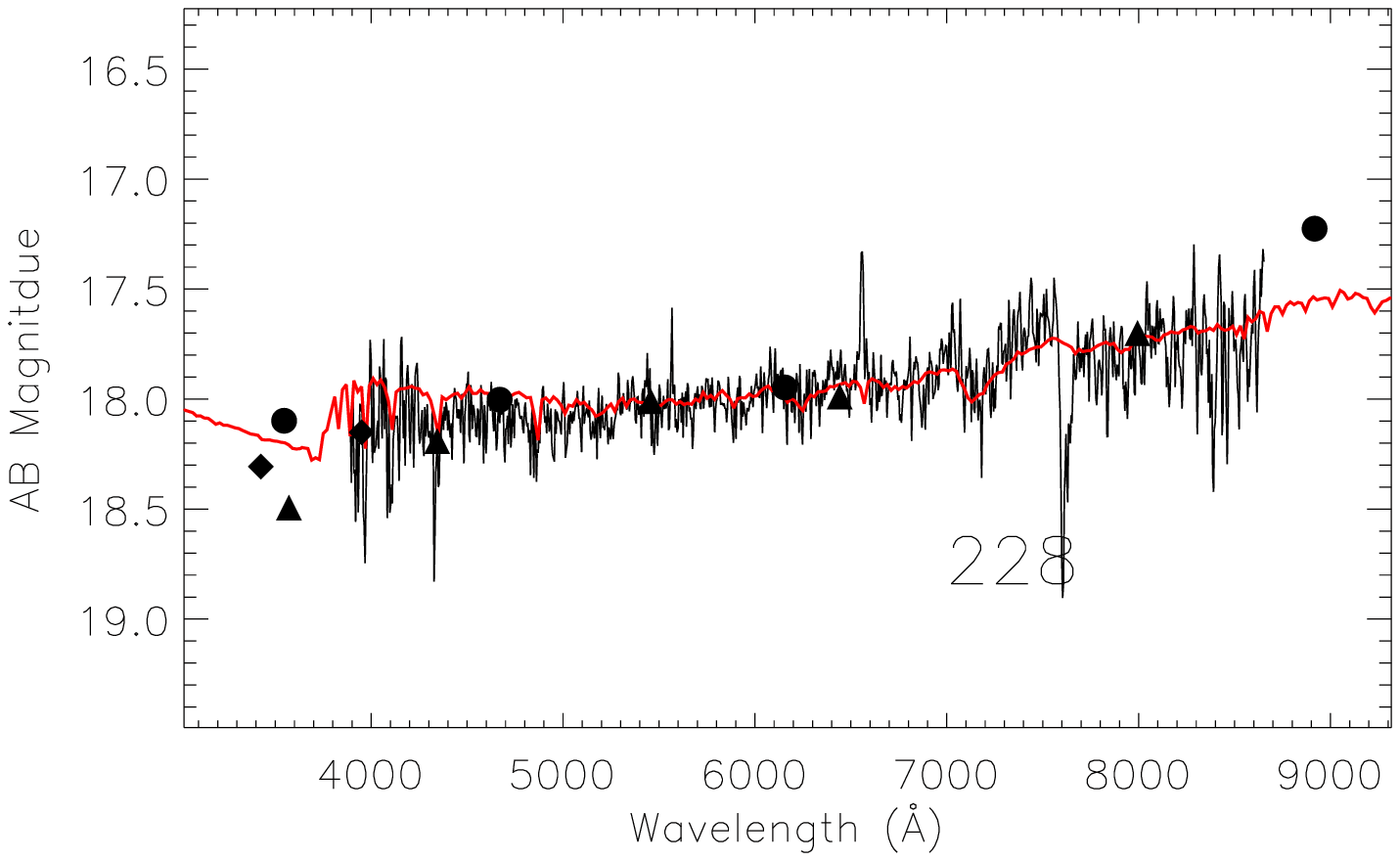}
  \includegraphics[width=.5\linewidth]{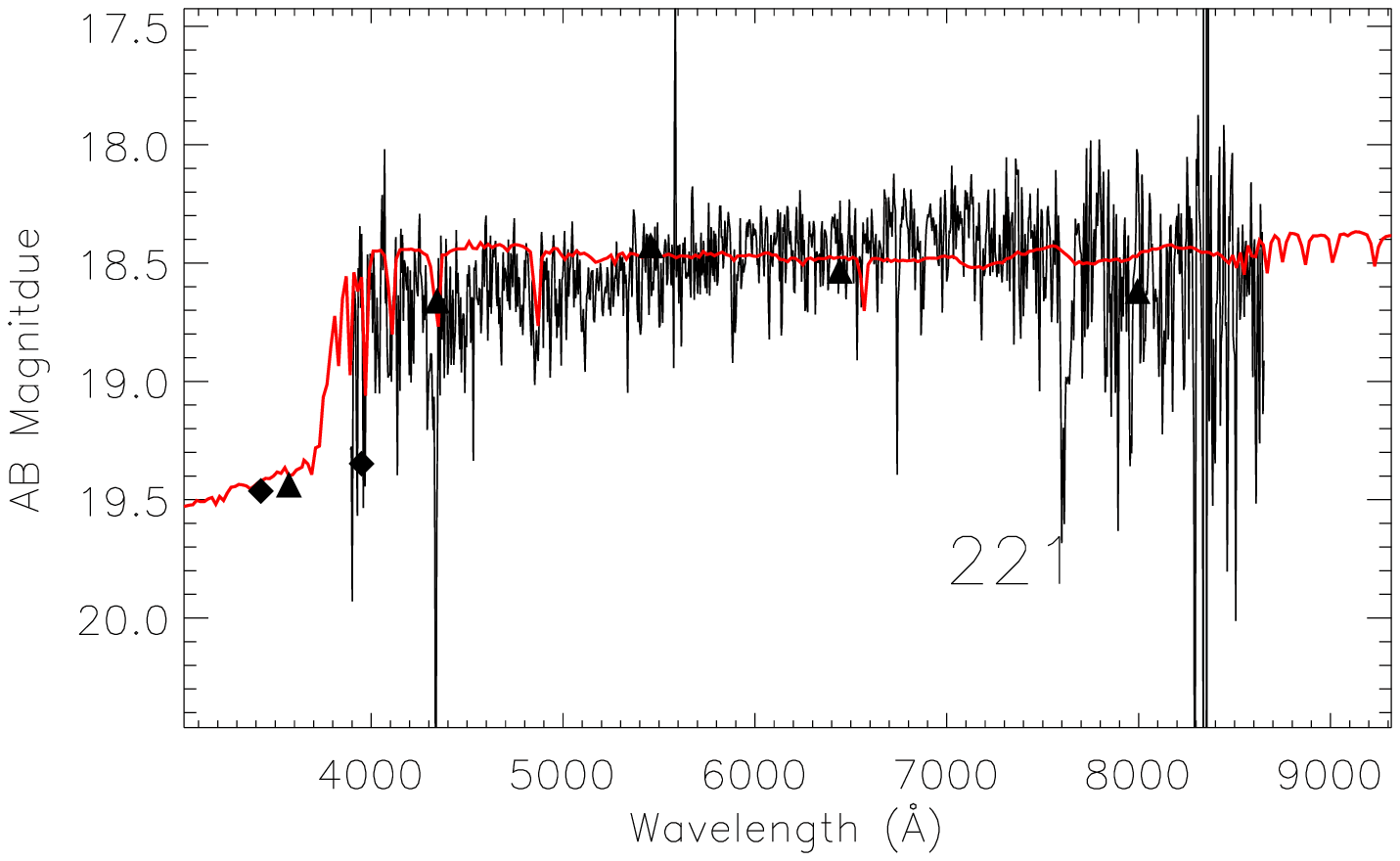}
\end{subfigure}
\begin{subfigure}
  \centering
 \includegraphics[width=.5\linewidth]{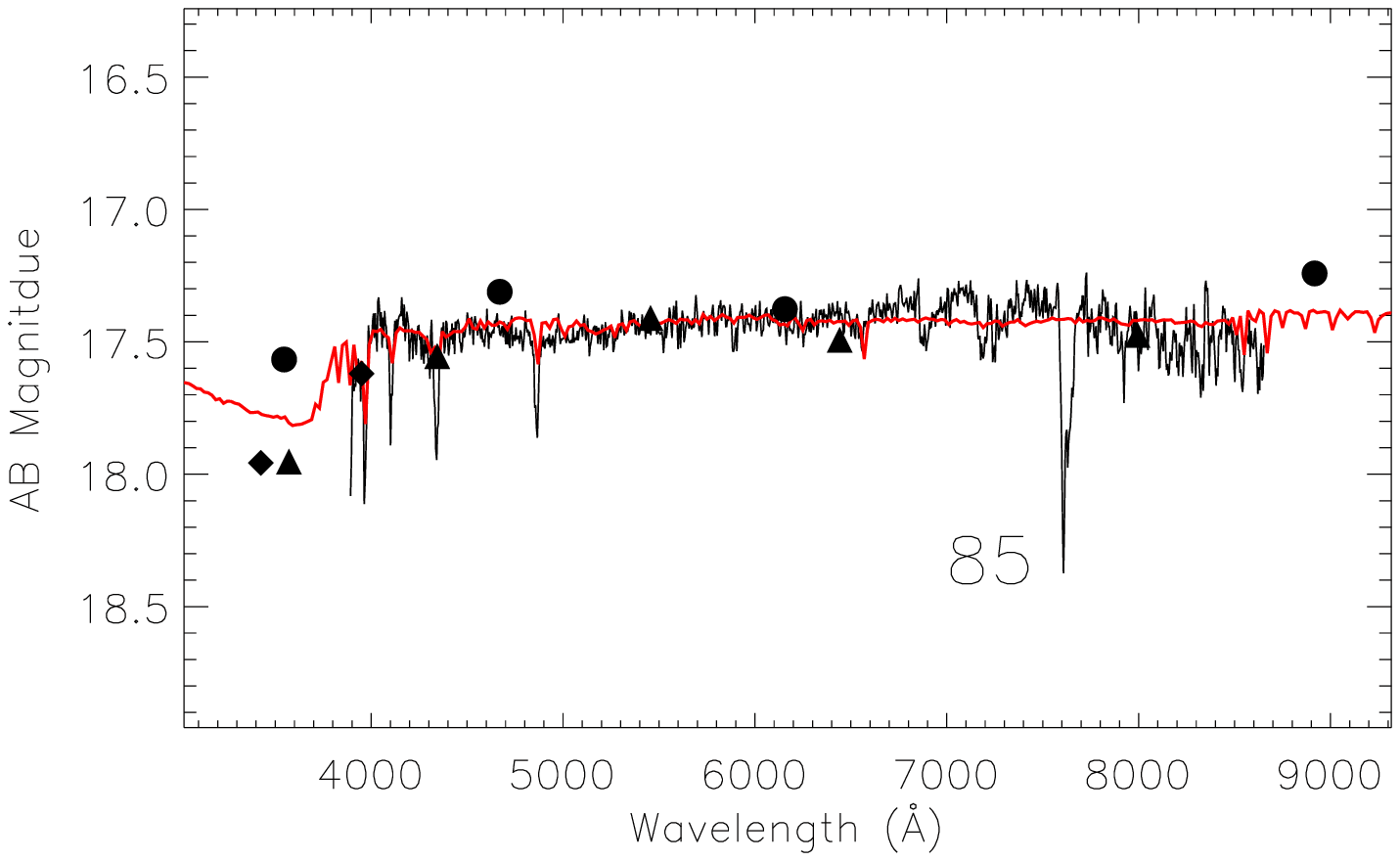}
 \includegraphics[width=.5\linewidth]{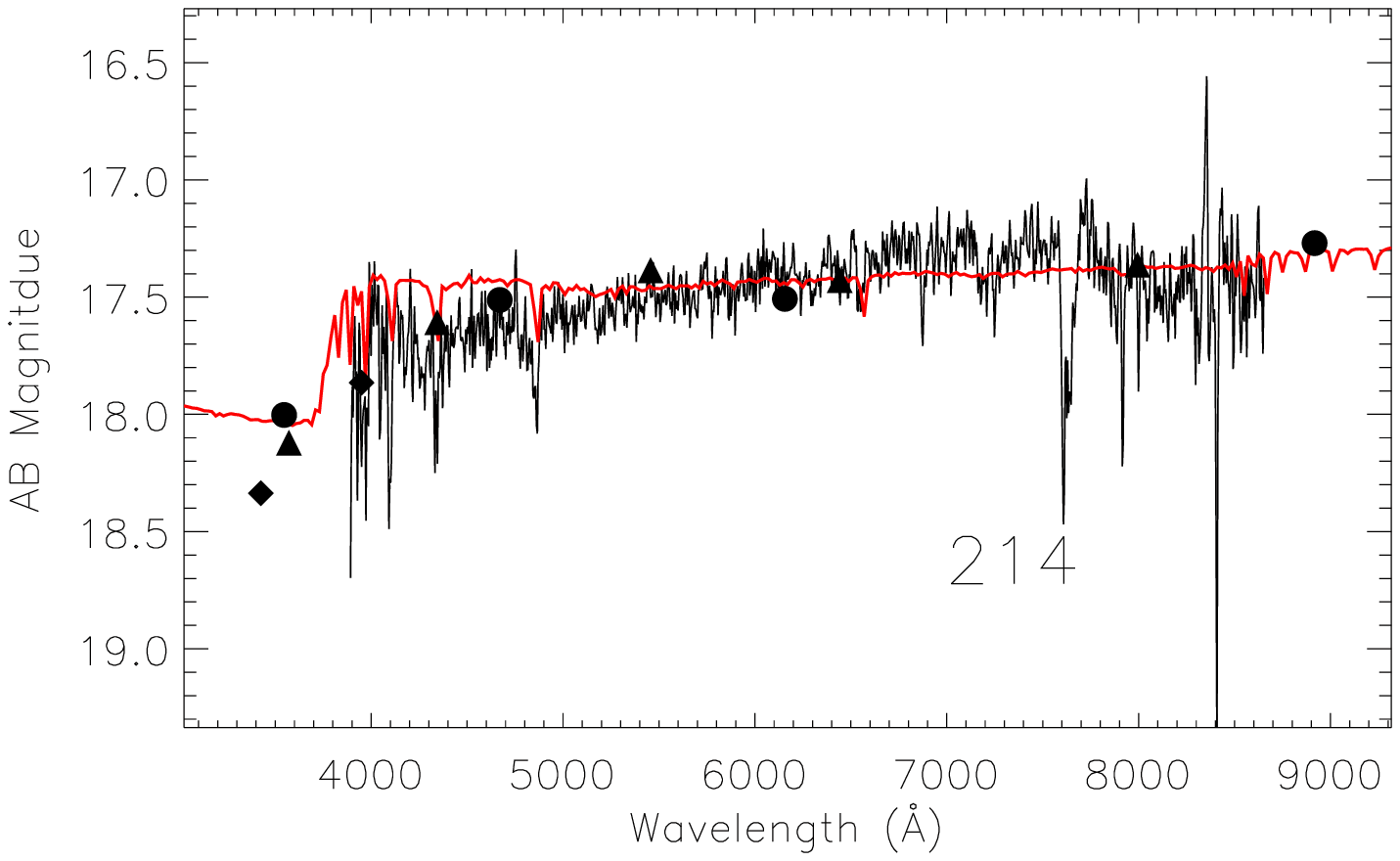}
\end{subfigure}
\caption{-Continued.}
\label{fig13}
\end{figure}
\begin{figure}
\begin{subfigure}
  \centering
  \includegraphics[width=.5\linewidth]{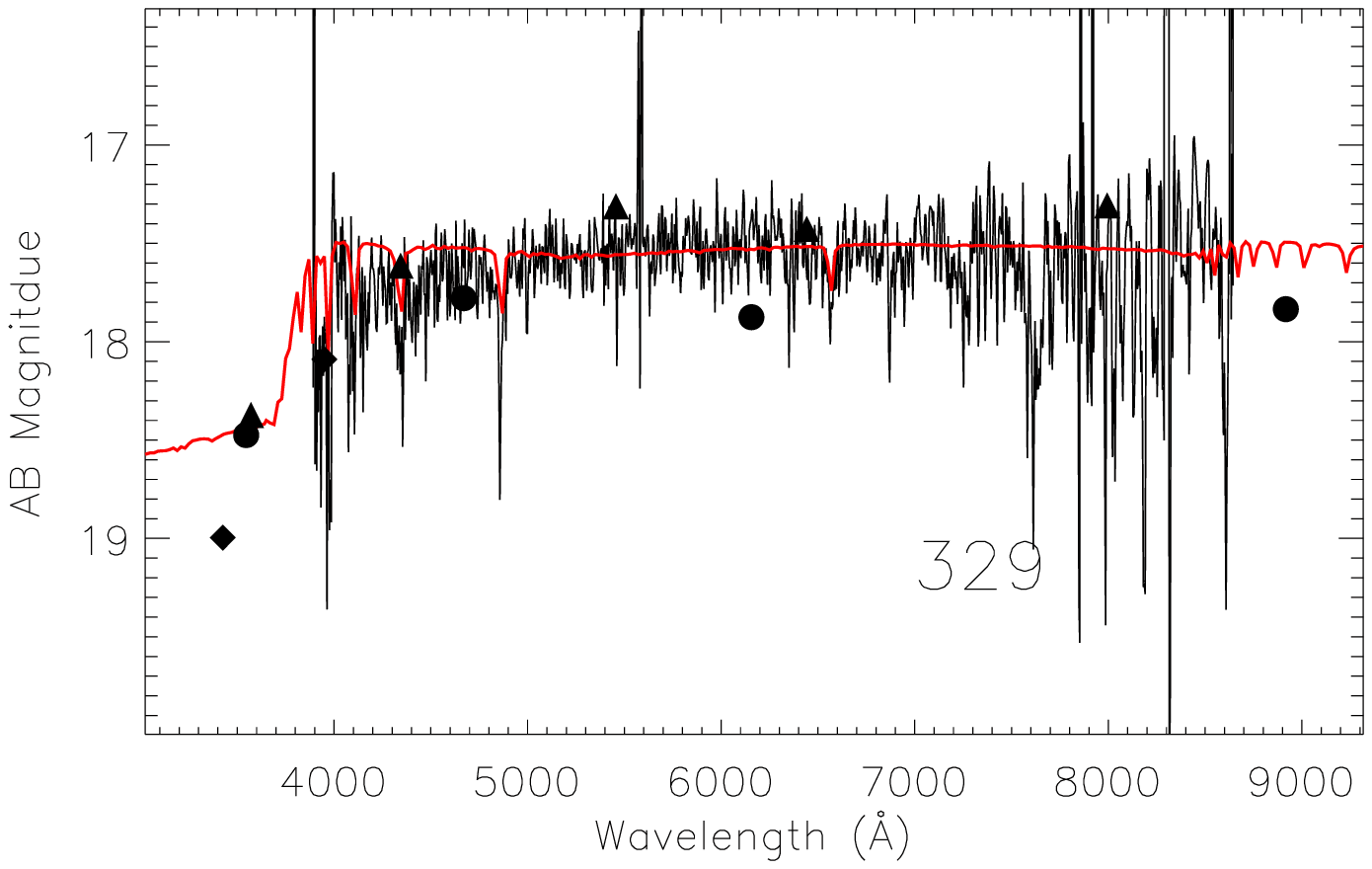}
\end{subfigure}
\caption{-Continued.}
\label{fig14}
\end{figure}

\begin{figure}
  \centering
  \includegraphics[angle=0,scale=0.8]{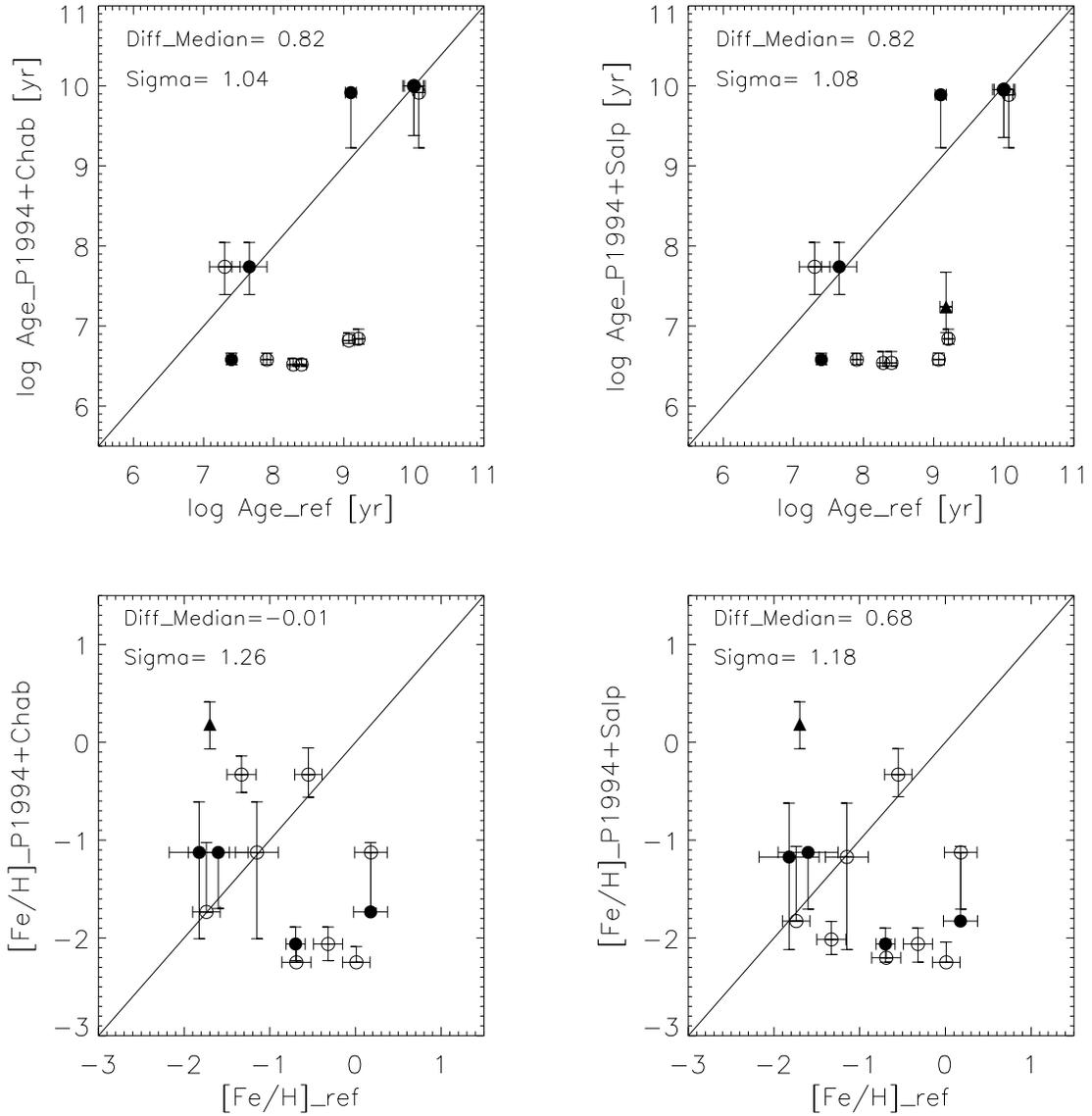}
  \caption{Comparisons of the our spectrum-SED fitting results of
    Table~\ref{t2.tab}, but for the parameters derived with the Padova 1994 track
    + \citet{chab} IMF (left panels) and \citet{salp} IMF (right panels) and
    those from \citet{bea15} (open circles), the filled circles are
      from \citet{fan14} and the filled triangles are from \citet{sha10}.}
  \label{fig15}
\end{figure}

\begin{figure}
  \centering
  \includegraphics[angle=0,scale=0.8]{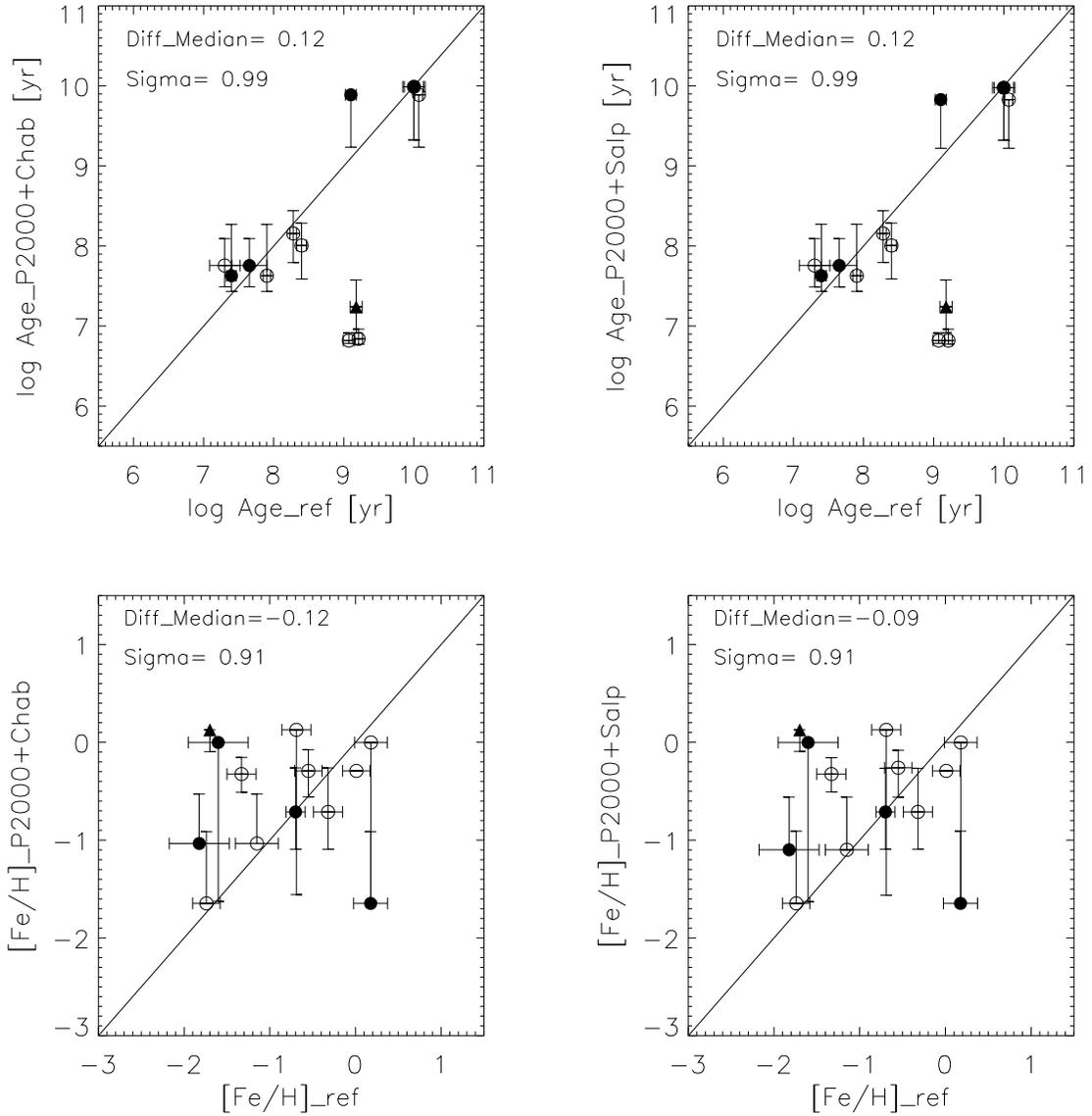}
  \caption{Comparisons of the our spectrum-SED fitting results of
    Table~\ref{t2.tab}, ages and metallicities derived with the Padova 2000
    track + \citet{chab} IMF (left panels) and \citet{salp} IMF  (right panels) and
    those from \citet{bea15} (open circles), the filled circles are
      from \citet{fan14} and the filled triangles are from \citet{sha10}.}
  \label{fig16}
\end{figure}


\begin{figure}
  \centering
  \includegraphics[angle=0,scale=1.2]{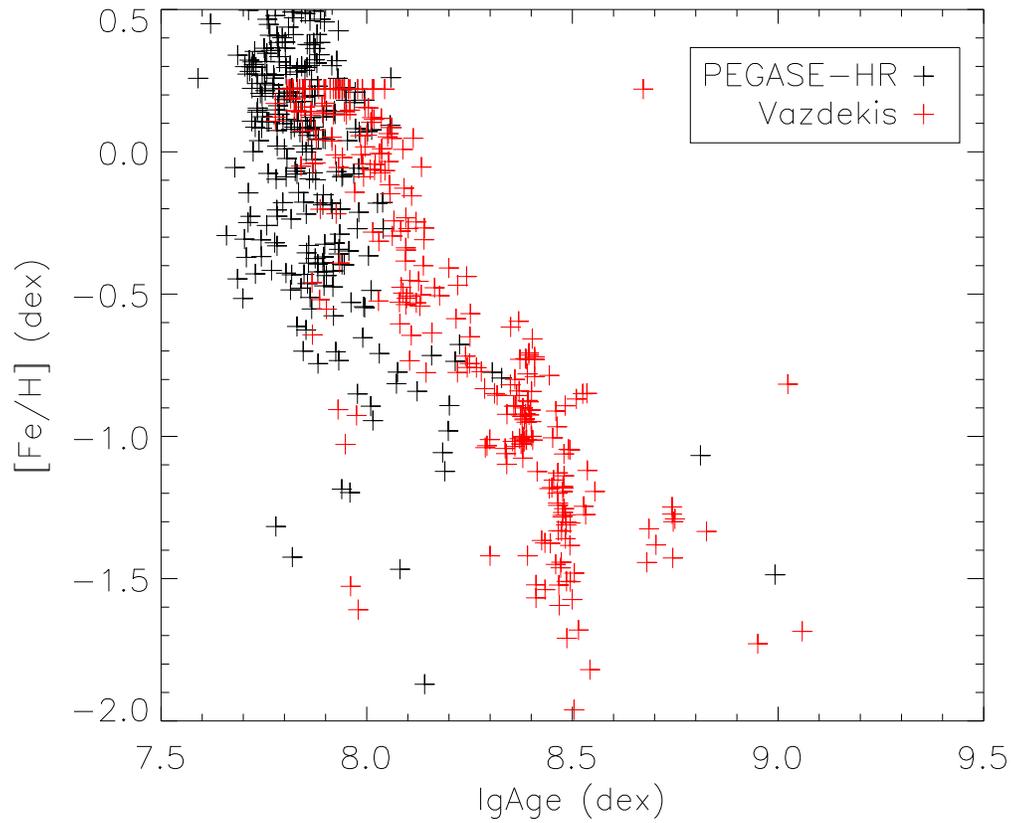}
  \caption{The age-metallicity degeneracy testing in our full-spectrum fitting with
    {\sc pegase-hr} models (black crosses) and Vazdekis models (red crosses).}
  \label{fig17}
\end{figure}

\begin{figure}
  \includegraphics[angle=0,scale=0.5]{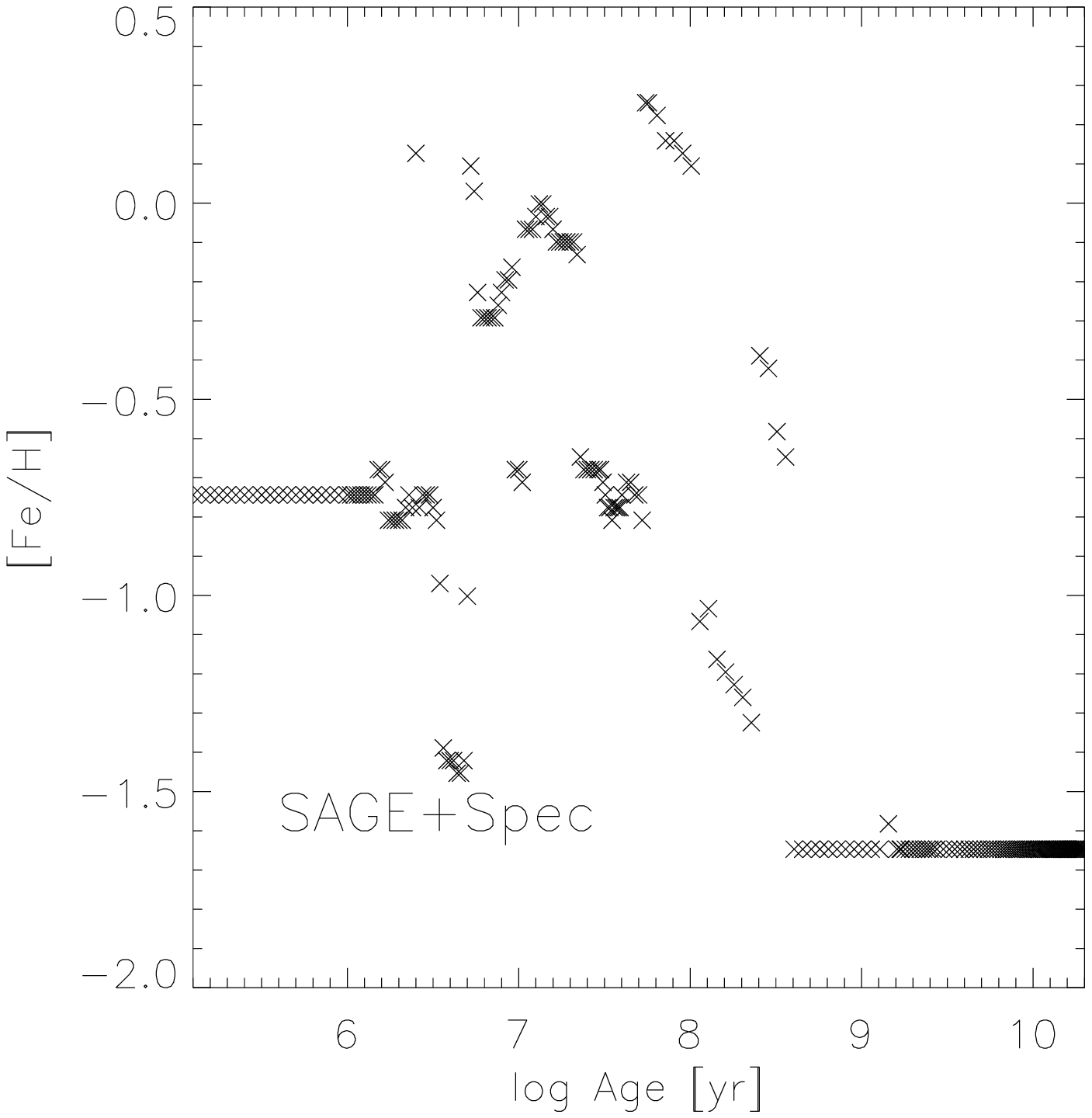}
  \includegraphics[angle=0,scale=0.5]{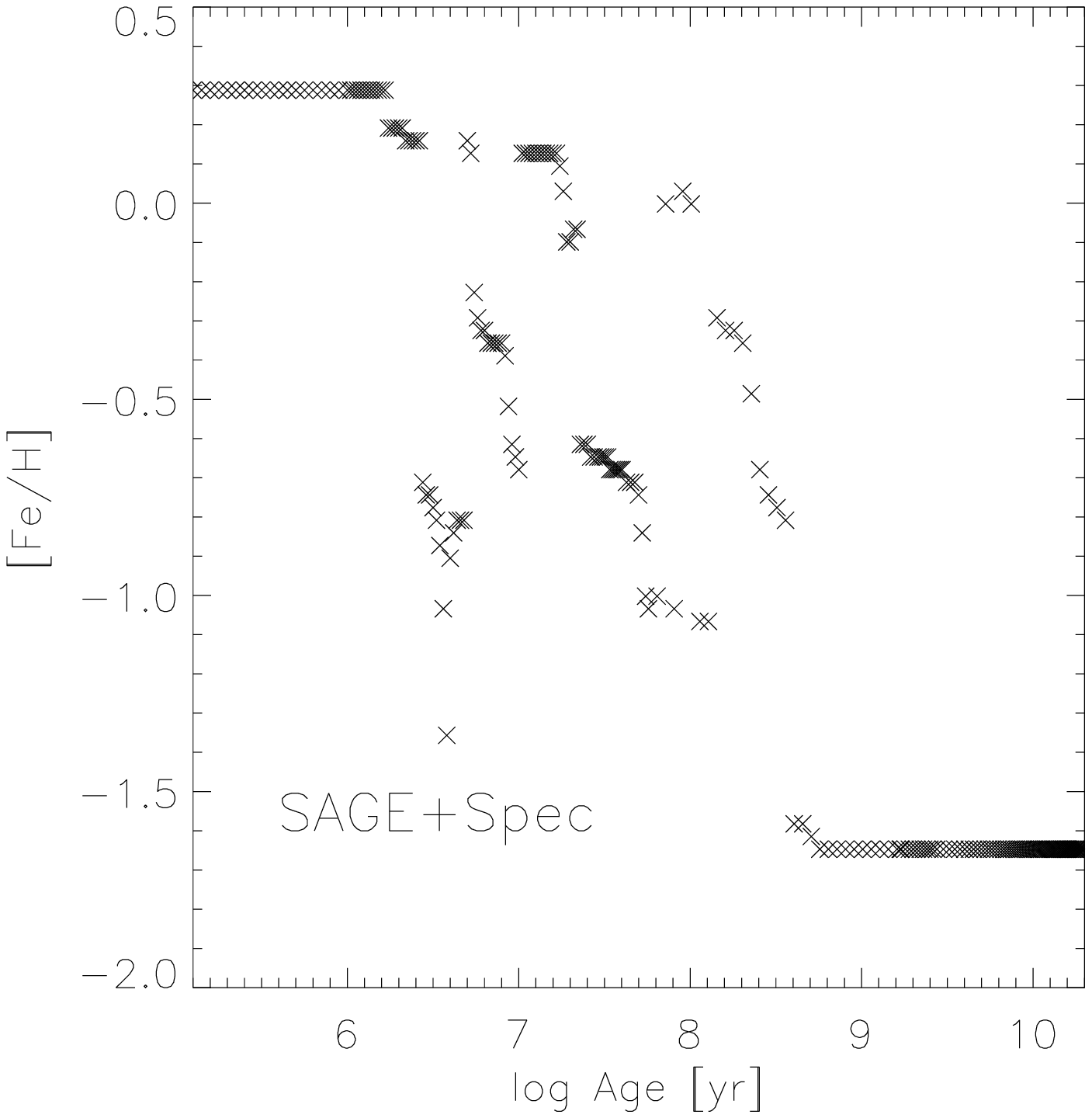}
  \includegraphics[angle=0,scale=0.5]{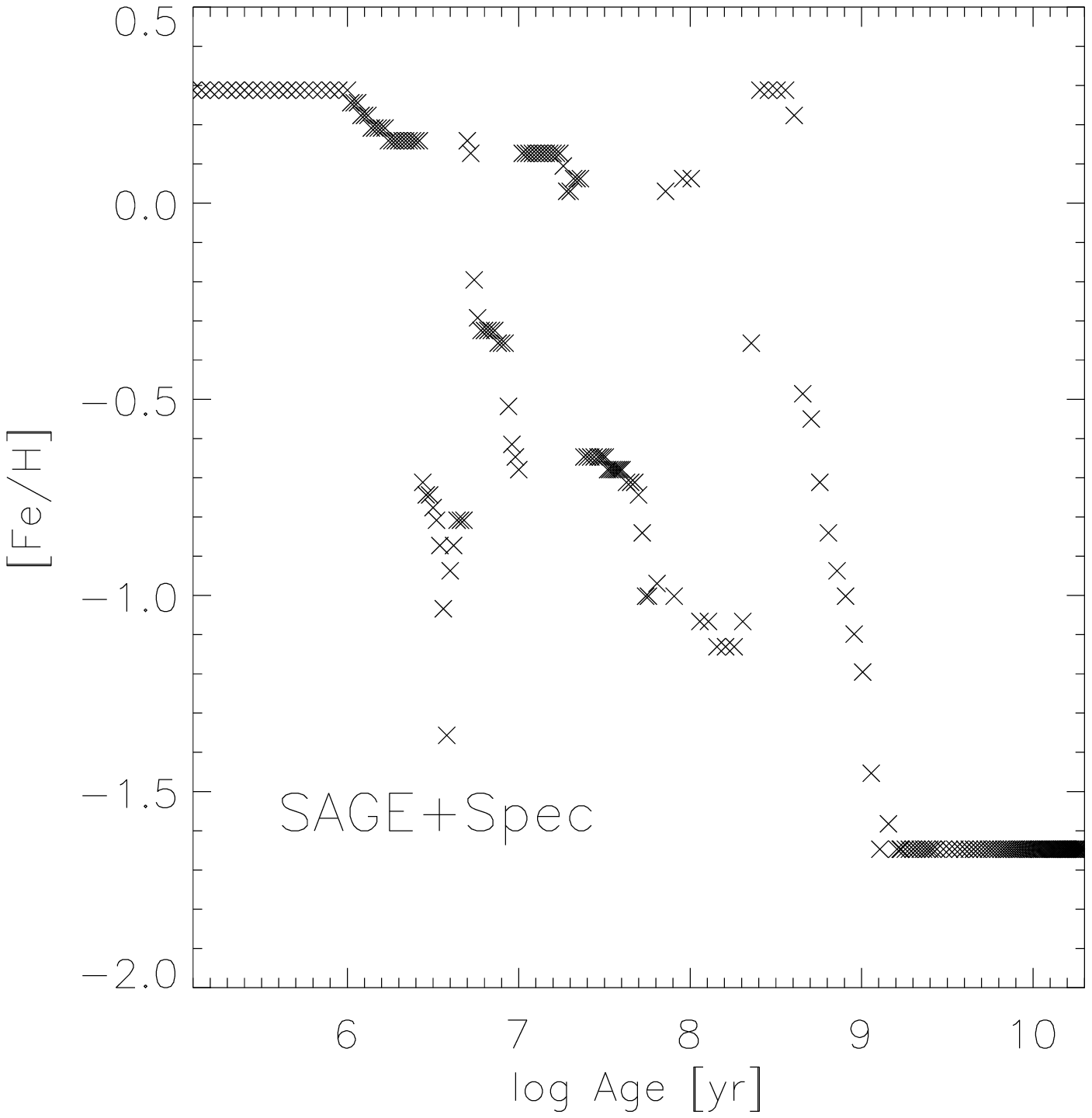}
  \caption{The age-metallicity degeneracy testing in our the fitting with
    Padova 2000 track and \cite{chab} IMF of BC03 models. We show the different
    degeneracy effect in the full-spectrum (top left), SED+spectrum
    (top right) and SAGE photometry+spectrum(bottom left).}
  \label{fig18}
\end{figure}

\begin{figure}
  \centering
  \includegraphics[angle=0,scale=0.75]{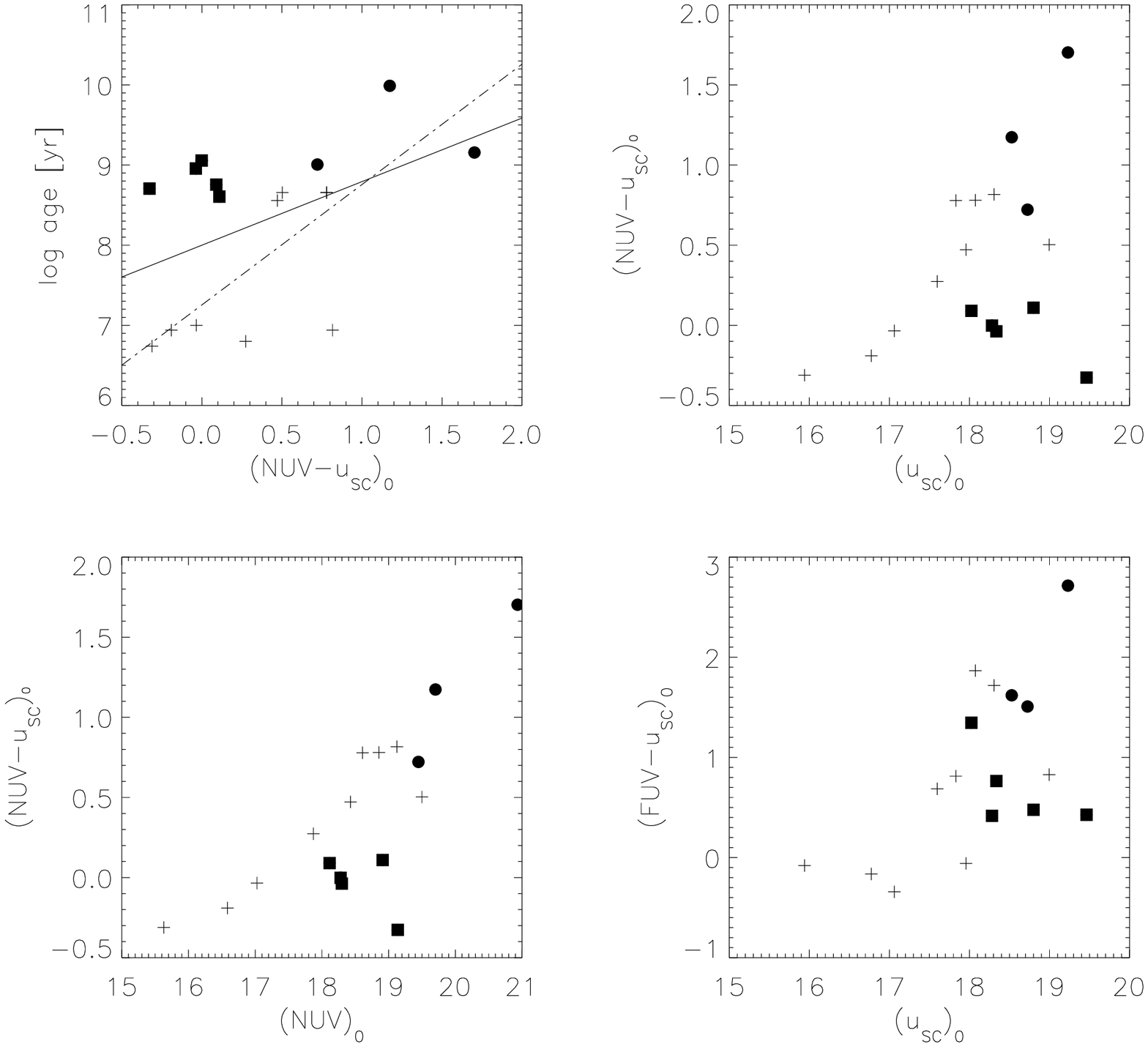}
  \caption{{\it Top left:} The relations between GALEX $\rm (NUV-u_{SC})_0$ color and the
    age (SAGE photometry+observed spectrum, fitted with Padova 2000
    evolutinary track and \cite{chab} IMF of BC03 models) for our sample star
    clusters; {\it Top Right:}  $\rm (NUV-u_{SC})_0$ color versus SAGE $\rm
    (u_{SC})_0$;  {\it Bottom Left:}  $\rm (NUV-u_{SC})_0$ color
    versus GALEX $\rm NUV_0$;
    {\it Bottom Right:} GALEX $\rm 
    (FUV-u_{SC})_0$ color versus SAGE $\rm (u_{SC})_0$. The filled circles are clusters
    older than 1 Gyrs; the black squares (ID: 245, 70, 94, 221 and 214)
    are further away from the bulk correlation which are considered to
    be with different characters with the bulk; the rest are
    crosses. The solid line is the linear fitting for all the data set
    while the dashed-dotted line is the fitting excluded the five
    black squares, which improves the correlation coefficients
    significantly (from 0.44 to 0.76).}.
  \label{fig19}
\end{figure}

\begin{figure}
  \centering
  \includegraphics[angle=0,scale=0.6]{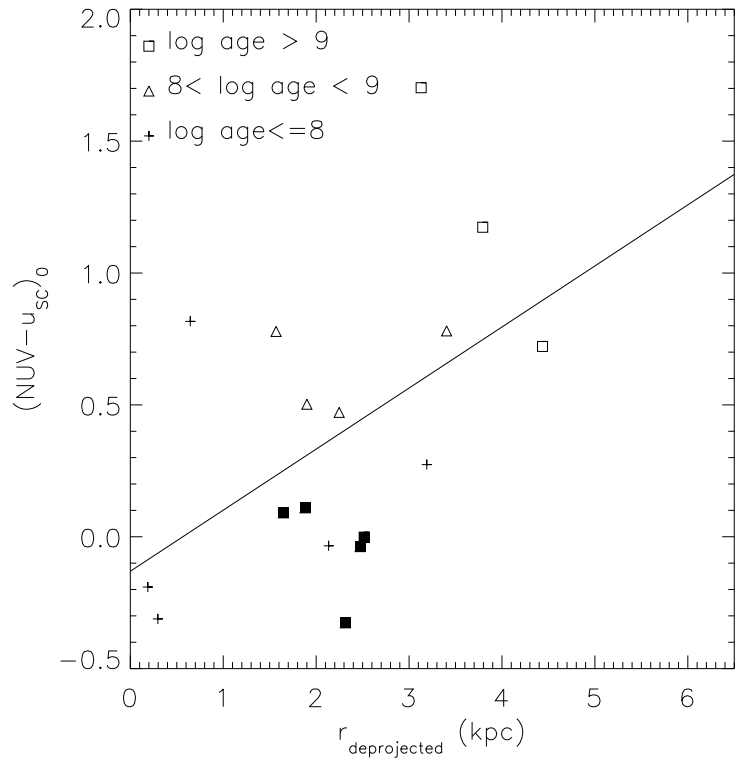}
  \caption{The GALEX $\rm (NUV-u_{SC})_0$ color versus clusters’
    deprojected distance to M33 center. The sample clusters are shown
    with different symbols for different age subgroups. The solid line
    is the linear fitting for all the clusters. The black squares
      (ID: 245, 70, 94, 221 and 214) are the sources further away from the bulk
      correlation in Figure~\ref{fig19}}. 
  \label{fig20}
\end{figure}

\label{lastpage}
\end{document}